\documentclass[a4paper,fleqn,usenatbib]{mnras}

\usepackage{newtxtext,newtxmath}
\usepackage[T1]{fontenc}
\usepackage{ae,aecompl}

\usepackage{graphicx}	% Including figure files
\usepackage{amsmath}	% Advanced maths commands
\usepackage{amssymb}	% Extra maths symbols

\newcommand{\Msun}{M_{\odot}}

\newcommand{\mdot}{\dot{M}}

\newcommand{\ergs}{erg~s$^{-1}$}
\newcommand{\cmmt}{cm$^{-2}$}
\newcommand{\cmt}{cm$^{-3}$}

\newcommand{\Mbh}{\mbox{$M_{\rm BH}$}}

\newcommand{\aion}{\mbox{$\alpha_{\rm ion}$}}

\newcommand{\fnrepeat}[1]{$^{\ref{#1}}$}

\newcommand{\dd}{\textrm{d}}
\newcommand{\amin}{\mbox{$a_{\rm min}$}}
\newcommand{\amax}{\mbox{$a_{\rm max}$}}
\newcommand{\ngas}{\mbox{$n$}}
\newcommand{\mic}{\mbox{$\mu$m}}
\newcommand{\Tsub}{\mbox{$T_{\rm sub}$}}
\newcommand{\tsub}{\mbox{$t_{\rm sub}$}}
\newcommand{\Tgr}{\mbox{$T_{\rm grain}$}}

\newcommand{\HeII}{\mbox{He\,{\sc ii}}}
\newcommand{\NV}{\mbox{N\,{\sc v}}}
\newcommand{\NeVIII}{\mbox{Ne\,{\sc viii}}}
\newcommand{\CIV}{\mbox{C\,{\sc iv}}}
\newcommand{\Tbb}{\mbox{$T_{\rm BB}$}}

\title[The origin of the BLR in AGN]{Dust inflated accretion disc as the origin of the Broad Line Region in Active Galactic Nuclei}

\author[A.~Baskin and A.~Laor]{
	Alexei Baskin\thanks{Contact e-mail: AB --  \href{mailto:alexeiba@soreq.gov.il}{alexeiba@soreq.gov.il}, AL -- \href{mailto:laor@physics.technion.ac.il}{laor@physics.technion.ac.il}}\thanks{Present address: Soreq Nuclear Research Center, Yavne~8180000, Israel}
	and Ari Laor\footnotemark[1]
	\\
	Physics Department, Technion -- Israel Institute of Technology, Haifa~3200000, Israel
}

\date{Accepted XXX. Received YYY; in original form ZZZ}

\pubyear{2017}

\begin{document}
\label{firstpage}
\pagerange{\pageref{firstpage}--\pageref{lastpage}}
\maketitle

\begin{abstract}
The Broad Line Region (BLR) in AGN is composed of dense gas ($\sim 10^{11}$~cm$^{-3}$)
on sub-pc scale, which absorbs about 30 per cent of the ionising continuum. The outer size
of the BLR is likely set by dust sublimation, and its density by the incident radiation pressure compression
(RPC).
But, what is the origin of this gas, and what sets its covering factor (CF)? 
\citet{CH11} suggested that the BLR is a failed dusty wind from the outer accretion disc.
We explore the expected dust properties, and the implied BLR structure. 
We find that graphite grains sublimate only at $T\simeq 2000$~K at the predicted density of 
$\sim 10^{11}$~cm$^{-3}$, and therefore large graphite grains ($\ge 0.3$~\mic) survive down to the observed size of the BLR, $R_{\rm  BLR}$.
The dust opacity in the accretion disc atmosphere is $\sim 50$ times larger than previously assumed, and 
leads to an inflated torus-like structure, with a predicted 
peak height at $R_{\rm  BLR}$. The illuminated surface of this torus-like
structure is a natural place for the BLR.
The BLR CF is mostly set by the gas metallicity, the radiative accretion efficiency, a dynamic configuration,
and ablation by the incident optical-UV continuum.
This model predicts that the BLR should extend inwards of $R_{\rm  BLR}$ to the disc radius where the 
surface temperature is  $\simeq 2000$~K, which occurs at $R_{\rm in}\simeq 0.18 R_{\rm  BLR}$. 
The value of $R_{\rm in}$ can be tested by reverberation mapping of the higher 
ionisation lines, predicted by RPC to peak well inside $R_{\rm  BLR}$. 
The dust inflated disc scenario can also be tested based on the predicted response of $R_{\rm  BLR}$
and the CF to changes in the AGN luminosity and accretion rate.
\end{abstract}


\section{Introduction}\label{sec:intro}

The AGN characteristics of UV peaked continuum emission, with broad emission lines, 
is found over a vast range of bolometric luminosity, $L_{\rm bol}$, from 
$\sim~10^{40}$~erg~s$^{-1}$, observed in the nearest AGN 
(e.g.\ NGC~4395, \citealt{Moran99}), to $\sim 10^{48}$~erg~s$^{-1}$ in AGN at a redshift of $z\sim 3$ \citep{Richards06}. 
Why is accretion onto a massive Black Hole (BH) commonly associated with broad emission lines?
Furthermore, why are these broad lines always associated with gas at an ionisation parameter of $U\sim 0.1$,
a typical density of $n\sim 10^{11}$~cm$^{-3}$, and a covering factor CF $\sim 0.3$?\footnote{CF is the fraction of the ionising continuum converted to line emission. It equals the geometric CF, if the ionising radiation is isotropic.} Part of these questions 
may have been answered. A value of $U\sim 0.1$ is naturally expected for photoionized gas compressed
by the incident AGN radiation pressure (\citealt*{paperII, Stern16}). At $U\sim 0.1$, dust embedded in the 
photoionized gas suppresses heavily
the line emission, and most of the ionising radiation is absorbed by the dust and reemitted in the IR 
(\citealt{LaorDraine93, NetzerLaor93, Ferguson97}; \citealt*{paperI}, \citealt{paperII}). However, 
since no dust grains survive below 
\begin{equation}
R_{\rm sub}=0.2L_{46}^{1/2}~{\rm pc}, 
\end{equation}
where $L_{\rm bol}=10^{46}L_{46}$~erg~s$^{-1}$ (\citealt{LaorDraine93}; see refined calculations below), 
the dust sublimates and the gas becomes an efficient line emitter, only at $R<R_{\rm sub}$. Thus, even if photoionized gas is present on all scales in AGN, 
efficient line emission is found only at $R<R_{\rm sub}$. Indeed, 
reverberation mapping (RM), which measures directly the size of the 
Broad emission Line Region (BLR), $R_{\rm BLR}$, yields 
\begin{equation}
R_{\rm BLR}=0.1L_{46}^{1/2}~{\rm pc}, 
\label{eq:R_BLR}
\end{equation}
over the entire observed range of AGN luminosities, $L_{\rm bol}=10^{40} - 10^{48}$~erg~s$^{-1}$ (\citealt{Kaspi05, Kaspi07}).
Thus, dust sublimation provides a natural mechanism for the outer radius of the BLR.
RM of the near-IR K band emission confirms the presence of hot dust at
$\sim 2R_{\rm sub}=4R_{\rm BLR}$ (\citealt{Suganuma06, Koshida14}), validating that the BLR is bound by hot dusty 
gas.\footnote{We use the term dusty gas, rather than dust, as dust is always associated with gas.} 

The universal density of the gas at the BLR, irrespective of luminosity, is naturally explained by 
the radiation pressure compression of gas at the dust sublimation radius.
The required radiative flux in order to sublimate a spherical blackbody grain, is 
$F= 4\sigma \Tsub^4$, (the factor of 4 is the emitting/absorbing area). 
The grain sublimation temperature at the BLR density is
$\Tsub\simeq 2000$~K (see Section~\ref{sec:dust_prop}), which is achieved when the incident flux is $F=3.63\times 10^9$~erg~s$^{-1}$cm$^{-2}$. The radiation pressure at $R_{\rm sub}$ is therefore at a fixed
value of
\begin{equation}
F(R_{\rm sub})/c=0.12~T_{2000}^4~{\rm erg~cm}^{-3},
\end{equation}
where $T_{2000}\equiv \Tsub/2000$~K. 

Compression of the gas by the incident radiation pressure (RPC) gives $P_{\rm gas}=P_{\rm rad}$, i.e.
\begin{equation} 
2nkT= F/c
\end{equation}
where $n$ is the ionised gas electron density, and $T$ its temperature (e.g.\ \citealt{paperI}). 
Photoionized 
gas is typically at $T\sim 10^4$~K, so its density is $n=F/2kTc \sim {\rm few}\times 10^{10}$~cm$^{-3}$. 
This naturally explains the typical value of $n$ at $R_{\rm BLR}$, irrespective of the AGN luminosity.
So, dust sublimation naturally gives $R_{\rm BLR}$, and RPC then gives $U$ and $n$. But, where does the BLR gas come from?  and why does it extend over a CF$\sim 0.3$?

Careful simultaneous modelling of RM results and the gas kinematics suggests the BLR lies in a geometrically thick disc configuration, which is optically thick for the H$\beta$ line (\citealt{Pancoast14, Grier17}).

Since AGN are powered by accretion of gas, most likely in a thin disc configuration (e.g., \citealt{Shields78, Capellupo16}), the accretion disc (AD) is the most plausible source for the BLR gas. But, what mechanism lifts
the gas from the disc to produce the required CF of the BLR?
A natural mechanism is the underlying disc radiation pressure. This led to the suggestion of a UV-driven disc outflow 
(e.g., \citealt*{Shlosman85, Murray95, Proga04}).
However, the required disc launching radius for UV driving is $\sim 10^2$ gravitational radii, $R_{\rm g}=GM/c^2$, while the range
of broad line widths of $v=$1000-10,000~km~s$^{-1}$ indicates $R_{\rm BLR}\sim 10^3-10^5R_{\rm g}$ (in Kepllerian motion
$v/c=\sqrt{R_{\rm g}/R}$). 
Alternatively, magnetic fields may be able to lift the gas, by either launching and accelerating 
a wind (\citealt*{Emmering92, Lovelace98}), launching a disc outflow
which is then driven by radiation pressure \citep{Konigl94}, or by the simultaneous effect of both radiation pressure
and a magnetocentrifugal torque (\citealt{Everett05, Keating12}). However, given the somewhat divine attributes of magnetic fields (being omnipotent and working in mysterious ways), their significance cannot yet be firmly established. 

An interesting new suggestion was made by \citet[hereafter CH11]{CH11}, who suggested that the disc 
outflow is launched by radiation pressure on a dusty disc atmosphere. This naturally solves the compact size problem
of the disc UV launching mechanism, for the following reason. The disc surface temperature scales as $R^{-3/4}$. 
The required disc surface temperature for a dusty wind is $\sim 25$ times smaller compared to a UV line driven wind 
($\sim 2000$~K versus $\sim 50,000$~K). As a result,  
the implied radius is 
$\sim 25^{4/3}\sim 70$ times larger, i.e.\ $\sim 10^4R_{\rm g}$ instead of $\sim 10^2R_{\rm g}$, consistent 
with $R_{\rm BLR}$. 
At a small enough radius the disc atmosphere becomes too hot to allow grains to survive, and this radius 
will form the innermost radius of the BLR, $R_{\rm in}$, as a dusty wind cannot be launched at $R<R_{\rm in}$.
As discussed by CH11 (see below), AD models give $R_{\rm in}\propto L^{1/2}$, and an absolute value which is
a few times smaller than $R_{\rm sub}$. 

Thus, a dusty wind is launched at $R>R_{\rm in}$, and if in addition $R<R_{\rm sub}$,
the dust sublimates once it gets exposed to the ionising radiation, the
gas becomes an efficient line emitter, and the observed BLR emission is produced.
In contrast, dusty wind launched at $R>R_{\rm sub}$ remains dusty, an inefficient line emitter, 
which radiates mostly in the IR \citep{paperI}. At $R_{\rm in}< R <R_{\rm sub}$, once the dust 
sublimates, the dust driving by radiation pressure stops, and the outflow 
may fall back to the disc, forming a failed wind (CH11).

The same mechanism of dust sublimation therefore sets both the inner and outer radii of the BLR. 
Dust is required in the disc atmosphere to feed gas into the BLR, and its sublimation there 
forms the innermost BLR radius. Once the dusty gas is exposed to the
ionising radiation, the dust needs to sublimate, to allow efficient line emission, 
and the largest radius where this sublimation occurs sets the 
outer radius of the BLR.

The purpose of this paper is to explore whether the radiation pressure driving,  
provided by the local disc emission, is large enough
to produce the required CF $\sim 0.3$. This value is suggested by the typical equivalent width (EW) of the BLR emission 
lines (\citealt{Korista97, Maiolino01, Ruff12, paperII}, section 3.2.1 there). 
The vertical extent of the dusty atmosphere depends on the dust opacity, which is set by the local grain properties (grain size distribution and composition),
and the peak wavelength of the local disc emission. Both properties change with radius. 
The advantage of the proposed mechanism 
is that it can be calculated from first principles. Our purpose here is to explore the derived value of 
the CF, given the AD parameters, the gas metallicity, and the grain properties set
by the local ambient conditions. 

As we show below, the dust opacity leads to an inflated disc structure, which forms a torus-like BLR. The illuminated face of the inflated structure
forms the BLR, and the back side forms a dusty torus. The structure and emission properties of an obscuring dusty 
torus received much attention (see a thorough review in the introduction of \citealt{Chan16}).
These torus models are all of an externally illuminated structure, in contrast with the structure discussed here, which
is vertically supported by the underlying local AD emission. The torus structure, derived here, resides at about the
dust sublimation radius, and produces very hot dust emission at $\lambda\simeq 2-3$~\mic. This is in contrast with the typical 
torus models, which are placed at a larger range of radii, selected to produce the bulk of the IR emission at
3 -- 30~\mic. The IR observations suggest two distinct continuum emission components in the IR. One
producing a 3~\mic\ bump, which corresponds to dust close, or at, the sublimation temperature, and another component 
which peaks at $\lambda >10$~\mic\ (e.g., \citealt{Wills87, Mor11, Hernan16, Symeonidis16}). These
components were interpreted by \citet{Mor12}
as the hot pure-graphite dust which is just outside the BLR, and the clumpy torus
and narrow line region (NLR) dust, which reside on larger scales and produce the mid-IR emission. 
Our study here of the dust inflated AD is relevant only for the hot, blackbody like, pure-graphite 
dust component suggested by \citet{Mor12}.

We note in passing that a similar inner rim of sublimating dust, is predicted and 
possibly observed, in nearby circumstellar and protoplanetary discs (e.g., \citealt*{Dullemond01, Natta01, Isella05}).

The paper is organised as follows. In Section~\ref{sec:analytic_estimates} we provide analytic estimates of the size and CF of a BLR produced by a dust inflated disc structure, in Section~\ref{sec:dust_prop} we calculate the effects of sublimation on the dust properties, and in Section~\ref{sec:solutions} we calculate
the implied CF. The results of the calculations and their implications are discussed in Section~\ref{sec:discussion}, and the conclusions are summarised in Section~\ref{sec:conclusions}.

\section{Some analytic estimates} \label{sec:analytic_estimates}
\subsection{The inner and outer radius of the BLR}

Below we compare the ratio of the inner BLR radius $R_{\rm in}$, set by dust sublimation in the
disc atmosphere, to the outer BLR radius, $R_{\rm out}$, set by the sublimation of the largest 
dust grains exposed to the luminosity of the central source. 

The effective temperature $T_{\rm eff}$ at the disc surface at a radius $R$ satisfies
\begin{equation}
\sigma T_{\rm eff}^4 = \frac{3}{8\pi}\frac{GM\mdot}{R^3} , 
\label{eq:flux1}
\end{equation}
where $M$ is the BH mass, and $\mdot$ is the accretion rate. This gives in convenient units
\begin{equation}
T_{\rm eff}(R)=88 (M_8\mdot_1)^{1/4}R_{\rm pc}^{-3/4}~{\rm K} \ ,
\label{eq:T_R}
\end{equation}
and
\begin{equation}
R(T_{\rm eff})=0.016 (M_8\mdot_1)^{1/3}T_{2000}^{-4/3}~{\rm pc},
\label{eq:R_2000}
\end{equation}
where $M_8$ is the BH mass in units of $10^8M_{\odot}$, $\mdot_1$ is the accretion rate in units
of $M_{\odot}~{\rm yr}^{-1}$  ($6.3\times 10^{25}$~gr~s$^{-1}$), $R_{\rm pc}$ is the radius in pc,
and $T_{2000}=T_{\rm eff}/2000$~K.
The value of $M\mdot$ can be measured directly from the observed luminosity density, through the relation
\begin{equation}
M_8\mdot_1=1.4L_{\rm opt, 45}^{3/2}\ ,
\label{eq:mmdot}
\end{equation}
where $L_{\rm opt, 45}$ is $\lambda L_{\lambda}$ at 4861\AA\ in units of $10^{45}$~erg~s$^{-1}$ (e.g., \citealt{DavisLaor11}). 
As we show below, dust can survive down to a radius where $T\simeq 2000$~K. Therefore, 
the innermost disc radius which can have a dusty atmosphere, $R_{\rm in}=R(T_{2000}=1)$, 
is given by
\begin{equation}
R_{\rm in}=0.018L_{\rm opt, 45}^{1/2}~{\rm pc} .
\label{eq:Rin}
\end{equation}
This radius is a function of $L$ only, and scales as $L^{1/2}$, just as $R_{\rm BLR}$ 
(eq.~\ref{eq:R_BLR}).
Since $L_{46}\simeq L_{\rm opt, 45}$ 
\citep{Richards06}, we get a fixed ratio
\begin{equation}
R_{\rm in}/R_{\rm BLR}= 0.18.
\label{eq:RinRblr}
\end{equation}

The sublimation radius of dust above the AD, $R_{\rm out}$, is estimated assuming the dust is exposed to the integrated disc emission, $L_{\rm bol}$, and that the emission is isotropic.
We also assume the dust absorption and emission efficiencies are unity; and that the dust grains are spherical, illuminated from one side, isothermal, and radiate isotropically. The dust temperature is then of a blackbody, $T_{\rm BB}$, which satisfies
\begin{equation}
\sigma T_{\rm BB}^4 = \frac{L_{\rm bol}}{16\pi R_{\rm out}^2}~. 
\label{eq:flux2}
\end{equation}
The effects of a non isotropic $L_{\rm bol}$, and of the grain size dependence of the absorption and 
emission efficiencies are discussed in Section~\ref{sec:dust_prop}, where we show that the blackbody approximation is valid for large enough grains. The above expression yields
\begin{equation}
R_{\rm out}=0.15 L_{46}^{1/2}T_{2000}^{-2}~{\rm pc} ,
\label{eq:Rout}
\end{equation}
and for T$_{\rm eff}=2000$~K, we get the following ratio, 
\begin{equation}
R_{\rm out}/R_{\rm BLR}= 1.6\ .
\end{equation}
We note that the commonly used value for the dust sublimation radius is 
$R\simeq 1.3 L_{46}^{1/2}$~pc (\citealt{Barvainis87}), which is almost a factor of 10 larger than $R_{\rm out}$. 
This results from the combined effect of using small graphite grains (0.05~$\mu$m), which are significantly hotter than a blackbody; and from the use of a lower sublimation temperature of 1500~K, which is relevant for densities much lower than the BLR gas
densities (see Section~\ref{sec:dust_prop}). 
 
The above derivation implies that $R_{\rm out}/R_{\rm in}\sim 10$, irrespective of the
system parameters. However, the relation $L_{\rm bol}\simeq 10L_{\rm opt}$, or
$L_{46}\simeq L_{\rm opt, 45}$, is valid only on average \citep{Richards06}. 
This relation is also not generally valid for simple thin AD model spectral energy distributions (SEDs)
\citep{DavisLaor11}, which predict ratios in the range of $\sim 1-100$.

It is therefore useful to estimate  
$R_{\rm out}/R_{\rm in}$ independently of the relation between $L_{\rm bol}$ and $L_{\rm opt}$. 
Equating the sublimating illumination flux (eq.~\ref{eq:flux2}), to the sublimating local disc
flux (eq.~\ref{eq:flux1}), and further using the relation 
\begin{equation}
L_{\rm bol}=\epsilon \mdot c^2 ,
\end{equation}
where $\epsilon$ is the accretion radiative efficiency, gives
\begin{equation}
\frac{\epsilon \mdot c^2}{16\pi R_{\rm out}^2}  =\frac{3}{8\pi}\frac{GM\mdot}{R_{\rm in}^3}\ .
\end{equation}
If the value of $\mdot$ on both sides is the same (see discussion below), we get, 
\begin{equation}
\frac{R_{\rm out}}{R_{\rm in}}=\sqrt{\frac{\epsilon R_{\rm in}}{6R_{\rm g}}}\ .
\end{equation}
Using typical values $M_8=\mdot_1=1$, $\epsilon\simeq 0.1$, the expression for $R_{\rm in}$ (eq.~\ref{eq:R_2000}), 
and 
\begin{equation}
R_{\rm g}=GM/c^2=4.79\times 10^{-6}M_8~{\rm pc}~\ ,
\end{equation}
gives $R_{\rm out}/R_{\rm in}=7.46$, comparable to the ratio of $1.6/0.18=8.9$ derived above. 

The above expression for $R_{\rm out}/R_{\rm in}$ does not assume 
a specific $L_{\rm bol}/L_{\rm opt}$, and
implies that if $R_{\rm in}<6\epsilon^{-1} R_{\rm g}$, then $R_{\rm out}<R_{\rm in}$, which essentially
implies the BLR will not be able to form. 
Specifically, the innermost disc radius where a dusty wind can be 
launched, is larger than the dust sublimation radius outside the disc. The dusty wind will remain dusty,
even at the smallest radius where it is launched. The gas will not become an efficient line emitter, and the 
BLR will not form. 
It also implies that in
low $\epsilon$ systems, and in systems where $R_{\rm in}/R_{\rm g}$ is small (i.e.\ relatively
cold AD), the BLR is expected
to span a smaller range of radii. Both prediction are further discussed below (Section~\ref{sec:discussion}).

\subsection{The expected CF of the BLR} \label{sec:sub:expected_CF}

Below we estimate the expected maximal height that static dusty disc atmosphere can reach, 
when it is supported vertically by the local AD radiation pressure. 

A radiative flux $F$, exerts a pressure $F/c$, and provides
an acceleration 
\begin{equation}
a_{\rm rad}(R)=\frac{F\kappa}{c} 
\end{equation}
on material with opacity $\kappa$. The radiative force is countered by the vertical
component of gravity, which provides an acceleration of 
\begin{equation}
a_{\rm BH}(R,z)=GM\frac{z}{(R^2+z^2)^{3/2}}\ , 
\label{eq:a_gz}
\end{equation}
where $z$ is the height above the disc plane. Below we make the approximation that  
$z/R\ll 1$, which gives $a_{\rm BH}(z)\simeq GMz/R^3$. A static solution, 
i.e.\ $a_{\rm rad}=a_{\rm BH}$,  yields a height
\begin{equation}
H=\frac{F\kappa}{c}\times \frac{R^3}{GM}.
\end{equation}
The local AD flux is
\begin{equation}
F=\frac{3}{8\pi}\frac{GM\mdot}{R^3} ,
\label{eq:ADflux}
\end{equation}
which implies 
\begin{equation}
H=\frac{3}{8\pi}\frac{\mdot \kappa}{c} .
\label{eq:H_mdot}
\end{equation}
The disc height is set by $\mdot$ and $\kappa$ only, with no direct dependence on $R$ and $M$. 
This solution is the same as the
\citet{SS73} inner disc solution, where radiation pressure and electron scattering dominate.
However, in contrast with the electron scattering opacity of ionised gas, 
$\kappa=0.4$~cm$^2$~gr$^{-1}$, which is a constant, here $\kappa$ is the dust opacity, which rises with frequency. This leads to a rising $H$ with decreasing $R$, as the disc
gets hotter, and the peak emission frequency rises.
The hottest dust is expected to reach $T\sim 2000$~K before sublimation, and the corresponding Planck
mean opacity for blackbody emission at this temperature is $\kappa \sim 50$~cm$^2$~gr$^{-1}$ (see below).
The largest $H$ is therefore   
\begin{equation}  
H=4.06\times 10^{-3}\kappa_{50}\mdot_1~{\rm pc} ,
\label{eq:H1}
\end{equation}
where $\kappa=50\kappa_{50}~{\rm cm}^2{\rm gr}^{-1}$. Using the value of $R_{\rm in}$ (eq.~\ref{eq:Rin}),
gives
\begin{equation}  
H/R=0.23\kappa_{50}\mdot_1 L_{46}^{-1/2}\ .
\label{eq:CF}
\end{equation}
Adopting
\begin{equation}
L_{46}=5.67\epsilon \mdot_1\ ,
\label{eq:epsilon} 
\end{equation}
which applies if $\mdot$ in the BLR and at the centre are the same,
we get
\begin{equation}  
H/R=0.041\kappa_{50}\epsilon^{-1}L_{46}^{1/2}\ .
\label{eq:CF1}
\end{equation}
So, for a typical quasar with $L_{46}=1$, and a plausible $\epsilon\simeq 0.1$, 
we get $H/R\simeq 0.4$, comparable to the value required to get CF $\sim 0.3$.

The above estimate of $H/R$ ignores some critical effects. The illumination,
and consequent sublimation, of the dust by the central continuum source,
which reduces $H$ significantly. The bulk of the BLR gas is located at $5.5R_{\rm in}$ 
(eq.~\ref{eq:RinRblr}), rather than at $R_{\rm in}$, which reduces $H/R$. 
In addition, the accretion disc at $R_{\rm BLR}$ is colder than at
$R_{\rm in}$, leading to a lower $\kappa$, and thus a lower $H$. 
These effects lead to a significantly lower CF. 

However, other effects increase the CF. 
First, the above estimate for $\kappa$ is for Solar metallicity
gas ($Z=Z_\odot$). Various line ratios from the BLR indicate $Z/Z_\odot\sim 2{\rm -}10$  
(\citealt{HamannFerland99, Dietrich03, Simon10}). Second, as pointed out by CH11, the underlying disc radiation pressure may lead to a dynamic failed wind solution, rather than a static solution.
A failed dusty wind reaches a height where its vertical velocity is zero, 
which is higher than the static solution height, where the acceleration is zero. 
Third, the incident central source continuum can ablate the surface layer of the disc 
and also produce a sheared layer wind (\citealt*{Pier95, Namekata14}).
Fourth, further vertical support may be provided by the radiative transfer of the 
incident ionising continuum, converted to  
thermal IR emission, which also diffuses upwards through the torus, increasing the vertical component
of the IR radiation pressure.
This is the mechanism which is generally invoked to provide the vertical support of a dusty torus, 
which resides at $R>R_{\rm BLR}$ (e.g., \citealt{Pier92, Honig06, Krolik07}; \citealt*{Dorod11}; \citealt{Wada12, 
Dorod12, Schartmann14, Chan16}).  

A dynamic solution for the BLR line emitting gas is indicated from a different argument.
In the static torus case, the BLR velocity field is purely circular. Such a velocity field 
produces double peaked emission line profiles. Such profiles are seen only when the
BLR emission line profiles are extremely broad (${\rm FWHM}>10,000$~km~s$^{-1}$, e.g., \citealt{Eracleous03}).
In most objects the line profiles are singly peaked, indicating a significant non-Keplerian contribution to the velocity field, which smoothes the emission line profile 
(e.g., \citealt*{Flohic12, Landt14, Pancoast14}). A vertical velocity component is expected 
from the failed wind of a dynamic BLR scenario. A failed wind may be
produced for the following reason, as suggested by CH11. The dusty gas is being pushed upwards by the large force 
multiplier provided by the dust, a process which is most significant at $R_{\rm in}$. 
The push upwards is maintained as long as the dust survives. At a certain
height the dust becomes exposed to the illumination by the central source. 
This inevitably leads to sublimation since $R_{\rm in}/R_{\rm sub}\simeq 0.1$.
The gas then looses the vertical driving force, but the wind continues ballistically
to a certain height, and falls back to the disc (unless it reaches the local escape speed and becomes unbound, as discussed in Section~\ref{sec:discussion}). Once the dust sublimates, the line emission efficiency is expected to jump, and the BLR emission is produced.

The above simple analytic estimates assume the central source emits isotropically. 
The expected angular dependence of the observed luminosity, $L_{\rm obs}$, for a Newtonian geometrically 
flat disc, is $L_{\rm obs}(\mu)=2L_{\rm bol}\mu$, where $\mu=\cos\theta$, and $\theta$ is the observer angle from the 
disc zenith. A stronger beaming may be produced if the inner disc is geometrically thick, and the outer AD
will then be shielded. The reverse effect of more isotropic emission
is produce by the relativistic Doppler beaming, which enhances the emission towards the disc plain
\citep{LaorNetzer89}. However, it is not clear that the thin disc solution applies for the innermost disc
\citep{LaorDavis14}. A shielded region allows a larger vertical extent where the dusty gas is accelerated upwards,
giving it a higher velocity, as it enters the unshielded region, looses the dust opacity, and 
gets photoionized.

\section{The dust properties and opacity} \label{sec:dust_prop}

In order to calculate the inflated disc structure, we first evaluate the expected dust properties,
i.e.\ grain size distribution and composition. These properties are used to calculate the dust $\kappa(\lambda)$, 
which is the key factor for the calculation of the expected CF. We note that the commonly used 
calculations of low temperature opacities \citep{Semenov03, Ferguson05}, include silicates but do
not include graphite grains. As we show below, the graphites are the dominant opacity source for
gas at $T\sim 2000$~K, and their neglect leads to an underestimate in the values of the Planck mean $\kappa$
by a factor of $\sim 40-50$.
	
For the sake of simplicity, we start with the normal Galactic ISM dust, described by the MRN dust model of graphite 
and silicate
grains, with a power-law size distribution, $dn/da\propto a^{-3.5}$, 
over the range of grain radii $0.005<a<0.25$~\mic\ \citep*{MRN77}. This distribution is modified close to the centre due to sublimation. The sublimation temperature, \Tsub\,
of a grain is set by the grain composition
and the ambient gas density, while the temperature of the grain, \Tgr, is set by the incident flux and its size and composition. 
Below we derive \Tsub\ of silicate and graphite grains 
for the BLR gas density. 
We then calculate the grain temperature as a function
of grain size, $\Tgr(a)$, for graphites and silicates. Since $\Tgr(a)$ increases with decreasing $a$,
while \Tsub\ is essentially independent of $a$,
the smallest grains sublimate first as one gets closer to the centre. We explore the effect of the implied modified grain size distribution
and composition on the value of $\kappa(\lambda)$, both in the UV, relevant for the gas line emission efficiency, 
and in the 
IR, relevant for the gas vertical support in the disc atmosphere. 

As we show below, since only the largest graphite grains survive close to the BLR, 
the assumed initial grains size distribution and composition is irrelevant for $\kappa$ at the BLR.
The UV opacity is strongly suppressed, while the near IR opacity is slightly enhanced. The major  
significant free parameter which controls $\kappa$ is the gas metallicity, $Z/Z_\odot$.

\subsection{\Tsub} \label{sec:T_sub}
	
The value of \Tsub\ is 
set by the saturation vapour pressure of graphite and silicate. When the gas pressure equals the saturation pressure, the rate of evaporation of atoms from the grain surface equals the rate of condensation of gas atoms back on the grain. This condition implies
\begin{equation}
n_{\rm x} k T = 6\times10^{11}T_{\rm sub}^{0.5} \exp\left(-\frac{T_0}{\Tsub}\right), \label{eq:P_sat}
\end{equation}
where $k$ is the Boltzmann constant, $n_{\rm x}$ is the number density of either C for graphite, or Si for silicate, $T$ is the gas temperature, and $T_0=81,200$~K for graphite, and $68,100$~K for silicate \citep{guhathakurta_draine89}. The value of $T_0$
depends on $a$ for very small grains ($a<0.005$~\mic), due to the larger reduction of the binding energy of the surface atoms 
\citep{guhathakurta_draine89}. However, this correction is not relevant here, as such small grains do not survive at our
region of interest. We therefore assume below that \Tsub\ is independent of $a$. 
The density $n_{\rm x}$ can be converted to the total gas density $n$ 
(approximated by the total H density) using the fractional abundance $f_{\rm x}\equiv n_{\rm x}/n$, where $f_{\rm x} = 10^{-3.44}$ and  $10^{-4.45}$ for C and Si, at $Z=Z_{\odot}$. Eq.~\ref{eq:P_sat} can be solved 
by iterations for the value of \Tsub\ as a function of $n$. A convenient form for iteration is obtained by rewriting eq.~\ref{eq:P_sat} in the form
\begin{equation}
X\exp(2X)=An^{-2}T_0 , \ \ {\rm where}\ \ X=\frac{T_0}{\Tsub}, \ \ 
A=\left(\frac{6\times10^{11}}{f_{\rm x} k T}\right)^2\ ,
\end{equation}
or equivalently
\begin{equation}
X = 0.5\ln (AT_0) -\ln n - 0.5\ln X\ .
\end{equation}
Substituting the values for graphite and silicate, and assuming $T=10^4$~K, which generally corresponds to the typical $T$ of photoionized gas near the H ionization front (e.g.\ \citealt{ferland99,paperII} for the BLR), we get the following
relations. For graphite
\begin{equation}
X = 68.002 -\ln n - 0.5\ln X \ , 
\end{equation}
and for silicate
\begin{equation}
X = 69.974 -\ln n - 0.5\ln X \ , 
\end{equation}
where we look for the solution for $X$ for a given $n$.
An approximate simple analytic solution for \Tsub\ as a function of $n$ is
\begin{equation}
\Tsub({\rm K}) = \frac{81,200}{66.003 -\ln n} , \label{eq:n_grap}
\end{equation}
for graphite, and 
\begin{equation}
\Tsub({\rm K}) = \frac{68,100}{67.957 -\ln n} , \label{eq:n_sil}
\end{equation}
for silicate. These expressions are accurate to better than 0.5 per cent for $1<n<10^{12}$~cm$^{-3}$.
	
\begin{figure}
\includegraphics[width=\columnwidth]{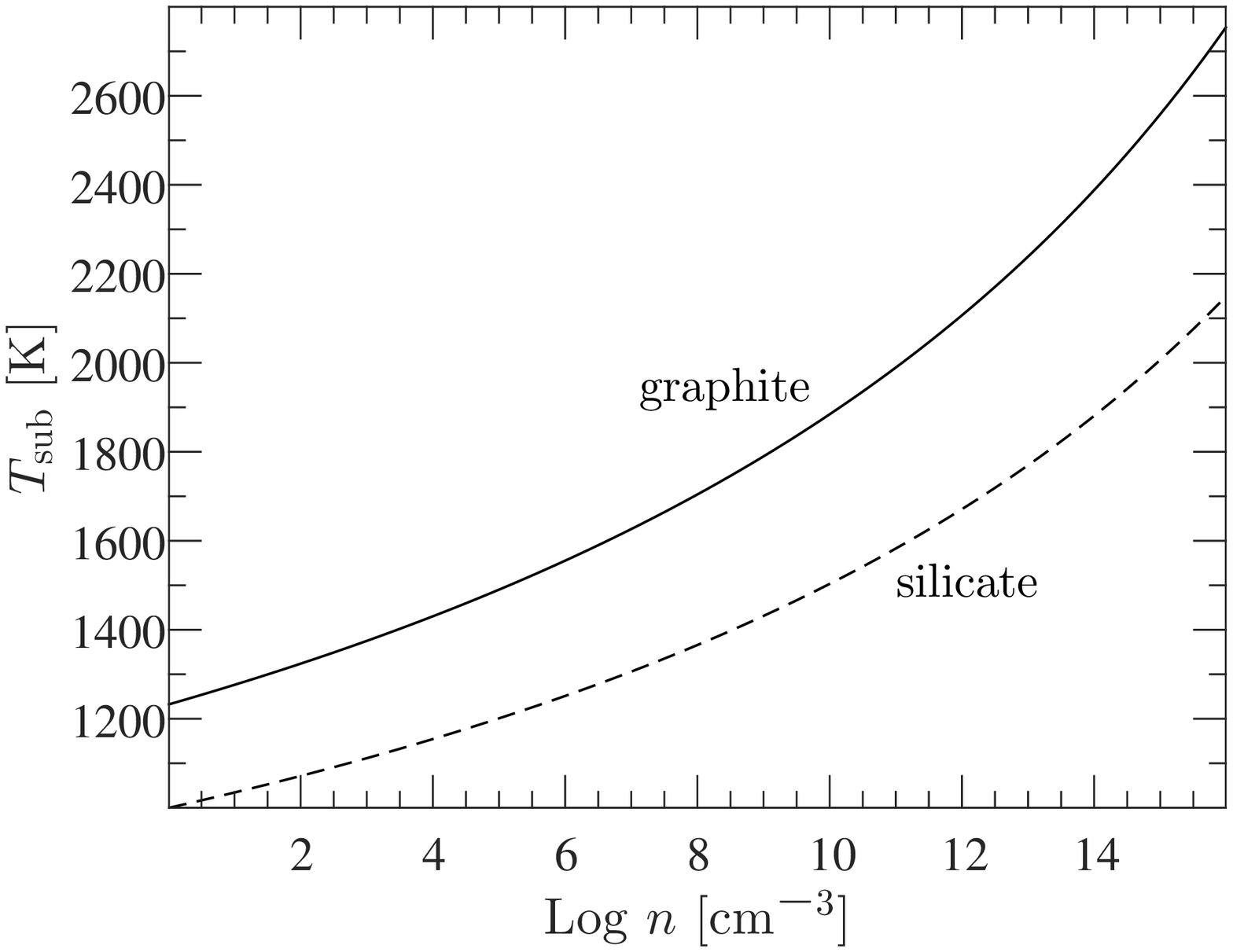}
\caption{The grain sublimation temperature as a function of gas density for graphite and silicate grains
(for $T=10^4$~K). The sublimation temperature of graphite is larger by 
$\sim 300-500$~K compared to that of silicate, as the graphite C atoms are more tightly bound than the Si atoms in silicates.
Thus, graphite grains sublimate at a smaller distance from the AGN continuum source.
At the BLR, $\ngas\sim10^{11}$~\cmt, which implies $\Tsub\simeq 2000$~K, rather than 1500~K, as commonly assumed in earlier studies.}
\label{fig:T_sub}
\end{figure}

Figure~\ref{fig:T_sub} presents a direct solution of eq.~\ref{eq:P_sat} for \Tsub,  as a function of \ngas, 
for graphite and silicate grains.  Note the logarithmic scale for \ngas, and linear scale for \Tsub, due to the
roughly logarithmic dependence of \Tsub\ on \ngas. Increasing \ngas\ from $1$ to $10^{14}$~\cmt\ increases \Tsub\ only by 
a factor of two (1200~K to 2400~K for graphite).
Graphite has \Tsub\ that is larger by $\sim 300$-500~K compared to silicate, as the C atoms in graphite are
more tightly bound (higher $T_0$) than the Si atoms in silicate. Thus, graphite grains can generally 
survive closer to the centre. As shown below, they can survive down to the BLR, in contrast to silicate grains, 
which all sublimate well outside the BLR. The typical density of the BLR is 
$\ngas\sim10^{11}$~\cmt, and we therefore adopt 
the corresponding $\Tsub=2000$~K (Fig.~\ref{fig:T_sub}) as the dust sublimation temperature. This temperature is significantly
higher than $\Tsub\simeq 1500$~K assumed in earlier studies (e.g., \citealt{Barvainis87, Schartmann2005, Nenkova2008}),
which applies for $\ngas\sim10^{5}$~\cmt\ gas. Thus, $\Tsub\simeq 1500$~K is more relevant for molecular 
clouds in the ISM, or NLR gas in AGN.

One should note that eq.~\ref{eq:P_sat} above assumes neutral gas \citep{guhathakurta_draine89}. This likely provides
a good approximation for the gas at the relevant parts of the disc atmosphere, where the temperature is below
2000~K, so dust can survive. In contrast, the BLR gas is photoionized, and both the gas ions and the grains are likely 
positively charged.
This may reduce the condensation rate of ions on the grain surface, and thus lower \Tsub, if the 
grains are sufficiently positively charged to repel most ions (i.e.\ $eV\gg kT$, where $V$ is the grain
potential).
		
\subsection{The dust sublimation timescale} \label{sec:dust_sub}

In Section~\ref{sec:T_sub} we derived \Tsub\ above which dust cannot survive, but how fast is the sublimation process? 
Can the dust shield itself effectively from sublimation?
The optical depth of dust to absorption and scattering is $\tau\sim \Sigma/10^{22}~{\rm cm}^{-2}$ in the visible, and $\tau\sim \Sigma/10^{21}~{\rm cm}^{-2}$ in the UV,
where $\Sigma$ is the H column density (e.g.\ \citealt{LaorDraine93}). Since the bulk of the heating in AGN is by optical-UV
continuum, and $\Sigma\ge 10^{24}$~cm$^{-2}$ gas is likely present along some lines of sight 
(a Compton thick absorber), then a dusty cloud will be self-shielded
by $\tau\sim 10^2-10^3$. Can such dust survive down to small radii, where otherwise $\Tgr\gg \Tsub$ without shielding?

The answer is generally negative. Self-shielding is maintained only on
the sublimation time scale, \tsub\ of the grains at the cloud surface, which are directly exposed to the AGN illumination. 
As we show below, \tsub\ is generally very short 
compared to other relevant time-scales at $R_{\rm BLR}$. As a result, grains cannot effectively exist in regions where
$\Tgr\gg \Tsub$, regardless of the value of $\tau$. 

The sublimation time-scale is evaluated as follows. In a steady-state, the gas pressure equals the saturation pressure, i.e.\ the evaporation rate, say of the C atoms for graphite grains, equals their condensation rate, given by
\begin{equation}
\dd N_{\rm C} / \dd t \simeq n_{\rm C,sat} v_{\rm C} 4\pi a^2,
\end{equation}
where $n_{\rm C,sat}$ is the density of C atoms in gas at the saturation pressure,
and $v_{\rm C}=\sqrt{8kT/\pi m_{\rm C}}$ is the C atoms average thermal velocity, where $m_{\rm C}$ is the mass of a C atom. A grain sublimates when the evaporation rate $\gg$  condensation rate. Thus, a grain composed of $N_{\rm C}$ atoms will 
sublimate completely over a time-scale of
\begin{equation}
\tsub\equiv \frac{N_{\rm C}}{\dd N_{\rm C} / \dd t}  \approx \frac{4\pi}{3}a^3 \frac{\rho_{\rm grain}}{m_{\rm C}} \Bigg/  n_{\rm C,sat} v_{\rm C} 4\pi a^2, \label{eq:t_sub}
\end{equation}
where  $\rho_{\rm grain}=2.26$~g~\cmt\ is the density of graphite. 
In a convenient form, 
\begin{equation}
\tsub= 2.8 a_{0.1}n_{10}^{-1} ~{\rm days}
\label{eq:t1_sub}
\end{equation}
where $a=0.1a_{0.1}$~\mic, $n=10^{10}n_{10}$~\cmt, and we adopt $T=10^4$~K and $n_{\rm C,sat}=10^{-3.44}n$.  

\begin{figure}
\includegraphics[width=\columnwidth]{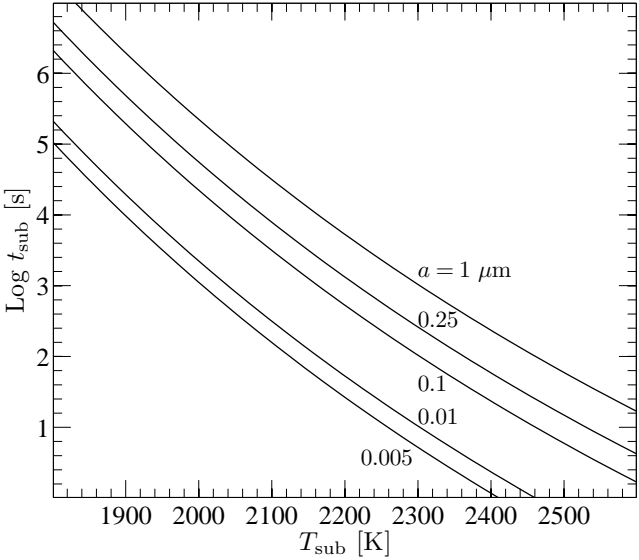}
\caption{The sublimation time-scale as a function of \Tsub\ for a graphite grain of various radii $a$, embedded in gas with $T=10^4$~K. For $\Tsub=2000$~K, relevant for grains at the BLR density, the time-scale is typically a few hours. It reaches $\sim 3$ days for the largest grain assumed here, of $a=1$~\mic. This time-scale is shorter than the light crossing time of the BLR. Thus, sublimation is effectively instantanous for most grain sizes.}
\label{fig:time_sub}
\end{figure}

Figure~\ref{fig:time_sub} presents $\tsub$ as a function of \Tsub\ for graphite grains of different
$a$ values. To derive this relation we first use eq.~\ref{eq:P_sat} to derive $n$ that corresponds to the 
assumed \Tsub, and then use eq.~\ref{eq:t1_sub} to derive \tsub\ that corresponds to that 
$n$.
The value of \tsub\ increases linearly with $a$ (eq.~\ref{eq:t1_sub}) and decreases roughly exponentially with \Tsub\ (eq.~\ref{eq:P_sat}). For the adopted value of $\Tsub=2000$~K, the smallest grains ($a=0.005$~\mic) are destroyed effectively instantaneously (15 min), while typical large grains ($a=0.1$~\mic) are destroyed in a 
few hours. Even the largest grains assumed here ($a=1$~\mic) sublimate in $\tsub\simeq 3$~days,  which is 
still shorter than the light crossing time for the BLR in most AGN. Thus, the hottest dust in AGN generally cannot shield itself from sublimation, as grains at the cloud surface sublimate 
effectively instantaneously when $\Tgr>\Tsub$.

The reverse process, of condensation back on the surface of grains at $T<\Tsub$, also occurs on the $\tsub$ 
timescale. It is therefore likely that when part of the grains sublimate, the sublimated material will condense
back on the cooler $T<\Tsub$ grains, and the total mass in the grains will not change significantly.	
	
\subsection{\Tgr} \label{sec:T_grain}

The value of \Tgr\ for a grain at a given distance from an illuminating UV source, depends on the grain size
and composition (e.g.\ \citealt{LaorDraine93}). This occurs since the 
wavelengths of the incident radiation generally satisfies $\lambda_{\rm in}<2\pi a$, which ensures an absorption efficiency 
close to unity for all grains. In contrast, the grain emission generally peaks at $\lambda_{\rm em}>2\pi a$, leading to an emission efficiency which typically is well below unity, and scales like a negative power of $a$ \citep{draine11}. Large enough grains can achieve 
$\lambda_{\rm em}\le 2\pi a$, when they are hot
enough, which leads to an emission efficiency which is also close to unity. In such a case, the grain will be at the local blackbody temperature, $\Tgr=T_{\rm BB}$. In smaller grains, the emission efficiency is smaller than unity, and as a result 
$\Tgr>T_{\rm BB}$.

The value of \Tgr\ is derived by equating the grain cooling rate $C$ to the grain heating rate $H$. 
The grain cools through thermal emission, which implies
\begin{equation}
C=4\pi a^2\int \pi B_{\lambda}(\Tgr)Q_{\rm abs}(a,\lambda)\dd\lambda,
\end{equation}
where $B_{\lambda}$ is the Planck function, and $Q_{\rm abs}$ is the absorption coefficient\footnote{The values of $Q_{\rm abs}(a,\lambda)$ are provided in machine readable format in the following url which is maintained by B.~Draine: \url{http://www.astro.princeton.edu/~draine/dust/dust.diel.html}.}. The grain heats due to absorption of the continuum source radiation, which we approximate by a point source. This yields
\begin{equation}
H= \pi a^2 \int \frac{2\mu L_{\lambda}}{4\pi R^2}Q_{\rm abs}(a,\lambda)\dd\lambda,
\label{eq:Heat}
\end{equation}
where $L_\lambda$ is the isotropic continuum luminosity density (here, we neglect heating by the local AD radiation, which is generally negligible). We adopt the SED $L_\lambda$ from \citet{paperII}, a standard AGN SED which peaks close to the Lyman edge, 
with an intermediate ionizing slope value of $\aion=-1.6$. We also assume that the continuum source is a flat Newtonian AD, with the simple $2\mu$ 
angular dependence for the local flux. We solve numerically by iterating over the value of \Tgr\ until the equality $C=H$ is satisfied.

\begin{figure}
\includegraphics[width=\columnwidth]{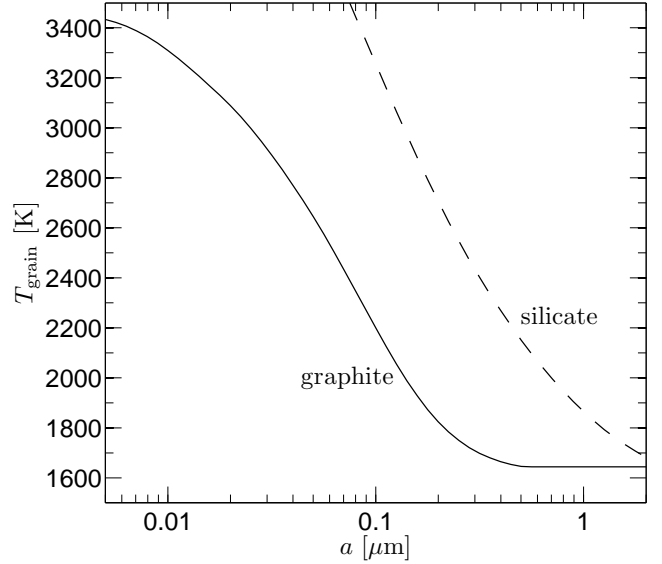}
\caption{The grain temperature as a function of grain radius for graphite and silicate grains. The illuminating continuum has $L_{\rm bol}=10^{45}$~\ergs. The grains are located at $R_{\rm BLR}=0.1L_{\rm bol, 46}^{1/2}$~pc and an inclination angle of $\mu=0.1$. The small silicate grains are hotter than graphite grains, at a given $a$, 
by $\sim 1000$~K, because of their lower emission 
efficiency in the near IR. Note that for graphites, $\Tgr<\Tsub\simeq2000$~K  is achieved in the BLR 
only for grains with $a\ga 0.1$~\mic. For silicates, $\Tgr>\Tsub\simeq1600$~K for all grains, and thus
none of the silicate grains survive at $R_{\rm BLR}$.}
\label{fig:T_gr_cos0.1}
\end{figure}

Figure~\ref{fig:T_gr_cos0.1} presents the dependence of \Tgr\ on $a$ for graphite and silicate grains at a given distance. We assume a continuum source with $L_{\rm bol} = 10^{45}$~\ergs, and grains located at a distance
$R=R_{\rm BLR} = 0.1 L_{\rm bol, 46}^{1/2}$~pc (\citealt{Kaspi07}, and the relation $L_{\rm bol}=3L_{1350}$). 
We also assume $\mu=0.1$, a plausible value for the BLR (see below).
As expected, the graphite grain temperature decreases with increasing $a$, up to $a\approx 0.5$~\mic, 
above which $\Tgr\simeq1650$~K, which is independent of $a$. This occurs since the peak emission wavelength satisfies 
$\lambda_{\rm em}<2\pi a$, the emission efficiency reaches unity, leading to $\Tgr=T_{\rm BB}$ for $a> 0.5$~\mic.
As noted above, at the BLR, $n\simeq 10^{11}$~cm$^{-3}$, which gives  $\Tsub=2000$~K. Thus, as Fig.~\ref{fig:T_gr_cos0.1}
shows, only relatively large grains with $a\ga0.1$~\mic, have $\Tgr<\Tsub=2000$~K and avoid sublimation at the BLR. In addition, the small silicate grains are hotter than the small graphite grains, at a given
$a$, by about 1000~K. This occurs because of the significantly lower radiative efficiency of hot silicates grains
(being nearly transparent in the near IR) compared to hot graphite grains (being nearly black), in contrast with their similar and high absorption opacity to the incident UV (\citealt{draine_lee84, LaorDraine93}). In addition, in silicates 
$\Tsub\simeq 1600$~K at $n\sim10^{11}$~\cmt\ (Fig.~\ref{fig:T_sub}). 
As a result, even the largest silicate
grains assumed here ($a\sim 1$~\mic) do not reach $\Tgr<\Tsub$. We therefore do not expect any silicates at $R_{\rm BLR}$.

To summarise, silicates do not exist at the BLR because of the combination of two effects. First, their binding energy is lower, 
their evaporation rate is therefore higher, leading to a lower \Tsub\ at a given 
ambient pressure. Second, they are significantly more transparent in the near IR, their emission efficiency 
is therefore lower, leading to a higher equilibrium temperature at a given distance.
Thus, large ($a\ga0.1$~\mic) graphite grains will be the only constituent of a UV dust-driven outflow at $r\sim R_{\rm BLR}$.

\begin{figure}
\includegraphics[width=\columnwidth]{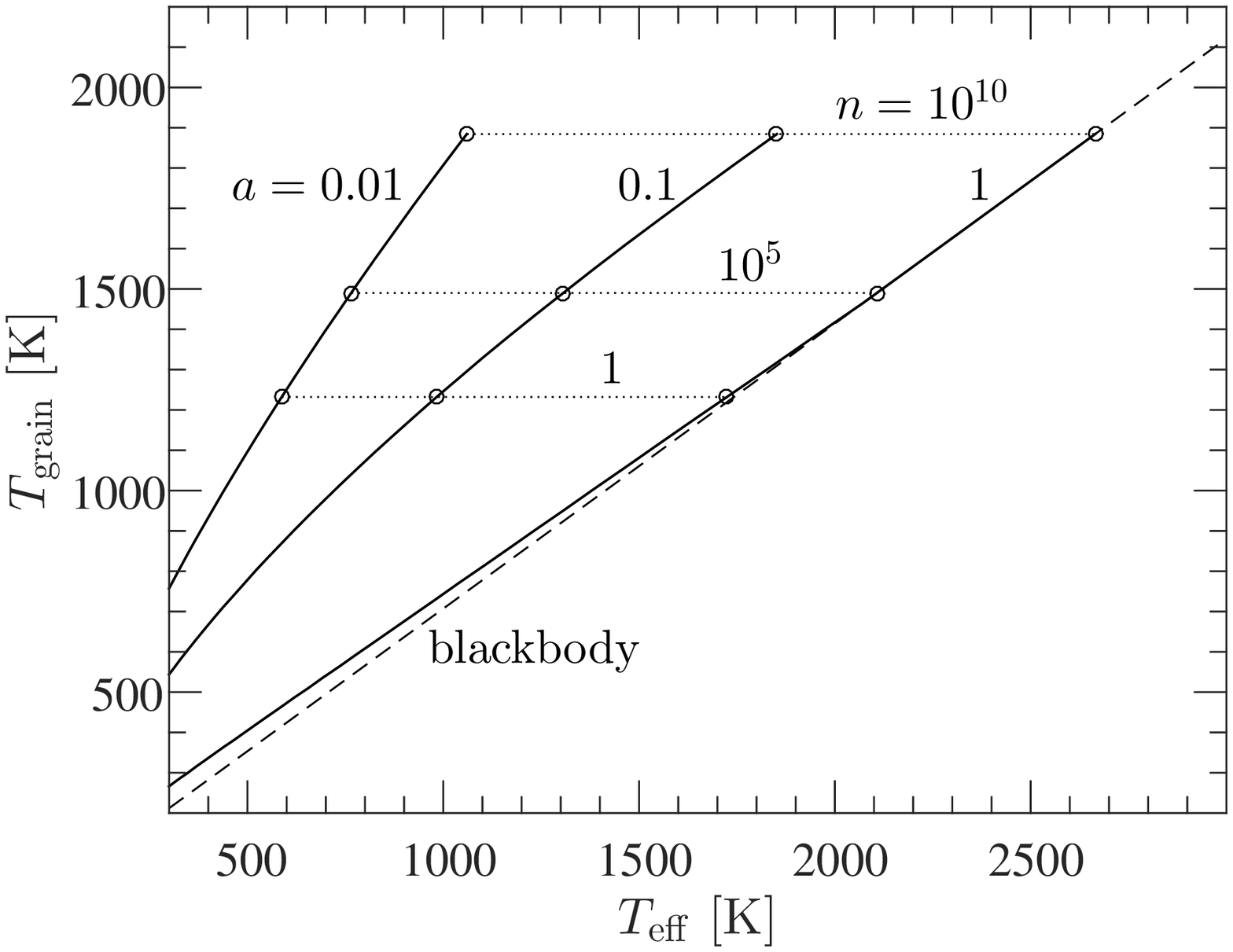}
\caption{The graphite grain temperature as a function of the AGN flux, as measured by $T_{\rm eff}$, for various grain sizes. 
The $a=1$~\mic\ grain \Tgr\ deviates from
$T_{\rm BB}$ ($=4^{-1/4}\,T_{\rm eff}$; denoted by a dashed line) only at the lowest 
$T_{\rm eff}$, where the grain emission efficiency falls below unity. As the grains get smaller,
\Tgr\ becomes larger, due to the decreasing emission efficiency.
The value of \Tsub\ for various gas densities is marked by the horizontal dotted lines. 
Sublimation can occur in the range $T_{\rm eff}=580-2670$~K, which corresponds to 
$R\sim 1-20R_{\rm BLR}$.}
\label{fig:Tgr_vs_T_eff}
\end{figure}

Figure~\ref{fig:Tgr_vs_T_eff} provides a general relation between the temperature of a graphite grain
of a given size, and the flux incident on it, i.e.\
$2\mu L_{\rm bol}/4\pi R^2$, as measured by $T_{\rm eff}$.  The largest grain ($a=1$~\mic)
generally follows $\Tgr=T_{\rm BB}$, where $T_{\rm BB}=4^{-1/4}T_{\rm eff}$ (eq.~\ref{eq:flux2}). 
At the lowest $T_{\rm eff}$, one gets $\Tgr>T_{\rm BB}$. This results from the drop in the emission efficiency, due to the lower
\Tgr, which leads to $\lambda_{\rm em}>2\pi a$. For a $a=0.1$~\mic\ grain, the emission efficiency
is always well below unity, leading to $\Tgr/T_{\rm BB}\simeq 1.5-2$ at all $T_{\rm eff}$
probed here. For a $a=0.01$~\mic\ grain, $\Tgr/T_{\rm BB}\simeq 3$ at all distances. 
The horizontal dotted lines in Fig.~\ref{fig:Tgr_vs_T_eff} mark \Tsub\ for various values of $n$. Thus, for
$n=10^{10}$~cm$^{-3}$, the $a=1$~\mic\ grains sublimate only at $T_{\rm eff}=2670$~K
(which gives \Tgr=1890~K), while the $a=0.01$~\mic\ grains sublimate already at $T_{\rm eff}=1070$~K 
(which also gives \Tgr=1890~K). In low density $n=1$~cm$^{-3}$ gas, the destructions
of the $1$ and $0.01$~\mic\ grains occur at $T_{\rm eff}=1720$ and $580$~K, respectively. 
Since $R\propto T_{\rm eff}^{-2}$, the $a=0.01$~\mic\ grains in low density gas
are destroyed at a radius $\sim 20$ times larger than the radius for the $a=1$~\mic\ grains
at $n=10^{10}$~cm$^{-3}$.  The graphite grain size distribution is modified by sublimation as far out as $\sim 20R_{\rm BLR}$. As noted above,
the silicate grain size distribution will be modified by sublimation at even larger distances.

\begin{figure}
\includegraphics[width=\columnwidth]{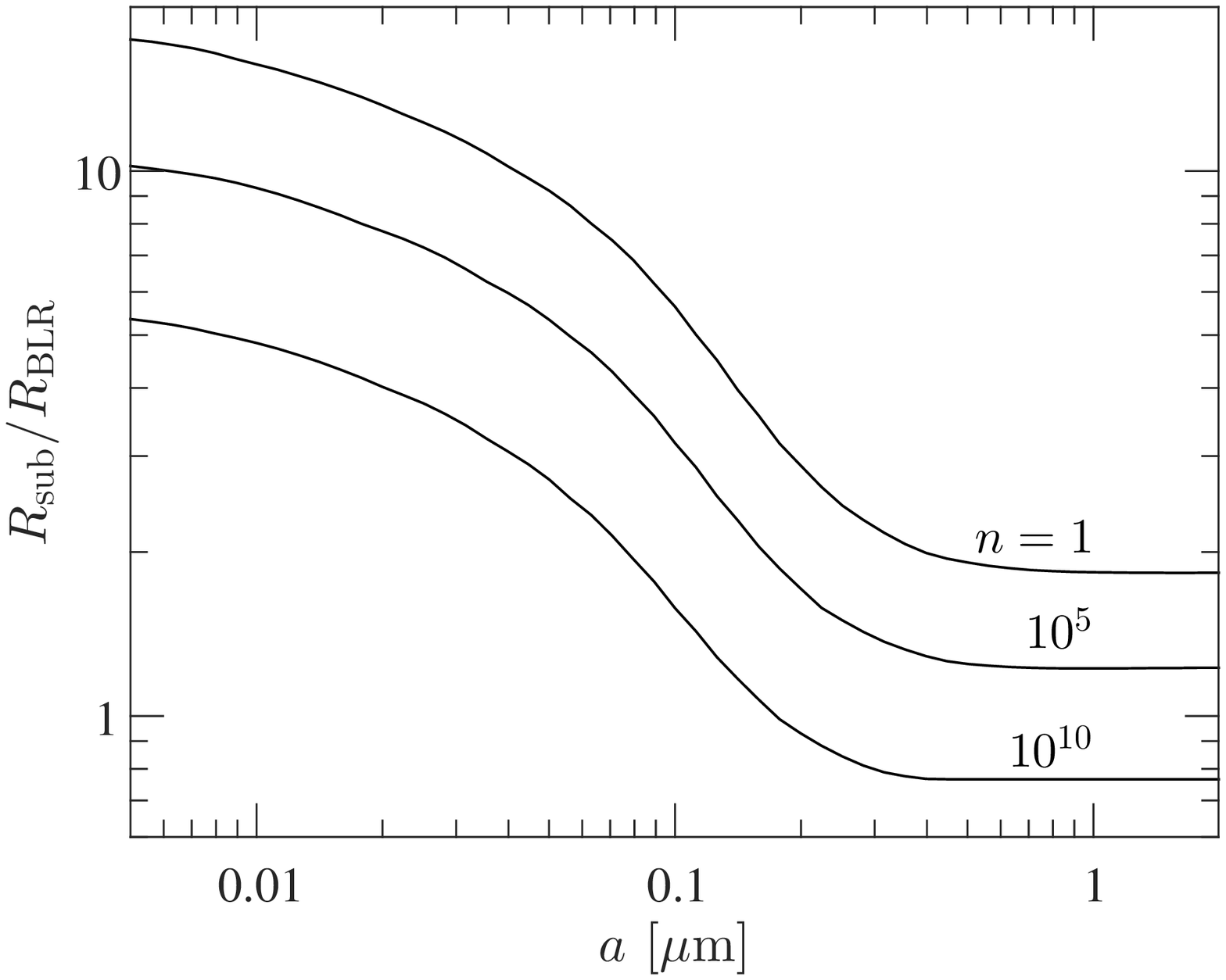}
\caption{The sublimation radius, measured in units of $R_{\rm BLR}$, versus $a$, for graphite grains 
at different gas densities (i.e.\ different \Tsub), for $\mu=0.1$.
Grains with $a>0.2$~\mic\ can survive in dense gas ($n\ge 10^{10}$~cm$^{-3}$) at the BLR. 
All grains survive, essentially irrespective of $n$, once $R>20R_{\rm BLR}$.
Since $R/R_{\rm BLR}$ sets the value of $T_{\rm eff}$, the above solution applies in all AGN, independent of 
$L_{\rm bol}$.
}
\label{fig:Rsub_vs_a}
\end{figure}

The radiative flux at a given distance satisfies $F(R) \propto L_{\rm bol}/R^2$. Since
 $R_{\rm BLR}\propto L_{\rm bol}^{1/2}$, then $F(R_{\rm BLR})$ has a universal value
at $R_{\rm BLR}$.
One can therefore use $R/R_{\rm BLR}$ as a measure of the local flux, or equivalently
a measure of $T_{\rm eff}$ (for a given $\mu$).  
Figure~\ref{fig:Rsub_vs_a} presents the sublimation radius $R_{\rm sub}$ in units of $R_{\rm BLR}$,
as a function of $a$ for a range of $n$ values (for $\mu$=0.1). 
One can see that in $n=10^{10}$~cm$^{-3}$ gas at $R_{\rm BLR}$, only $a>0.2$~\mic\ grains survive 
(i.e.\ their $R_{\rm sub}/R_{\rm BLR}<1$). In contrast, gas with $n<10^5$~cm$^{-3}$
can retain all its grains only at $R>10R_{\rm BLR}$. However, such a low density gas at say $10R_{\rm BLR}$
will have a ionization parameter $U\sim 1000$, which gives $T> 10^5$~K, and leads to thermal sputtering of the grains.

\subsection{$\kappa$}\label{sec:dust_op}
	
The dust pressure opacity $\kappa$ is a function of $\lambda$, and its functional form is set by the dust model, i.e.\ the grain composition, size distribution, and the
dust to gas mass ratio $f_{\rm d}$. The flux weighted mean value of the opacity is given by the expression
\begin{equation}
\kappa=f_{\rm d} \frac{\int \alpha_\lambda F_\lambda \dd\lambda }{F_{\rm bol}} \ ,
\end{equation}
where $F_\lambda$ is the incident flux density, $F_{\rm bol}=\int F_\lambda \dd\lambda$ is the integrated flux, and $\alpha_\lambda$ is given by
\begin{equation}
\alpha_{\lambda}= \int_{a_{\rm min}}^{a_{\rm max}} \pi a^2 \sum_i Q_{\rm pr}^i(a,\lambda)\dd n_i(a)\ , \label{eq:alpha_lambda}
\end{equation}
where $Q_{\rm pr}^i(a,\lambda)=Q_{\rm abs}^i(a,\lambda)+(1-g^i(a,\lambda))\times Q_{\rm scat}^i(a,\lambda)$ 
is the radiation pressure coefficient representing the fraction of the incident radiation pressure flux
transferred to the grain, $g^i(a,\lambda)$ is the mean $\mu$ of the scattered light ($=0$
for isotropic scattering), and $i$ is the grain composition, either graphite or silicate.
The number density of grains with radii in the interval $[a,a+\dd a]$ is parametrised as a power-law 
\begin{equation}
\dd n_i(a) = A_i n_{\rm H} a^{\beta} \dd a \quad (\amin\leq a\leq \amax), \label{eq:gr_size_distr}
\end{equation}
where $A_i$ and $\beta$ are free parameters, and \amin\ and \amax\ are the minimum and the maximum grain radii. 
Finally, $f_{\rm d}$ is the grain mass per H mass, given by

\begin{equation}
f_{\rm d} = \frac{4\pi}{3 m_{\rm H}}\frac{a_{\rm max}^{\beta+4}}{\beta+4} \left[1-\left(\frac{a_{\rm min}}{a_{\rm max}}\right)^{\beta+4}\right] \Sigma_i A_i \rho_i,
\end{equation}
where $\rho_i=2.26$ and 3.3~gr~\cmt\ for graphite and silicate, respectively. The `standard' model of dust is adopted to be the MRN dust, which assumes $\amin=0.005$~\mic, $\amax=0.25$~\mic\ and $\beta=-3.5$. Following \citet{draine_lee84}, we adopt $f_{\rm d}=0.01$ and the ratio $A_{\rm sil}/A_{\rm gra}=1.12$ for the MRN dust mixture. The dust/H mass ratio is assumed to scale linearly with $Z$ (\citealt*{Issa90, Draine07, Remy14}), i.e.
\begin{equation}
f_{\rm d}=0.01 \frac{Z}{Z_{\odot}}, \label{eq:fd_vs_Z}
\end{equation}
in all models explored below.
	
\begin{figure}
\includegraphics[width=\columnwidth]{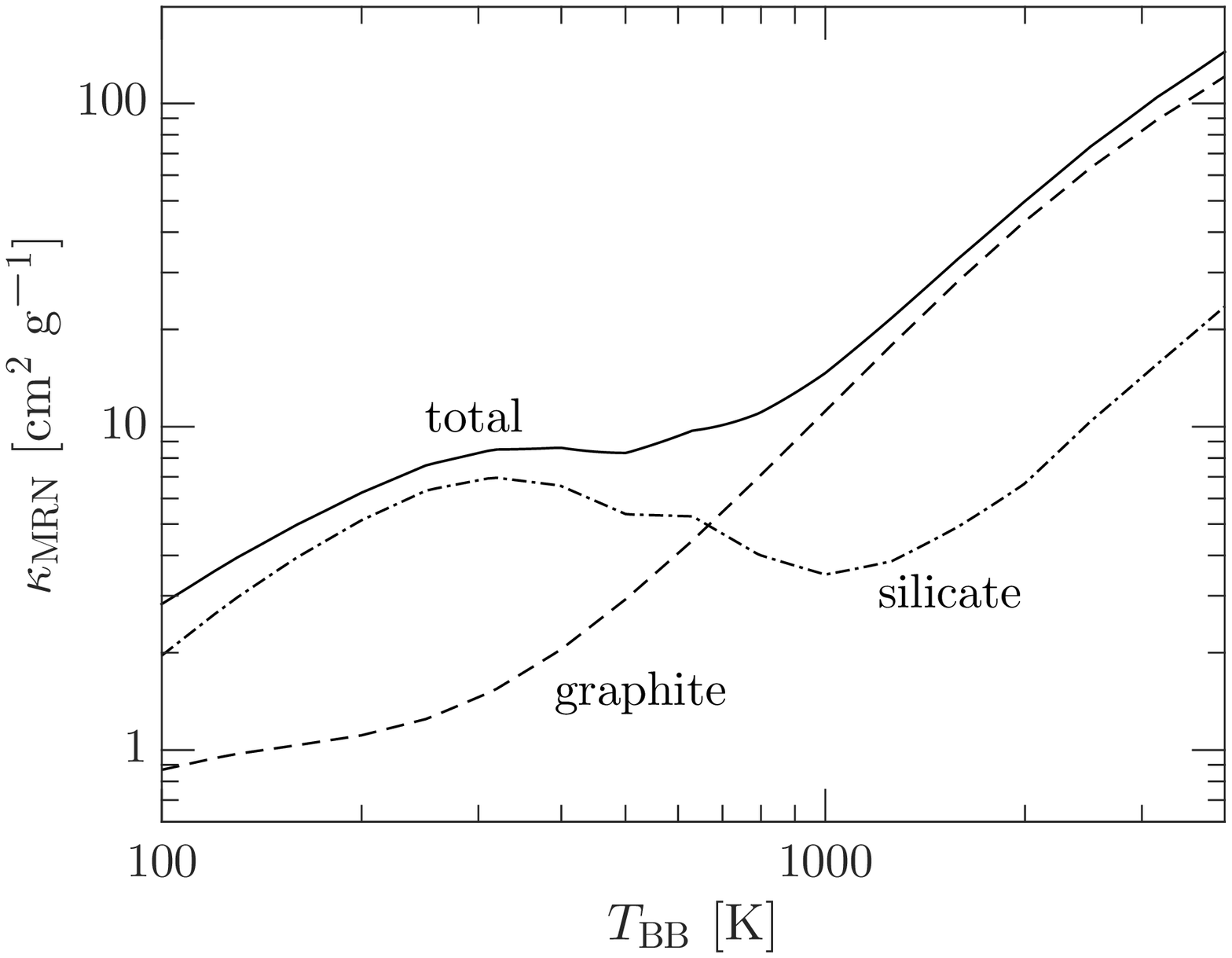}
\caption{The dependence of the Planck mean pressure opacity $\kappa_{\rm MRN}$ on 
$T_{\rm BB}$. The dashed lines show the individual contribution
of the silicate and the graphite grains. The silicates dominate the opacity at $T_{\rm BB}\sim 300$~K, while at 
$T_{\rm BB}> 1000$~K graphites dominate the opacity.
As a result, the sublimation of the silicate grains has no 
significant effect on the hottest
dust opacity.}
\label{fig:kappa_MRN_vs_T}
\end{figure}

Figure~\ref{fig:kappa_MRN_vs_T} presents the Planck mean opacity of a MRN dust, $\kappa_{\rm MRN}$, as a function of 
$T_{\rm BB}$, and the contribution of graphites and silicates. The opacity is evaluated using the values of $Q_{\rm pr}(a,\lambda)$ from \citet{draine_lee84}, \citet{LaorDraine93} and \citet{weingartner_draine01}. The value of $\kappa_{\rm MRN}$ increases with $T_{\rm BB}$ as the blackbody peak emission shifts to shorter $\lambda$, where the dust opacity is higher. Interestingly, the relative contribution of silicates and graphites changes drastically with $T_{\rm BB}$. At $T_{\rm BB}\sim 300$~K the silicates heavily dominate the opacity. This occurs because of the strong silicate opacity features at 9.7~$\mic$ and 18~$\mic$, which match the peak emission at 
$\lambda \sim 10~\mic$  for
$T_{\rm BB}\sim 300$~K, while the graphite opacity falls steeply with $\lambda$ in this wavelength range.  
In  contrast, at $T_{\rm BB}\sim \Tsub=2000$~K, the graphites dominate the opacity, as silicates become transparent at $\lambda<3~\mic$, while graphites remain black (see figs~2 and 3 in \citealt{LaorDraine93}). The relation between $\kappa$ and $T_{\rm BB}$ is explored further in Appendix~\ref{sec:appendix}.

Thus, we have the `fortunate' coincidence that graphite not only has a higher \Tsub\, and a lower \Tgr\ at a given $R$, which allows the graphite to survive much closer to the centre than silicate, it also dominates the dust opacity. So, the fact that the dust looses all the silicates close to the centre, has almost no effect on the dust opacity. 

At the temperature range, $500<T_{\rm BB}<2500$~K, which is relevant for the calculations below, 
$\kappa_{\rm MRN}$ of graphite only is well fit by the expression 
\begin{equation}
\kappa=43 T_{2000}^{1.94}\times Z/Z_{\odot}~{\rm cm}^2~{\rm gr}^{-1}\ . \label{eq:kappa}
\end{equation}
If the grain size distribution extends to larger grains, with $a_{\rm max}=1$~\mic, the grain mean opacity is somewhat larger, but shows a more gradual increase with temperature. Specifically, 
\begin{equation}
\kappa=54 T_{2000}^{1.16}\times Z/Z_{\odot}~{\rm cm}^2~{\rm gr}^{-1} . 
\label{eq:kappa1}
\end{equation}
These expressions are useful for the analytic estimates of $H(R)$, as further discussed below.

The Planck mean $\kappa_{\rm MRN}$ derived here is consistent with the Planck mean dust opacity 
presented by Draine (2011, fig.~23.12 there), for the updated models of the ISM dust. However, 
the \citet{draine11} and our results for $T>500$~K 
are significantly different from the Planck mean opacity
results of \citet{Semenov03} and \citet{Ferguson05}, and earlier references cited in these papers.
At $T=2000$~K we get $\kappa\sim 50~{\rm cm}^2~{\rm gr}^{-1}$, in contrast with
$\kappa\sim 1~{\rm cm}^2~{\rm gr}^{-1}$ in the above calculations. This results from the inclusion
of only silicate grains in the earlier calculations. Indeed, our results including only the silicate
grain opacity, are consistent with the earlier calculations, which yield a Planck mean opacity of
$\kappa\sim 3-6~{\rm cm}^2~{\rm gr}^{-1}$ at $T\sim 200-600$~K. The lack of
inclusion of graphite grains in the earlier calculations, leads to a significant underestimate
of $\kappa$ at $T>600$~K, which can reach a factor of $\sim 50$ in dense gas where graphites
can survive at $T\sim 2000$~K. Such conditions are likely relevant for AD in various systems, 
so graphites should not be ignored, unless they are not expected  to exist (e.g.\ \citealt{Anderson17}).

The MRN dust model assumes a power-law size distribution, with a minimal grain size $\amin=0.005$~\mic. As shown above,
with decreasing $R$ the smallest grains sublimate (Fig.~\ref{fig:Rsub_vs_a}), and as a result the value of 
$\amin$ increases. How does this affect the dust UV opacity? As was discussed by \citet{NetzerLaor93},
in photoionized dusty gas the dust UV opacity strongly suppresses
the ionising continuum and the resulting line emission of the gas. The drop in the dust UV opacity with decreasing $R$, as shown below,
reduces the 
ability of the dust to suppress the line emission of photoionized dusty gas.

\begin{figure}
\includegraphics[width=\columnwidth]{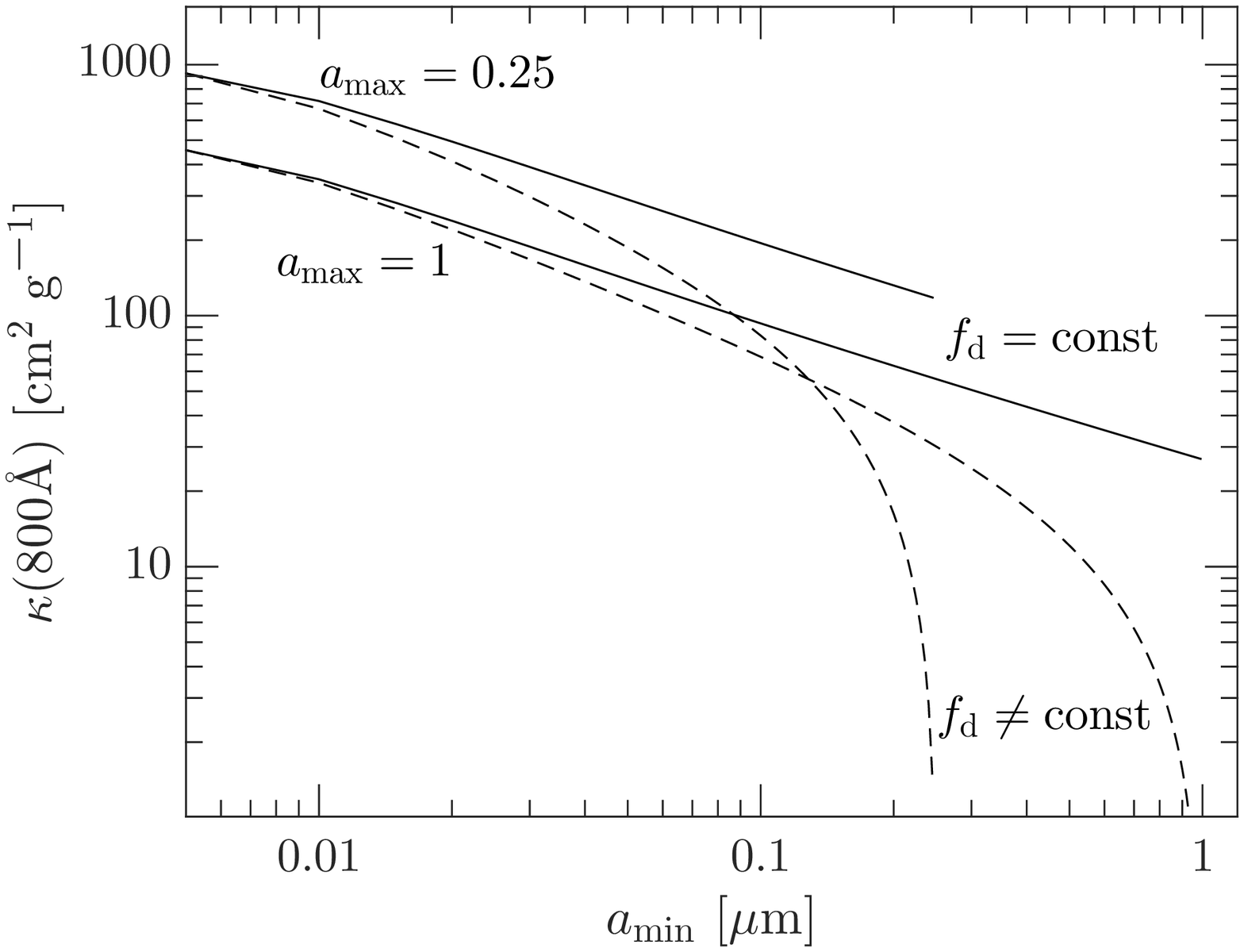}
\caption{The dust absorption opacity at $\lambda=800$~\AA\ as a function of \amin. Two cases are assumed: (i) the small grains that sublime condense on the larger grains, thus keeping $f_{\rm d}=0.01$ constant (solid line); 
(ii) the mass of sublimed grains is lost, thus lowering $f_{\rm d}$ (dashed line). The value of $\kappa$ has a strong dependence on \amin, and drops by about a order of magnitude when only the largest grains remain, in particular for the $\amax=1~\mic$ case.}
\label{fig:kappa_800_vs_amin}
\end{figure}

Figure~\ref{fig:kappa_800_vs_amin} presents the dependence of the dust UV opacity,
at $\lambda=800$~\AA, on the value of \amin, for $\amax=0.25~\mic$ (MRN dust), and for
$\amax=1~\mic$. The larger $\amax$ may be more relevant to the dense BLR clouds, 
where grains may be able to grow further. In the later case, 
$\kappa$ is smaller at a given \amin\ value, as most of the dust mass (for the MRN grain size distribution)
resides in the largest grains, which have a lower cross section per unit mass.
The sublimated grain material may either condense back on the larger grains,
so the fraction of mass in the grains remains constant (denoted by $f_{\rm d}=$~const in Fig.~\ref{fig:kappa_800_vs_amin}), or it may remain 
in the gas phase ($f_{\rm d}\ne $~const). In both cases, $\kappa$ drops steeply with increasing \amin, as 
most of the UV absorption opacity is contributed by the smallest grains, which have the highest cross section
per unit mass. 
The value of $\kappa$ drops by a factor of $\sim 8$ for $\amax=0.25~\mic$, and by a factor of $\sim 17$ for $\amax=1~\mic$, when only the largest grains remain (for $f_{\rm d}=$~const, 
and obviously a larger drop for $f_{\rm d}\ne $~const).

In photoionized dusty gas, with MRN dust, the dust opacity dominates the gas opacity for the ionising UV,
and the overall line emission of the gas is suppressed by about an order of magnitude
(e.g.\ \citealt{paperI}). The order of magnitude drop in $\kappa$, when only the
largest grains remain, means that the dust suppression
of the ionising UV radiation will no longer be dominant, in particular when  $\amax=1~\mic$, and the
gas becomes an efficient line emitter. But, what happens to $\kappa$ in the near IR, when only large grains are left?  Is the ability of the dust to lift the gas from the disc is also eliminated? 

\begin{figure*}
\includegraphics[width=2\columnwidth]{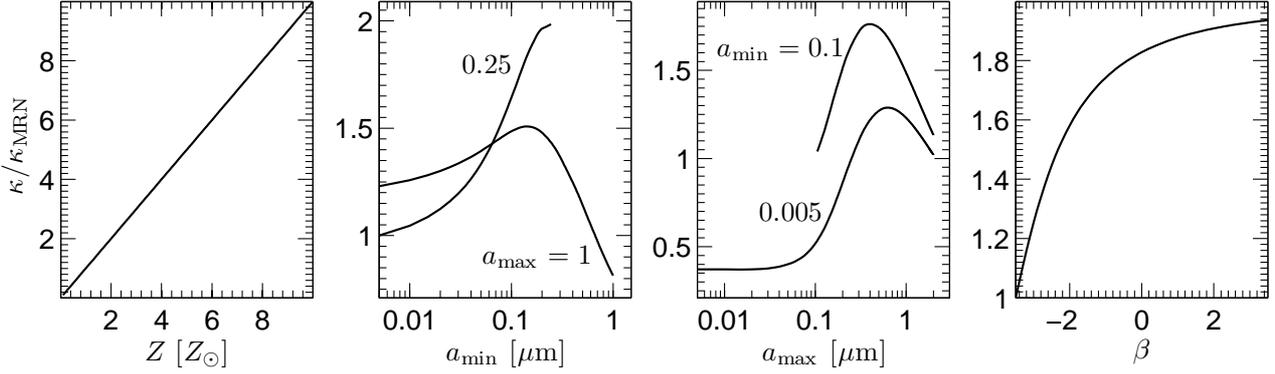}
\caption{The effect of $Z$ and the grain size distribution on the Planck mean 
$\kappa$ for $T_{\rm BB}=2000$~K. The opacity is normalized by the Planck mean $\kappa_{\rm MRN}$
of an MRN dust with graphite only.  In each panel, only one parameter is varied, while the other three are fixed at the MRN values: $\amin=0.005$~\mic, $\amax=0.25$~\mic, $\beta=-3.5$ and $Z=Z_{\odot}$.
The opacity increases linearly with $Z$, which is the dominant parameter in setting $\kappa$.
The grain size-distributions have a small effect on the near-IR $\kappa$, as the opacity is mostly
a grain volume effect, which is fixed if the total dust mass is fixed. }
\label{fig:Rkappa}
\end{figure*}

Figure~\ref{fig:Rkappa} explores the effects of different values for \amin, \amax, $\beta$, and  $Z$, on the Planck mean $\kappa$ for $T_{\rm BB}=2000$~K. The value of $\kappa$ 
is normalized by $\kappa_{\rm MRN}$, calculated for MRN dust with graphite grains only. The left panel in Fig.~\ref{fig:Rkappa} 
shows the dependence of $\kappa/\kappa_{\rm MRN}$ on the metallicity $Z$. Since we assume a linear dependence of
$f_{\rm d}$ on $Z$ (eq.~\ref{eq:fd_vs_Z}), there is a simple linear relation $\kappa/\kappa_{\rm MRN}=Z/Z_{\odot}$. The other three
panels show the effect of a modified grain size distribution, where in all cases we assume a fixed $f_{\rm d}$. The second panel shows the effect of increasing \amin, for  
\amax $=0.25~\mic$ or 1~$\mic$. In contrast with the UV opacity (Fig.~\ref{fig:kappa_800_vs_amin}), here $\kappa$ increases
as \amin\ increases, by up to a factor of $\sim 2$. A similar effect is seen when \amax\ decreases (third panel). 
Only when all grains are smaller
than 0.1~\mic\ there is a decrease in $\kappa$, by a factor of $\sim 2.5$. The increase in $\beta$ (last panel),
which corresponds to an increase in the fraction of large grains, also leads to an increase in $\kappa$, again by a factor
of $\sim 2$. Therefore, the Planck mean $\kappa$ for $T_{\rm BB}=2000$~K, is nearly independent of the grain size distribution, and is mostly set by the gas metallicity.

The weak dependence of $\kappa$ on the grain size distribution occurs for the following reason.
For most grains, the absorption efficiency in the near IR is $Q\simeq 2\pi a/\lambda$ (e.g.\ fig.~2 in \citealt{LaorDraine93}). 
Therefore, the absorption cross section
$\sigma=\pi a^2Q \propto a^3$, i.e.\ as the grain volume. The total absorption of a distribution of grains is then 
proportional to the total volume, i.e.\ to the total grain mass, regardless of how it is distributed. 
In the UV, $Q\sim 1$, thus $\sigma\propto a^2$, the integrated cross section over the grain size distribution
scales with the integrated surface area, and thus $\kappa$ scales as the area/mass, which 
is dominated by the smallest grains.

At $R\sim R_{\rm BLR}$, we therefore have the fortuitous situation that since only the largest graphite grains
survive, the UV opacity is reduced significantly and the gas becomes an efficient line emitter, yet the IR opacity
is not reduced (or even somewhat enhanced, for a fixed $f_{\rm d}$). Thus, the dust can provide significant 
vertical support for the dusty disc atmosphere, and potentially allow it to subtend a large enough solid angle, and absorb enough of the ionising radiation, to produce the observed BLR emission.

\section{Solutions for the BLR torus height}  \label{sec:solutions}
	
Below we describe two analytic solutions for the vertical structure of the BLR dusty torus, a static and a dynamic solution.
We then provide a numeric solution for the dynamic case.

\subsection{The static solution}\label{sec:static_model}
\subsubsection{The BH gravity and the local disc emission}

As in Section~\ref{sec:sub:expected_CF} we estimate the thickness of the disc, $H$, by equating the local radiation pressure to the local gravity. We begin again by evaluating an analytic approximation to $H$ by assuming $H\ll R$ (see also eq.~\ref{eq:H1}); and then we alleviate the assumption and derive an exact solution of $H$. Assuming the disc emits locally as a blackbody, the radiation pressure is given by the Planck mean opacity of $\kappa$.
 Since the disc surface effective temperature follows $T_{\rm eff}\propto R^{-3/4}$, while  $\kappa\propto T^{1.16}$
(for $a_{\rm max}=1$~\mic, eq.~\ref{eq:kappa1}), and $H\propto \kappa$, we get $H\propto R^{-0.87}$
(or $H\propto R^{-1.46}$ for $a_{\rm max}=0.25$~\mic). This
implies a rather steep rise of the CF$\propto R^{-1.87}$ inwards. 
More quantitatively, using eq.~\ref{eq:T_R} for $T_{\rm eff}(R)$, and eq.~\ref{eq:kappa1} for 
$\kappa(T_{\rm eff})$, with a linear dependence on $Z$, we get
\begin{equation}
\kappa=1.44(M_8\mdot_1)^{0.29}R_{\rm pc}^{-0.87}Z/Z_{\odot}\ ,
\end{equation}
which yields using eq.~\ref{eq:H_mdot}
\begin{equation}
H=1.17\times 10^{-4}M_8^{0.29}(\mdot_1)^{1.29}R_{\rm pc}^{-0.87}Z/Z_{\odot}~{\rm pc}.
\label{eq:H_mdot1}
\end{equation}

\begin{figure}
\vspace*{0.6cm}
\includegraphics[width=\columnwidth]{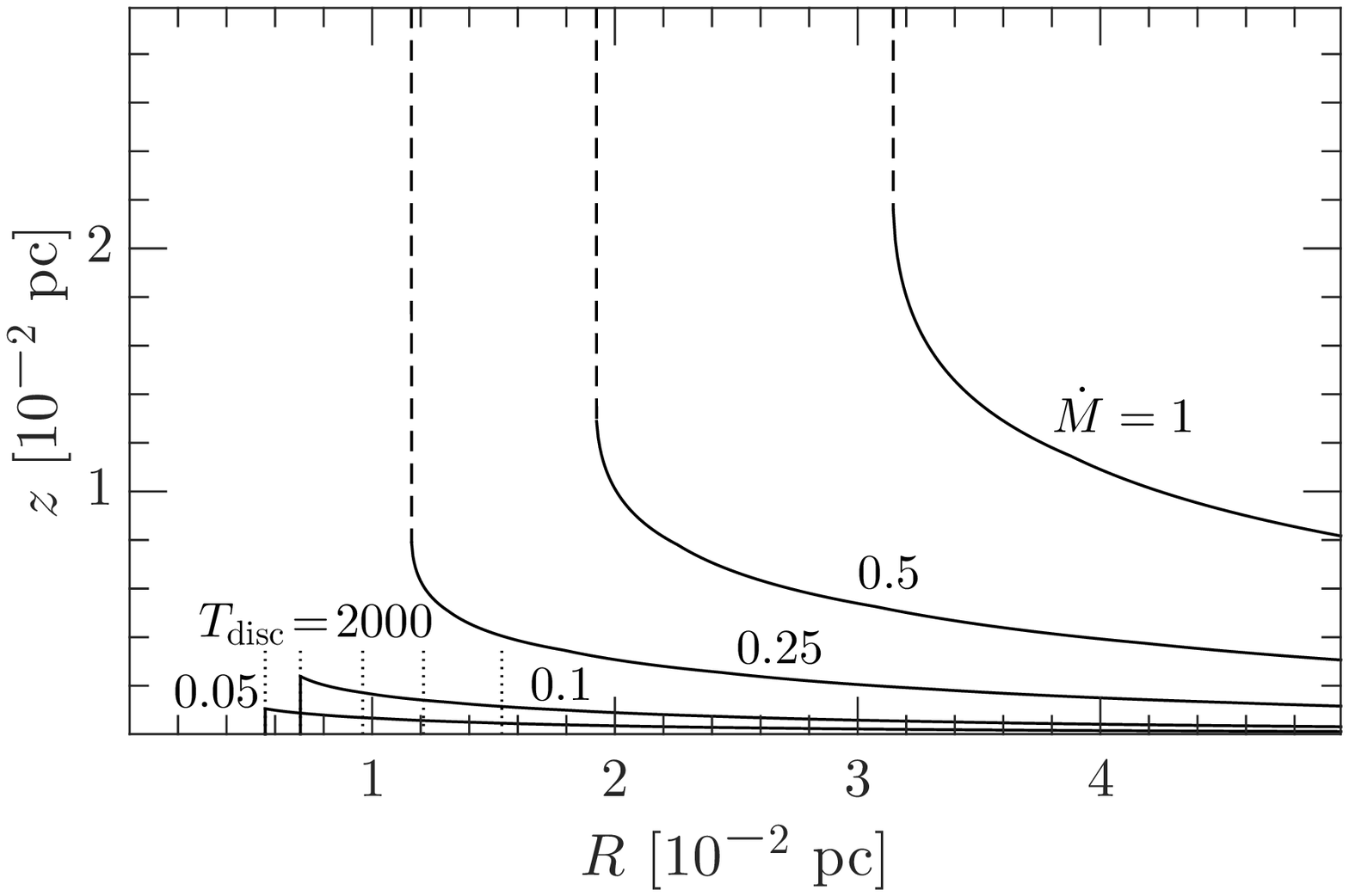}
\caption{The inflated AD structure for different values of $\mdot_1$. 
The solid curve is the static solution height, $H(R)$, set by the condition that $a_{\rm rad}=a_{\rm BH}$.
The dust is assumed to have $Z/Z_{\odot}=5$, and $\amax=1$~\mic. The solution becomes unbound, 
i.e.\ $a_{\rm rad}>a_{\rm BH}$ for all $z$, below a radius $R$ depicted by a transition to a dashed vertical line,
which occurs at $H/R=1/\sqrt{2}$. The disc $R_{\rm in}$ (where $T_{\rm eff}=2000$~K), 
is marked by the short vertical dotted line (increasing $\mdot_1$ from left to right). Below 
this radius $a_{\rm rad}=0$, and the inflated structure disappears. For $\mdot_1<0.25$ the inflated structure
extends down to $R_{\rm in}$. For $\mdot_1\ge 0.25$, a wind inevitably forms. 
The solution assumes no illumination by the central optical-UV source.}
\label{fig:static_solution_vs_Mdot}
\end{figure}

Figure~\ref{fig:static_solution_vs_Mdot} presents the derived $H(R)$ (solid line) for the
 specific case of an AD with $M_8=1$, and five different values of $\mdot_1$ in the range 0.05--1.
For $\epsilon=0.1$, these accretion rates correspond to $L_{46}=0.028-0.57$, or luminosity
in Eddington units $\dot{m}=0.023-0.45$, where 
\begin{equation} 
\dot{m}\equiv \frac{L}{L_{\rm Edd}}=0.8L_{46}M_8^{-1}\ .
\end{equation} 
The disc atmosphere is assumed to be dusty down to $R_{\rm in}$ 
(where $T_{\rm eff}=2000$~K, eq.~\ref{eq:Rin}), indicated by the five short vertical dotted lines 
for the five values of $\mdot_1$. At $R<R_{\rm in}$ the disc atmosphere is dustless,
and the solution for $H$ returns to the thin, gas pressure supported, vertical structure \citep{SS73}.
For the dust opacity we use an MRN dust, composed of
only graphite grains, with $a_{\rm max}=1$~\mic, and metallicity $Z/Z_\odot=5$,
which produces $\kappa_{50}=5.3$ at $R_{\rm in}$.

The static solution for $H(R)$ in Fig.~\ref{fig:static_solution_vs_Mdot} uses the exact expression (eq.~\ref{eq:a_gz}), rather than the analytic approximation for $a_{\rm BH}(R,z)=GMz/R^3$, 
which is valid for $H/R\ll 1$.
The exact expression implies that $a_{\rm BH}$ reaches a maximum, $a_{\rm BH, max}$, at 
$H=R/\sqrt{2}$. A large 
enough value of $\mdot_1$ leads to $a_{\rm rad}>a_{\rm BH, max}$ at some radius in the dusty region
($R>R_{\rm in}$), below which
a static solution is no longer possible. This situation occurs for the three 
higher $\mdot_1$ models, as indicated by the dashed vertical lines in Fig.~\ref{fig:static_solution_vs_Mdot},
where the solution for $H(R)$ diverges. The two lowest $\mdot_1$ models of 0.05 and 
0.1, do not reach $a_{\rm rad}>a_{\rm BH, max}$ in the dusty region, the disc height remains finite,
and the inflated disc extends inwards down to $R_{\rm in}$. The condition $a_{\rm rad}>a_{\rm BH, max}$
may imply a wind, as further discussed below.

\subsubsection{The effect of the central optical-UV illumination on the solution for $H(R)$}

The discussion above of $H(R)$ includes only the vertical support of the local IR disc emission. 
However, once the dusty gas is lifted upwards, it gets exposed to 
the optical-UV radiation from the inner accretion disc. This radiation may sublimate the grains, and thus eliminate the opacity which supports the inflated disc structure. This radiation may also affect the grain dynamics, as discussed briefly below.

The grains are illuminated by the standard AGN SED (\citealt{paperII}), with an intermediate far UV to soft X-rays slope of $\aion=-1.6$. The required flux ($=\sigma T_{\rm eff}^4$) 
to reach $\Tgr=2000$~K for a grain of size
$a$, $F_{\rm sub, a}$, is presented in Fig.~\ref{fig:Tgr_vs_T_eff}. A grain at a given $R$ will therefore sublimate
once it is elevated to $z>z_{\rm sub}(R,a)$, where $z_{\rm sub}(R)$ is derived from the relation
\begin{equation}
F_{\rm sub, a}=\frac{2 L_{\rm bol} \mu}{4\pi (R^2+z_{\rm sub}^2)}\ , 
\label{eq:T_sub_simpl}
\end{equation}
where $\mu=z_{\rm sub}/\sqrt{R^2+z_{\rm sub}^2}$.
This expression assumes a flat disc $\cos \theta$ illumination, i.e.\ an observed luminosity $L_{\rm obs}=2\mu L_{\rm bol}$,
as expected from thin Newtonian disc emission. Relativistic
effects will enhance the luminosity close to disc plain (e.g.\ \citealt{LaorNetzer89}, fig.~8 there). 
But, the innermost disc may become geometrically thick, which will suppress the 
luminosity close to the disc plain due to self-shielding. In addition, the innermost thin disc emission 
appears to be generally missing (e.g.\ \citealt{LaorDavis14}), and it is not entirely clear what is 
the geometry of the extreme UV continuum source. We therefore use below the $\cos\theta$ dependence
as a simple example to follow. We also present the results for an isotropic illumination. 

The value of the sublimation $T_{\rm eff}$ of the $a=1$~\mic\ graphite grains can be derived analytically, 
as $\Tgr=T_{\rm BB}$ for such large grains (Fig.~\ref{fig:Tgr_vs_T_eff}). 
Using $T_{\rm eff}=4^{1/4}T_{\rm BB}$ (one side illumination of a spherical grain), implies $T_{\rm eff}=2828$~K.
The sublimation values  of $T_{\rm eff}$ for the $a=0.1$~\mic\ and $a=0.01$~\mic\ are 2023~K and 1152~K, respectively (Fig.~\ref{fig:Tgr_vs_T_eff}). 

A simple solution for $z_{\rm sub}$, based on eq.~\ref{eq:T_sub_simpl}, valid for  $z_{\rm sub}/R\ll 1$, is
\begin{equation}
z_{\rm sub, pc}=(21.7, 5.7, 0.60)\times L_{46}^{-1}R_{\rm pc}^3\ ,
\label{eq:z_sub}
\end{equation}
where the three coefficients are for the $a=1$, 0.1, and $0.01$~\mic\ grains, respectively.

\begin{figure}
\includegraphics[width=\columnwidth]{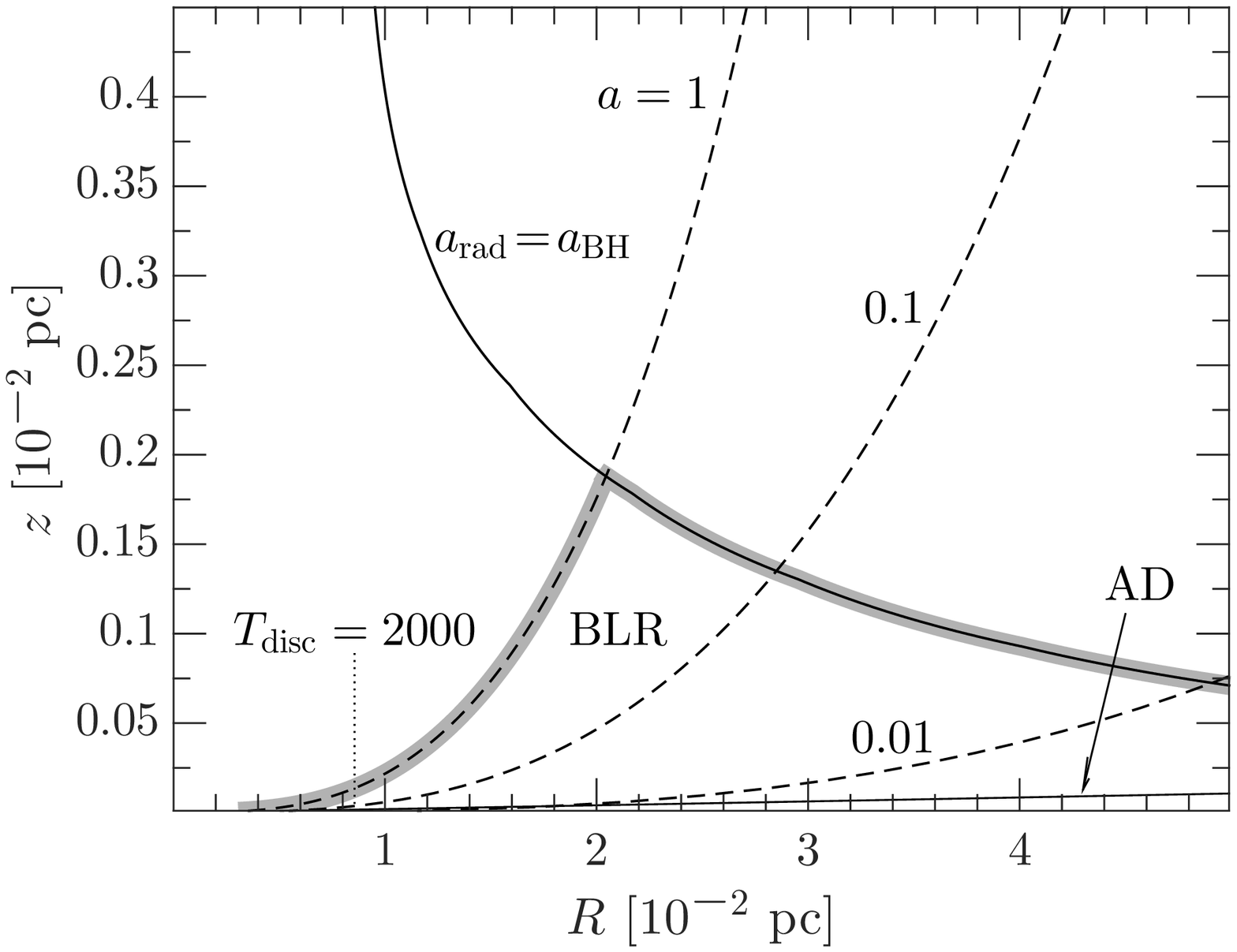}
\caption{The overall inflated AD structure (thick grey line), 
including the central source illumination. The thin solid line is the
static solution height, $H(R)$ for $\mdot_1=0.176$, which corresponds to $L_{46}=0.1$ for $\epsilon=0.1$.
The dust is assumed to have $Z/Z_{\odot}=5$, and $\amax=1$~\mic. The line marked AD is the surface for a dustless atmosphere supported by gas pressure only. The dashed lines designate the
sublimation height $z_{\rm sub}(R)$ of grains of $a=1$, 0.1, and $0.01$~\mic, illuminated by the central continuum source. At $R<2\times 10^{-2}$~pc, the disc surface is set by $z_{\rm sub}(R)$ 
of the $a=1$~\mic\ grains. 
At $z>z_{\rm sub}$, all grains sublimate, and the $H(R)$ solution is not valid. At $R>2\times 10^{-2}$~pc, the disc surface is set by
$H(R)$. The gas in the volume between the $a=1$~\mic\ and the $a=0.1$~\mic\ curves is an efficient line emitter,
and naturally sets the position of the BLR emission.}
\label{fig:static_solution}
\end{figure}

Figure~\ref{fig:static_solution} presents the solutions for $z_{\rm sub}(R)$ for grains with
$a=1$, $0.1$ and $0.01$~\mic, and also the $H(R)$ solution with no illumination. The results are for the
specific AD model with $M_8=1$, $L_{46}=0.1$ and $\epsilon=0.1$, which corresponds to $\mdot_1=0.176$. 
The dust opacity used for the $H(R)$ solution is of MRN dust composed of
only graphite grains, with $a_{\rm max}=1$~\mic\ and metallicity $Z/Z_\odot=5$.   
For these AD parameters, dust can survive in the disc atmosphere down to
$R_{\rm in, 2000}=0.009$~pc (eq.~\ref{eq:R_2000}, equals here 1800$R_g$), 
noted by the vertical dotted line in the plot.

No dust grains exists above $z_{\rm sub}(R)$ for the 
$a=1$~\mic\ grains, as these grains are large enough to be as cold as a local blackbody. 
Larger grains will also have $\Tgr=T_{\rm BB}$, regardless of their size, and will therefore also
be too hot to survive at $z>z_{\rm sub}$. As shown in Fig.~\ref{fig:Rsub_vs_a}, $T_{\rm BB}$ is already reached
at $a \simeq 0.3$~\mic, and therefore all $a\ge 0.3$~\mic\ grains sublimate above the $z_{\rm sub}(R)$ solution
for the $a=1$~\mic\ grains.\footnote{Note that the $a=1$~\mic\ sublimation curve starts slightly inwards of 
$R_{\rm in}$, as it corresponds to grains at the atmosphere surface ($\tau=0$) where the grains are somewhat
cooler than $T_{\rm eff}$, since the heating is from $2\pi$ and cooling is to $4\pi$.}

As shown above, the value of $\kappa$ for dust in the near IR remains unaffected, or enhanced,
as long as the largest grains survive. Therefore, the $z_{\rm sub}(R)$ solution for the
$a=1$~\mic\ grains (valid for $a>0.3$~\mic) represents the disc surface, above this height there is no dust
to support the gas against the BH gravity, regardless of the grain size. 
This sublimation height solution for the disc surface, is valid up to a radius $R_{\rm max}$, where 
$z_{\rm sub}(R_{\rm max})=H(R_{\rm max})$. At $R>R_{\rm max}$, the disc surface is set by the $H(R)$ 
solution. 

The dust therefore leads to an inflated disc structure, where the disc becomes thicker with decreasing
$R$ due to the increasing dust opacity with the rising surface temperature. This extends down to the radius 
where the illuminated dust sublimates. The resulting torus-like structure extends further inwards, 
getting thiner to avoid sublimation, until it reaches the innermost radius where the disc atmosphere
becomes too hot to harbour dust.

\subsubsection{The implied position and CF of the BLR} \label{sec:sub_static:sub:implied_RandCF}

The inner surface, defined by $z_{\rm sub}(R)$ of the $1$~\mic\ grains at $R<R_{\rm max}$, 
is directly illuminated by the central ionising continuum, and therefore serves
as a natural place for the BLR gas. Since the gas just below the surface still harbours the largest grains,
the dust near IR opacity is maintained, providing the vertical support, but the UV opacity is reduced 
by an order of magnitude (see Fig.~\ref{fig:kappa_800_vs_amin}), allowing efficient line emission.

Is the distance and CF of this inflated disc, torus-like, structure consistent with observations of the BLR?
Below we derive $R_{\rm max}$, $H(R_{\rm max})$, and the implied CF 
based on $H(R_{\rm max})/R_{\rm max}$.

To derive $R_{\rm max}$, we equate $z_{\rm sub}(R)$ (eq.~\ref{eq:z_sub}) with $H(R)$ (eq.~\ref{eq:H_mdot1}), 
which gives
\begin{equation}
  R_{\rm max}=0.044(L_{46}Z/Z_{\odot})^{0.26}M_8^{0.075}\mdot_1^{0.33}~{\rm  pc}\ . \label{eq:R_max}
\end{equation}
Since the directly illuminated smaller grains sublimate further outside, 
the coefficients for the $0.1$ and $0.01$~\mic\ grains are 0.062 and 0.11~pc, respectively. 

At $R_{\rm in}<R<R_{\rm max}$, the illuminated disc surface gas is free of dust.
The illuminated dusty gas which resides just below the disc surface (denoted by a grey thick line in Fig.~\ref{fig:static_solution}), and above the $z_{\rm sub}(R)$ curve for the $a=0.1$~\mic\ grains, 
can contain large grains, which have some effect on the line emission. In gas 
below the $a=0.1$~\mic\ sublimation curve, and above the $a=0.01$~\mic\ curve, line 
suppression becomes more significant, and below the $a=0.01$~\mic\ sublimation curve
line emission is heavily suppressed. 

Using the radiative efficiency $\epsilon$ and $L_{\rm bol}$, instead of $\mdot_1$ (eq.~\ref{eq:epsilon}) 
we get  
\begin{equation}
  R_{\rm max}=0.025L_{46}^{0.59}M_8^{0.075}Z/Z_{\odot}^{0.26}\epsilon^{-0.33}~{\rm  pc}\ , \label{eq:R1_max}
\end{equation}
which gives for the plausible values of $\epsilon=0.1$ and $Z/Z_{\odot}=5$, 
 \begin{equation}
R_{\rm max}=0.08L_{46}^{0.59}M_8^{0.075}~{\rm  pc}\ .
\label{eq:R2_max}
\end{equation}
The coefficients for the $0.1$ and $0.01$~\mic\ grains are 0.11 and 0.2~pc, respectively.
Thus, efficient line emission of directly illuminated gas extends  
out to 0.11~pc, and effectively disappears beyond 0.2~pc (for $L_{46}=1$,  
$M_8=1$). Note that the predicted dependence on $M_{\rm BH}$ is very weak.
These results are consistent with the observed relation of 
$R_{\rm BLR}=0.1L_{46}^{0.5}~{\rm  pc}$ (eq.~\ref{eq:R_BLR}).

If the grain size distribution extends only to $a_{\rm max}=0.25$~\mic\ (which sublimates at 
$T_{\rm eff}=2728$~K, rather than 2828~K for $a_{\rm max}=1$~\mic), then using 
$\kappa$ from eq.~\ref{eq:kappa}, leads to similar relations 
(see eqs.~\ref{eq:R_max} and \ref{eq:R1_max}), with
\begin{equation}
  R_{\rm max}=0.037(L_{46}Z/Z_{\odot})^{0.22}M_8^{0.11}\mdot_1^{0.33}~{\rm  pc}\ , 
\end{equation}
and
\begin{equation}
  R_{\rm max}=0.021L_{46}^{0.55}M_8^{0.11}(Z/Z_{\odot})^{0.22}\epsilon^{-0.33}~{\rm  pc}\ . 
\end{equation}
Given the similar solutions, we present below the results only for the $a_{\rm max}=1$~\mic\ case. 

The CF is set by $H_{\rm max}/R_{\rm max}$, which is
\begin{equation}
  H_{\rm max}/R_{\rm max}=0.040L_{46}^{-0.49}M_8^{0.15}\mdot_1^{0.67}(Z/Z_{\odot})^{0.51}\ ,
\end{equation}
or, using $\epsilon$ and $L_{46}$,
\begin{equation}
  H_{\rm max}/R_{\rm max}=0.013L_{46}^{0.18}M_8^{0.15}\epsilon^{-0.67}(Z/Z_{\odot})^{0.51}\ ,
\label{eq:HR11}
\end{equation}
which gives for $\epsilon=0.1$ and $Z/Z_{\odot}=5$, 
\begin{equation}
  H_{\rm max}/R_{\rm max}=0.14L_{46}^{0.18}M_8^{0.15}\ .
\label{eq:HR}
\end{equation}
How does this compare with observations?

For an isotropic ionising continuum emission, and an inflated structure with a half opening angle
$\theta_0$, the fraction of the ionising continuum intercepted by the surface is 
$\int_0^{\mu_0}d\mu=\mu_0$, where $\mu_0=\cos\theta_0$. Since the fraction derived for 
the observed BLR line strength is $\sim 0.3$, 
it implies $\theta_0=72.5\degr$, or $H_{\rm max}/R_{\rm max}=0.31$. However, for the 
non-isotropic ionizing continuum assumed here, which is used for the derivation of $H_{\rm max}/R_{\rm max}$,
the absorbed fraction is $\int_0^{\mu_0}2\mu d\mu=\mu_0^2$. This implies $\mu_0=0.55$,
a half opening angle of $\theta_0=57\degr$, or $H_{\rm max}/R_{\rm max}=0.65$.
Thus, for the plausible $\epsilon$ and $Z$ values used above, the static disc solution (eq.~\ref{eq:HR}) 
falls short by about a factor of 4.5 in making the disc thick enough for the estimated CF of the BLR. As we show below, a dynamic solution, and ablation of the disc surface,
are expected to increase the CF significantly.

\subsubsection{Isotropic illumination}

What is the effect of the isotropy of $L_{\rm bol}$ on the values of $R_{\rm max}$ and CF of the illuminated inflated disc structure? Below we derive these parameters when $L_{\rm obs}$ is isotropic.

In this case, $R_{\rm sub}$ is given by the relation
\begin{equation}
L_{\rm bol}/4\pi R_{\rm sub}^2=F_{\rm sub, a}\ , 
\end{equation}
which gives $R_{\rm sub}=0.15L_{46}^{1/2}$~pc, for
the $a=1$~\mic\ grains, and coefficients of 0.30 and 0.91~pc for the $a=0.1$ and $0.01$~\mic\ grains, respectively.
The bulk of the line emitting volume is then at $(0.15-0.3)\times L_{46}^{1/2}$~pc, which is
a factor of $\sim 2$ larger than $R_{\rm BLR}$. The disc thickness at $R_{\rm sub}$ 
is obtained from the  solution for $H(R)$ (eq.~\ref{eq:H_mdot1}), 
which gives
\begin{equation}
  H/R=4.3\times 10^{-4}L_{46}^{0.36}M_8^{0.29}\epsilon^{-1.29}Z/Z_{\odot}\ .
\end{equation}
The adopted parameter values ($\epsilon=0.1$, $Z/Z_{\odot}=5$, $L_{46}=1$, $M_8=1$) give
$H/R=0.042$, which is a factor $\sim 7$ smaller than the value of $H/R=0.3$ required for the isotropic illumination.

Thus, although dust can survive within the disc atmosphere down to $R_{\rm in}$, which is a factor of a few below 
$R_{\rm BLR}$ (eq.~\ref{eq:Rin}), isotropic illumination of the disc by the central optical-UV source, 
sublimates this dust out to $R_{\rm sub}$. The inflated disc structure therefore starts $R_{\rm sub}$ 
rather than at $R_{\rm in}$, leading to a smaller CF. Also, no significant line emission is expected from 
$R\simeq R_{\rm in}$, in contrast with RM results of the higher ionisation lines.

Self-shielding, either due to a flat disc $\cos\theta$ illumination, or as a self-obscuration 
effect of the outer disc from the ionising inner disc, yields
a smaller $R_{\rm max}$ and a larger CF.

\subsubsection{The effect of the AD self-gravity}\label{sec:ADself}

The derivation above includes only the $z$ component of the gravity of the BH. 
Is it valid to ignore the AD self-gravity?

The AD self-gravity provides an additional contribution, 
\begin{equation}
 a_{\rm disc}(R) = 2\pi G \Sigma_{\rm AD}(R)\ , \label{eq:f_disc}
\end{equation}
where $\Sigma_{\rm AD}(R)=\int \rho_{\rm AD}(R,z)dz$ is the column density of the AD,
and $\rho_{\rm AD}(R,z)$ is the AD gas density. The correct expression for the AD height is
derived from the condition 
\begin{equation}
 a_{\rm rad}(R) = a_{\rm BH}(R,z)+a_{\rm disc}(R)\ . \label{eq:f_grav}
\end{equation}
 The disc self-gravity modifies significantly the $H(R)$ solution when $a_{\rm disc}(R)>a_{\rm BH}(R,H)$, i.e.\
when 
\begin{equation}
\Sigma_{\rm AD}>M_{\rm BH}H/2\pi R^3
\label{eq:Mdlimit}
\end{equation}
(for $H/R\ll 1$). The disc mass is roughly
$M_{\rm disc}\sim \pi R^2 \Sigma_{\rm AD}$, and this criterion translates to $M_{\rm disc}> M_{\rm BH}\times H/2R$. 
Since $H/R\sim 0.5$ (to provide the required CF),
the accretion disc self-gravity becomes significant when $M_{\rm disc}\sim M_{\rm BH}$.
Can the AD be that massive at $R_{\rm BLR}$?

Likely not, for the following reason. The accretion time scale, $t_{\rm ac}(R)\equiv M_{\rm disc}(R)/\mdot$, satisfies $t_{\rm ac}\sim R/v_r$, 
where $v_r$ is the radial accretion velocity. For the above parameters we get $t_{\rm ac}\simeq 10^8$~yr,
which is of the order of the lifetime of a quasar.\footnote{If this applies, a self-gravitating disc at the BLR contains
enough mass to fuel a quasar for its lifetime, and no accretion from larger scales is required.}
Since $v_r\sim R/t_{\rm ac}$, such a long accretion time implies an extremely small radial velocity of $v_r\sim 
100$~cm~s$^{-1}$, or $v_r/v_{\rm Keppler}=3\times 10^{-7}$, where 
$v_{\rm Keppler}=\sqrt{GM_{\rm BH}/R}\sim 3\times 10^8$~cm~s$^{-1}$ is a typical value at the BLR. 
Is such a low $v_r$ plausible? In the $\alpha$-disc models
$v_r/v_{\rm Keppler}=\alpha\times (H/R)^2$ (e.g.\ \citealt*{Frank02}), which implies $\alpha\sim 10^{-6}$.
Typical values suggested are $\alpha\sim 0.1$, which implies a much larger $v_r$ and thus
$M_{\rm disc}\ll M_{\rm BH}$. Thus, a significant contribution from the AD self-gravity appears implausible.
However, since the value of $\alpha$ remains unknown, the required extremely small 
value cannot be excluded.

In the standard \citep{SS73} $\alpha$-disc solution, which does not include dust opacity,
the radius where self-gravity dominates (for AGN discs) is actually typically smaller than $R_{\rm BLR}$
(e.g.\ \citealt{LaorNetzer89}, eq.~18 there). This occurs for the following 
reason. In the absence of dust, the outer disc is extremely thin 
with $H/R\sim 10^{-3}$. Since $a_{\rm BH}\propto H$, it is correspondingly smaller, 
while $a_{\rm disc}$ is independent of $H$. Therefore, the condition 
$a_{\rm disc}>a_{\rm BH}$ already holds for $M_{\rm disc}/M_{\rm BH}\sim 10^{-4}$, which occurs 
at $R<R_{\rm BLR}$.

The condition to get an inflated disc when the disc self-gravity dominates,
is  $a_{\rm rad}> a_{\rm disc}$, or equivalently  
\begin{equation}
\frac{F}{\Sigma_{\rm AD}}>\frac{2\pi G c }{\kappa }\ ,
\end{equation}
which is simply the Eddington limit
\begin{equation}
\frac{L}{M}>\frac{4\pi G c }{\kappa }\ ,
\end{equation}
per unit area (the factor of two difference reflects the planar versus spherical geometry).\footnote{This may lead to the observed relation between the star formation rate per unit area ($\propto F$)
and gas mass per unit area ($\propto \Sigma_{\rm AD}$) 
in disc galaxies, a.k.a. the Kennicutt-Schmidt law \citep*{Thompson05}.}
Using the AD expression for $F$ (eq.~\ref{eq:ADflux}),
and $M_{\rm disc}\sim \pi R^2 \Sigma_{\rm AD}$, the condition for an inflated disc is
\begin{equation}
M_{\rm disc}/M_{\rm BH}<\frac{3}{16\pi}\frac{\kappa \mdot}{Rc}=\frac{H}{2R} ,
\end{equation}
which is identical to the condition derived above to get $a_{\rm BH}> a_{\rm disc}$.
So, we conclude that if the AD is massive enough for self-gravity to dominate, it will likely also 
not be inflated by radiation pressure.

Note that at large enough $R$ the disc self-gravity inevitably becomes dominant. 
This occurs since $a_{\rm rad}\propto \kappa F\propto R^{-3.87}$, while $a_{\rm disc}\propto 
\Sigma_{\rm AD} \propto (Rv_r)^{-1}$, so their ratio 
$a_{\rm disc}/a_{\rm rad}\propto R^{2.87} v_r^{-1}$ rises sharply with $R$, and self-gravity inevitably wins.

\subsection{The dynamic solution}\label{sec:dyn_model}
	
The above solution assumes a static disc atmosphere. However, if dust forms in the disc atmosphere intermittently, then once
it forms the local opacity jumps drastically, leading to a rapid acceleration upwards
of the dusty patch 
by the underlying AD thermal IR radiation pressure.
Once the static solution height is passed, deceleration starts. But, the gas will continue to
rise upwards until it stops, and then falls back to the disc.
The maximal hight $z_{\rm dyn}(R)$ reached by the dusty patch, is where the net work in the $\hat{z}$-direction equals zero. 
 
We find the dynamic solution by solving numerically the value of $z_{\rm dyn}(R)$ which satisfies 
the following equation
\begin{equation}
	W(R,z_{\rm dyn}) \equiv E_{\rm rad}(R,z_{\rm dyn}) - E_{\rm grav}(R,z_{\rm dyn}) = 0. \label{eq:work}
\end{equation}
The work done by the radiation pressure is
\begin{equation}
	E_{\rm rad}(R,z) = \int_{z_{\rm AD}(R)}^{\min[z,z_{\rm sub}(R)]} \sigma T_{\rm eff}^4(R)\frac{1}{c}\kappa[z^{\prime} ,T_{\rm eff}(R)]\dd z^{\prime} ,\label{eq:E_rad}
\end{equation}
where $z_{\rm AD}(R)$ is the disc height in the absence of dust, which generally satisfies 
$z_{\rm AD}(R)\ll z_{\rm sub}(R)$. The work done by gravity is
\begin{equation}
	E_{\rm grav}(R,z)  =  \frac{G M_{\rm BH}}{\sqrt{R^2+z_{\rm AD}^2}}-\frac{G M_{\rm BH}}
{\sqrt{R^2+z^2}} 
 + 
  2\pi G \Sigma_{\rm AD}[z-z_{\rm AD}]\ . 
\label{eq:E_grav}
\end{equation}

\subsubsection{An approximate analytic solution}

As in the analytic static solution, we assume $z_{\rm dyn}\ll R$, and keep terms of $z/R$ up to second order; neglect the 
disc self-gravity term (right hand side of eq.~\ref{eq:E_grav}); assume that $\kappa$ is independent of $z$, and 
$z_{\rm AD}(R)=0$.

Using the above approximations, eq.~\ref{eq:E_rad} gives
\begin{equation}
E_{\rm rad}(R,z) = \frac{3}{8\pi}\dot{M}\frac{G\Mbh}{R^3}\frac{1}{c}\kappa(R) \min[z,z_{\rm sub}(R)];
\end{equation}
and eq.~\ref{eq:E_grav} gives
\begin{equation}
E_{\rm grav}(R,z)=\frac{1}{2}G M_{\rm BH}\frac{z^2}{R^3}. \label{eq:E_grav_approx}
\end{equation}
Eq.~\ref{eq:work} is satisfied for
\begin{equation}
z_{\rm dyn}= \frac{3}{4\pi}\dot{M}\frac{1}{c}\kappa(R)\ , 
\end{equation}
for  $z_{\rm dyn}<z_{\rm sub}$, and
\begin{equation}
z_{\rm dyn}= \sqrt{2H(R)z_{\rm sub}} \ , 
\label{eq:z_dyn}
\end{equation}
for $z_{\rm dyn}>z_{\rm sub}$. 
Thus, $z_{\rm dyn}$ is simply twice the static solution $H(R)$ at $R>R_{\rm max}$,
while at $R<R_{\rm max}$, it is a geometric mean of $z_{\rm sub}$ and twice $H(R)$.

How is $R_{\rm max}$ modified in the dynamic solution? The maximum disc height occurs when
the two solutions above cross, i.e.\ where $2H(R)=\sqrt{2H(R)z_{\rm sub}}$, or
$z_{\rm sub}(R)=2H(R)$. Since $H\propto R^{-0.87}$ and
$z_{\rm sub}\propto R^3$, $R_{\rm max}^{3.87}$ increases by a factor of 2, and 
$R_{\rm max}$ by a factor of $\simeq 1.2$. The functional dependence on all the parameters remains unchanged
(eqs.~\ref{eq:R_max}--\ref{eq:HR}). Specifically, we get $R_{\rm max, pc}=0.1L_{46}^{0.59}M_8^{0.075}$
for $\epsilon=0.1$, and $Z/Z_{\odot}=5$ (eq.~\ref{eq:R2_max}).

How is the CF modified? Since CF$\propto H/R$ at $R_{\rm max}$, then the CF increases by a 
factor of $2/1.2=1.67$.
Specifically, we get
\begin{equation}
  H/R=0.23L_{46}^{0.18}M_8^{0.15}\ .
\label{eq:HR1}
\end{equation}
for $\epsilon=0.1$, and $Z/Z_{\odot}=5$ (eq.~\ref{eq:HR}), a factor $\sim 3$ short of the  
required value of $H/R\sim 0.65$.

\subsubsection{The numerical solution}

The numerical dynamic solution differs from the analytic approximation by calculating
$z_{\rm sub}$ as a function of $a$, which yields $a_{\rm min}(z)$. We then calculate
the dust Planck mean $\kappa(z)$ based on $a_{\rm min}(z)$, assuming that $\beta$ and $f_{\rm d}$ do not change. The value of $z_{\rm sub}$ is calculated
including also the local AD flux, in addition to the central source flux as a heating source for the 
dust (cf.\ eq.~\ref{eq:Heat}). As discussed above, the value of $\kappa(z)$ in the near IR is only weakly dependent on the grain size distribution,
and the value of $f_{\rm d}$ will also not change much, as most of the dust mass reside in the largest grains (Fig.~\ref{fig:Rkappa}).

The integration in eq.~\ref{eq:E_rad} is done numerically with a step size $\dd z= 3 \times 10^{-5}$~pc. At each $z$ we calculate $a_{\rm min}(z)$, and the corresponding  $\kappa (z)$. The AD height ${z_{\rm AD}(R)}$ is adopted from \citet{SS73}. 
In the explored range of $R$, ${z_{\rm AD}(R)}< \dd z$. Thus, our results are to a good approximation independent of 
${z_{\rm AD}(R)}$.
 
\begin{figure}
\includegraphics[width=\columnwidth]{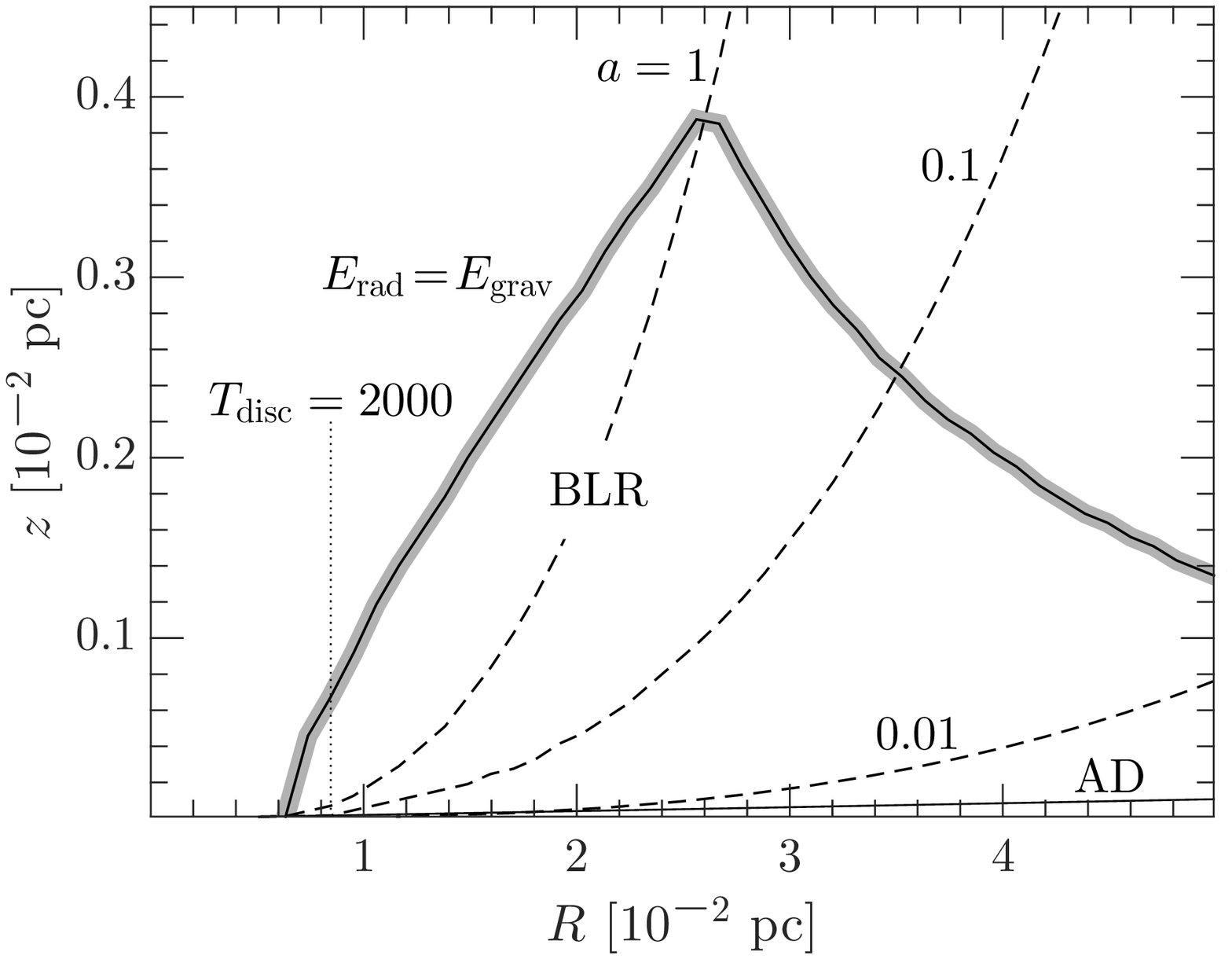}
\caption{Same as Fig.~\ref{fig:static_solution}, for a dynamic solution of the BLR vertical structure. The solution curve (thick grey line) is defined by $z(R)$ at which the total work executed on a parcel of gas from the AD surface (thin solid line) is zero. The sublimation curves (dashed lines) are as in Fig.~\ref{fig:static_solution}. 
Beyond the maximal disc height at $R>2.5\times 10^{-2}$~pc, the disc height is about twice the static solution height.
At $R<2.5\times 10^{-2}$~pc, $a_{\rm rad}=0$ above the highest dashed line, where all dust sublimates. However,
the dustless gas can climb up to the solid line, where it halts and falls back to the disc. 
The CF is about 1.7 larger than in the static solution, but it is still short by a factor of $\sim 3$ of the
observed CF of the BLR. }
\label{fig:dynamic_solution}
\end{figure}

Figure~\ref{fig:dynamic_solution} presents the solution $z_{\rm dyn}(R)$ of eq.~\ref{eq:work}. It also shows the sublimation curves $z_{\rm sub}(R)$ for $a=1$, $0.1$ and $0.01$~\mic. The curves are shifted slightly outward compared to $z_{\rm sub}$ of the static solution (Fig.~\ref{fig:static_solution}) due to the inclusion of grain heating by the local radiation of the AD. This contribution is small as the local disc $T_{\rm eff}\ll 2000$~K 
(eq.~\ref{eq:RinRblr}). The inflated disc structure starts at $R_{\rm in, 2000}=0.8\times10^{-2}$~pc and reaches
a maximal height of $z_{\rm dyn}\simeq0.39\times 10^{-2}$~pc, about twice the height of the static solution.
The value of $R_{\rm max}\simeq 2.5\times10^{-2}$~pc, is about 1.2 times the static solution (Fig.~\ref{fig:static_solution}). The changes in $H(R_{\rm max})$ and $R_{\rm max}$, compared to the static case, are about as expected from the above analytic estimates.

\subsection{The predicted dependence of $R_{\rm max}$ and the CF on the model parameters}

The numerical result above, for the model parameters $L_{46}=0.1$, $M_8=1$, $\epsilon=0.1$ and $Z/Z_{\odot}=5$,
shows that a dusty disc atmosphere produces an inflated structure with a peak position $R_{\rm max}$ within 
$\sim 20$ per cent of the observed $R_{\rm BLR}$, and with a CF which is a factor of $\sim 3$ too low (see a further discussion below). The predicted dependence of
$R_{\rm max}$ and CF on the model parameters allows to explore the validity of the dust inflated AD as
the source of the BLR. 

\begin{figure*}
\includegraphics[width=2\columnwidth]{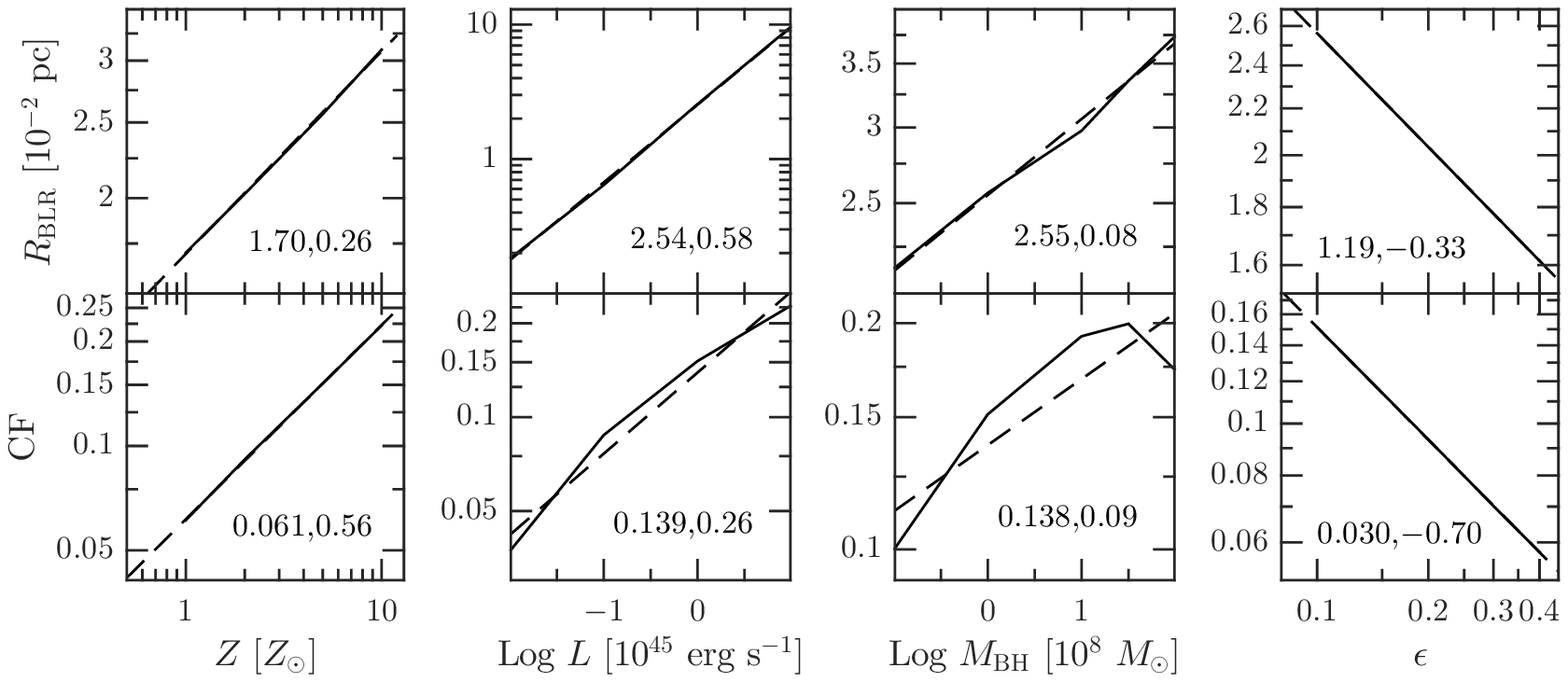}
\caption{The dependence of $R_{\rm BLR}$ (=$R_{\rm max}$) and CF, as derived from the numerical results,
on the model free parameters: $L$, $M_{\rm BH}$,
$\epsilon$ and $Z$. In each panel, one parameter is varied, while others are kept constant at $Z/Z_{\odot}=5$, 
$L_{46}=0.1$, $M_8=1$ and $\epsilon=0.1$. The numerical model results (solid line) are fitted by a power law (dashed line), i.e.\ $Y=aX^b$. The values of $a$ and $b$ are noted in each panel. The power-law fit produces a good match to the results for all parameters, with the values expected from the analytic solutions. The decrease in the CF for 
$\Mbh>10^{9.5}$~$M_{\odot}$ results from a corresponding low value of $\dot{m}\la 10^{-3}$ (see text).}
\label{fig:prop_vs_param}
\end{figure*}

Figure~\ref{fig:prop_vs_param} presents the dependence of $R_{\rm max}$ and CF on
$L_{46}$, $M_8$, $\epsilon$ and $Z/Z_{\odot}$, as derived from the numerical solution. In each panel we varied
one parameter, and held the other parameters fixed. The fixed values used are $Z/Z_{\odot}=5$, 
$L_{46}=0.1$, $M_8=1$ and $\epsilon=0.1$. We fit a power-law to each of the sets of runs, of the
form $Y=aX^b$, which provides an excellent fit in most cases. 
The best fit relation is
\begin{equation}
R_{\rm max}\propto L_{46}^{0.58} M_8^{0.08} \epsilon^{-0.33} (Z/Z_{\odot})^{0.26}\ ,
\end{equation}
which is almost identical to the expected dependence from the analytic static solution which yields 
(eq.~\ref{eq:R1_max}), 
\[
R_{\rm max}\propto L_{46}^{0.59} M_8^{0.075} \epsilon^{-0.33} (Z/Z_{\odot})^{0.26}\ .
\]
For the CF we get
\begin{equation}
\mbox{CF}\propto L_{46}^{0.26} M_8^{0.09} \epsilon^{-0.70} (Z/Z_{\odot})^{0.56}\ , \label{eq:CF2}
\end{equation}
versus the analytic solution of
\[
\mbox{CF}\propto L_{46}^{0.18} M_8^{0.15} \epsilon^{-0.67} (Z/Z_{\odot})^{0.51}\ .
\]
The one exception is the dependence of CF on $M_{\rm BH}$, which deviates
significantly from a simple power-law relation. The drop in the CF for $M_{\rm BH}=10^{10}M_{\odot}$ results
from the low $\dot{m}$ for this model, as further discussed below.

\section{Discussion} \label{sec:discussion}

\begin{figure*}
\includegraphics[width=2\columnwidth]{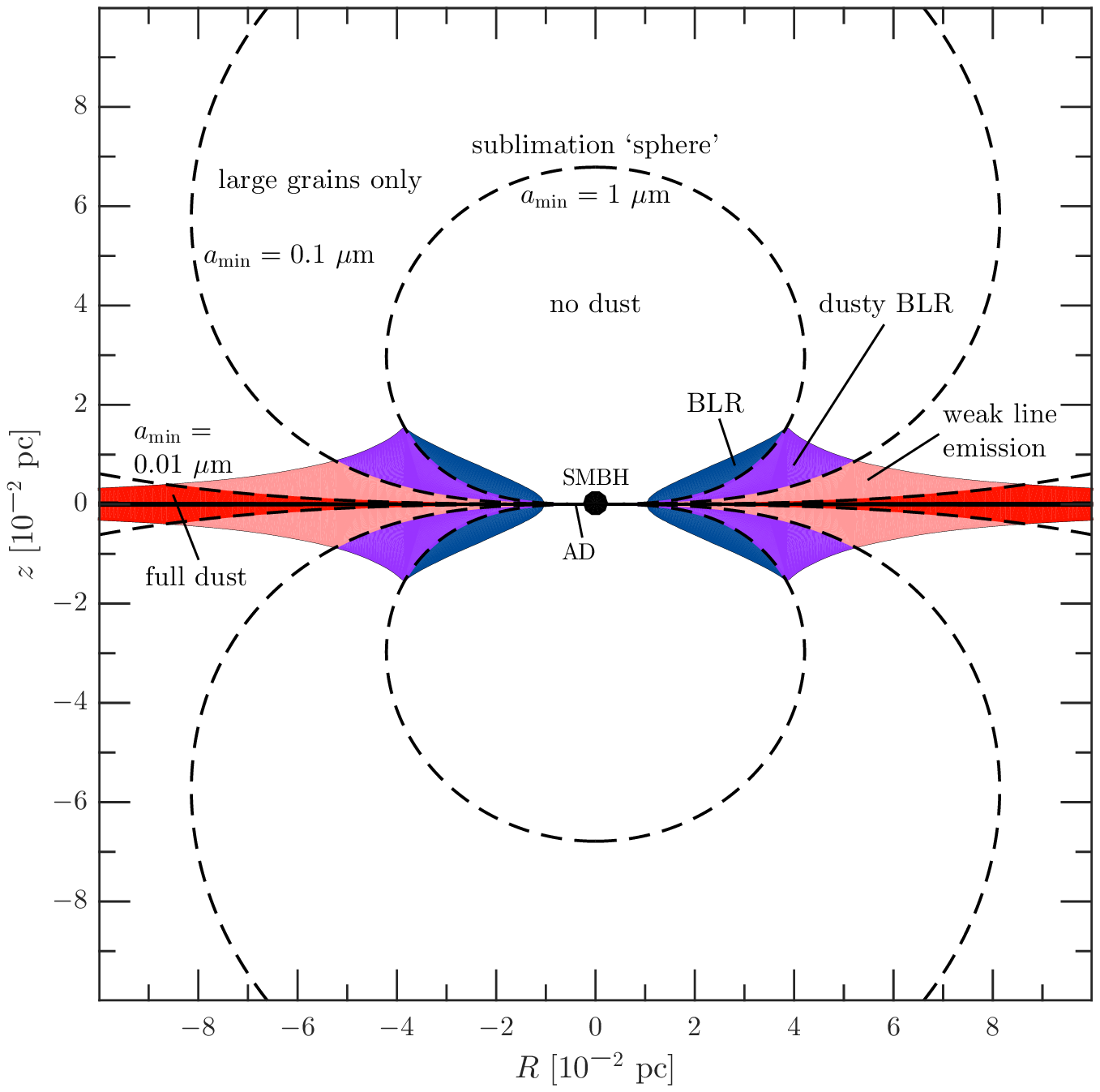}		
\caption{A side view of the dynamic solution disc profile, for $M_8=1$, $L_{46}=0.1$ and
$\epsilon =0.03$. The AD is inflated due to the dust opacity, leading to
a torus structure. The illuminated face is composed of dustless gas, and is a natural origin for 
the observed BLR emission. In the dusty BLR region, only large grains avoid sublimation, and given the reduced
dust opacity in the UV, this region will produce both unreddened line emission and IR emission. 
The back side of the torus is dusty and produces mostly hot dust emission.
The dynamics due to the illuminating optical-UV from the central source is not included.
This radiation pressure will likely drive a tangential sheared wind off the surface photoionized layer,
which will increase the CF of the BLR gas.}
\label{fig:BLR_illustr}
\end{figure*}

The dust inflated disc, or equivalently, the failed dusty disc wind proposed by CH11, 
provides a natural source for the BLR gas. 
Figure~\ref{fig:BLR_illustr} illustrates the global inflated disc structure, based on a dynamic solution with $M_8=1$, $L_{46}=0.1$ and $\epsilon=0.03$. This model allows to make some quantitative predictions, 
discussed below, which can be used to explore the validity of this mechanism.

\subsection{Earlier studies}

The suggestion that part of the broad line emission in AGN originates from the outer regions of the AD
was carefully explored in a series of papers by Collin-Souffrin and Dumont 
(\citealt{Collin87, Collin90, DC90a, DC90b, DC90c}). The AD is assumed to be thin, and the 
CF issue is addressed by assuming a hard
ionising source, which either resides high enough above the AD, or is scattered back on the disc from
an extended hot diffuse medium. Given the assumed hard ionising spectrum, and the relatively high density 
of the AD, this model produces only low ionisation lines. In order to get a non-negligible CF, the
height of the scattering medium needs to be a fair fraction of $R_{\rm BLR}$.
The high ionisation lines are assumed
to originate from shocks in a highly turbulent large scale diffuse medium (e.g. \citealt{Collin88}).
Although the above scenario is possible, it assumes the required components are there, and does not 
lead to robust testable predictions.

A closer look at the outer AD height profile, including the effect of the molecular gas opacity,
was carried out by \citet{Hure94}, which indeed revealed an inflated outer region. However, the
maximal height implies $H/R\sim 10^{-2}$, which is far too small. Indeed, the Planck mean 
opacity derived in that paper at 2000~K (figure 3 there), is about a factor of 100 smaller than
derived here based on the MRN dust model (see also Section~\ref{sec:dust_prop}).  

Similarly, \citet{Czerny17}, which explore the broad line profiles derived from the 
failed dusty wind model of CH11, note in passing the problem of the small CF,
which stems from the low dust opacity used. 

The high MRN dust opacity, or specifically the high opacity of graphite grains, which
survive at $T\simeq 2000$~K, is the crucial parameter. It leads to a $\sim 50$ times
larger dust opacity, and a significantly
inflated disc strucure, which naturally serves as the source of the BLR gas. 

A scattering medium may well be present on the BLR scale, which scatters some of the ionizing
radiation back on the AD. However, such a medium needs to have a rather fine tuned optical depth of
$\tau\sim 0.1-0.3$, so it scatters back enough ionising radiation to power the BLR, but still allows 
a direct view of the bulk of the continuum emission from smaller scales.

\subsection{The innermost dust emitting region}

Gas at $R_{\rm BLR}$, compressed by the incident radiation pressure, reaches a density of 
$n\sim 10^{11}$~\cmt\ (\citealt{paperII}). The sublimation temperature of graphites at this
density is $\sim 2000$~K. The coolest grains, which survive closest inwards, are the large 
($a>0.3$~\mic) graphite grains, and their thermal emission approaches a blackbody. 
Since the peak of $\nu B_{\nu}$ $(=\lambda B_{\lambda})$ for blackbody emission occurs 
at $h\nu=3.92kT$, we expect the hottest dust emission in AGN
to peak around $\lambda\simeq 1.8$~\mic, i.e.\ in the IR $H$ band. 

In the simplified earlier treatments of dust in AGN, dust either exists, or fully sublimates at
the sublimation radius (e.g., \citealt{Barvainis87, NetzerLaor93, Ferguson97,paperII}). Since
dust heavily suppresses the associated gas line emission, this simplified scenario leads to a sharp transition
between an IR emitting region (the so called dusty torus), to a line emitting region (the BLR).  
However, as discussed in \citet{LaorDraine93},
grains of different sizes and compositions, from the smallest silicates to the largest graphites,
sublimate over a large range of radii. Line suppression results from the high UV opacity of dust, 
mostly produced by the smaller ($a<0.1$~\mic) grains. As $R$ decreases, the value of $a_{\rm min}$ increases,
the dust UV opacity drops, and the line emission efficiency gradually rises. 

Since the dust IR opacity remains unchanged (or slightly enhanced) even when only the largest graphite 
grains survive, 
the gas is still supported vertically 
by the local thermal IR disc emission, forming an inflated disc structure. This structure is exposed to the
ionising continuum, and is an efficient line emitter.

Small grains have $T\sim 2000$~K at significantly larger radii. Thus,
some of the $\lambda\simeq 1.8$~\mic\ emission may come from scales $\sim 10$ times larger than the BLR,
if such small grains exist there, and absorb a significant fraction of the central optical-UV continuum.

\subsection{The innermost BLR}

In order to launch a dust driven outflow, the disc surface temperature
needs to satisfy $T_{\rm eff}<\Tsub$ to allow grains to exist. The value of \Tsub\
depends on $n$ at the disc atmosphere, which is not well constrained. However, the dependence 
is weak (eqs.~\ref{eq:n_grap} and \ref{eq:n_sil}), 
and taking the BLR $n\simeq 10^{11}$~cm$^{-3}$ gives $\Tsub=2000$~K. This indicates the BLR 
innermost radius, i.e. the smallest disc radius where a dusty wind can be launched, is
\begin{equation} 
R_{\rm in}=0.016(M_8\mdot_1)^{1/3}T_{2000}^{-4/3}~{\rm pc}, 
\end{equation} 
or $R_{\rm in}=0.018L_{46}^{1/2}$~pc. For $n$ in the range of $10^8-10^{14}$~cm$^{-3}$, one 
gets $\Tsub=1700-2350$~K (Fig.~\ref{fig:T_sub}), which occurs in the disc at 0.8--1.24$R_{\rm in}$,
i.e. a $\pm 0.1$ uncertainty in $\log R_{\rm in}$. The value of $R_{\rm in}$
is about a factor of 5.5 smaller than $R_{\rm BLR}$, based on RM of the Balmer lines.
RM of the \HeII\ lines gives a size which is a factor of $\sim 5$ smaller
than $R_{\rm BLR}$. The \HeII\ size is consistent with RPC modelling, which indicates that the \HeII\ 
emissivity extends 
down to about $0.2R_{\rm BLR}$, in contrast with the low ionisation lines, where RPC yields peak emissivity at 
$R_{\rm BLR}$. Gas further inwards will not emit \HeII\ efficiently,
if its density and ionisation structure is set by RPC.

RPC models predict that the peak emissivity of higher ionisation lines  occurs at even smaller $R$ 
(\citealt{paperII}, fig.~5 there). In particular,
the \NV\ emissivity of photoionized RPC gas peaks at $\sim 0.1R_{\rm BLR}$, and the \NeVIII\ emissivity peaks below  $0.03R_{\rm BLR}$. The RPC predictions of the various line emissivity as a function of
distance, do not address the origin of the photoionized gas, and just assume the gas is there.
RM of the higher ionisation lines can therefore be used to test the validity of the inflated disc 
 model for the origin of the BLR gas, as it predicts a clear cutoff in line emission below $R_{\rm in}$. 
If their response peaks well below $R_{\rm in}$, that will clearly exclude a dusty disc wind as the 
origin of the BLR gas, as it will indicate other processes feed gas into the AGN ionisation cone at
$R<R_{\rm in}$.

Note that the outermost radius of the BLR is set by the dust sublimation radius \citep{NetzerLaor93},
regardless of the origin of the dusty gas. The inflated disc structure leads to peak disc height
at $R_{\rm max}$ which is very close to $R_{\rm BLR}$ (Section~\ref{sec:sub_static:sub:implied_RandCF}),
but the more stringent test will be provided by the measured value of $R_{\rm in}$ from the higher ionisation lines.

\subsection{The position of the peak height}

The CF of the dusty gas is set by the joint effect of $\mdot$, which sets $H(R)$
of the dust inflated disc structure, and $L_{\rm obs}$, which sets $z_{\rm sub}(R)$ of 
the illuminated photoionized surface layer.
The value of the AD $T_{\rm eff}$ at $R_{\rm sub}$ also plays a role, as it sets the 
value of the dust $\kappa$, which also affects the value of $H(R)$.

If $L_{\rm obs}$ is isotropic, then there is effectively a vertical sublimation wall
at $R_{\rm sub}$, rather than $z_{\rm sub}(R)$ which rises with $R$.
In the isotropic illumination case, the failed dusty wind mechanism does not operate. A grain of a given size either exists at all 
heights above the AD ($R>R_{\rm sub}$), or does not exist ($R<R_{\rm sub}$). 
The AD inflated structure solution for $H(R)$ extends down to $R_{\rm sub}=0.15L_{46}^{1/2}$~pc, set by 
$a\ge 0.3$~\mic\ grains, i.e.\ down to $1.5R_{\rm BLR}$. Line emitting gas extends further out to 
$3R_{\rm BLR}$, set by the $a>0.1$~\mic\ grains sublimation, which keeps the dust UV opacity is low. 
The critical problem here is that the dustless AD at $R<1.5R_{\rm BLR}$ is very thin, 
with $H/R\sim 10^{-3}$, and will not be exposed to the ionising continuum. 
In contrast, RM of the \HeII\ ion indicates significant line emission extends down 
to $\sim 0.2R_{\rm BLR}$. We therefore conclude that an isotropic $L_{\rm obs}$ is excluded,
if the inflated disc is the origin of the BLR gas.

If $L_{\rm obs}$ is non-isotropic, then $R_{\rm sub}$ is set by $L_{\rm obs}(\mu)$. If $L_{\rm obs}\rightarrow 0$ then naively $R_{\rm sub}\rightarrow 0$. 
However, the dust is also heated by the underlying local AD emission, which sets a minimal radius
at $R_{\rm in}$.
 
The maximal height of the disc $H_{\rm max}$ occurs at $R_{\rm max}$ where the $z_{\rm sub}(R)$ and 
the $H(R)$ solutions cross.
For $L_{\rm bol}\propto \mu$, the maximal height occurs at $R_{\rm max}=0.8R_{\rm BLR}$ for the static 
solution, and  $R_{\rm max}=R_{\rm BLR}$ for the dynamic
solution, for $\epsilon=0.1, Z = 5, M_8=1$ and $L_{46}=1$. The parameter dependence is
$R_{\rm max}\propto L^{0.58}$ (Fig.~\ref{fig:prop_vs_param}), consistent with the observed relation 
$R_{\rm BLR}\propto L^{0.52-0.56}$, and the associated errors in various RM studies 
\citep{Kaspi05, Bentz09, Bentz13}. There is an additional predicted
very weak dependence of $R_{\rm max} \propto M_{\rm BH}^{0.08}$, likely too weak to be detectable.
There are additional predicted weak, but possibly detectable relations of $R_{\rm max} \propto Z^{0.26}$ and
$R_{\rm max} \propto \epsilon^{-0.33}$, where the main observational challenge is getting reliable estimates of
$Z$ and $\epsilon$ (see further discussion below).

\subsection{The CF}
\subsubsection{The amplitude problem}\label{sec:CFamp}

Although $R_{\rm max}$ agrees remarkably well with $R_{\rm BLR}$, both its value and its
luminosity dependence, the derived CF falls too short. About $\sim 30$ per cent of the estimated
ionising continuum needs to be absorbed by the photoionzed gas to explain the observed EW
of most lines (\citealt{Korista97, Maiolino01, Ruff12, paperII}). Since the isotropic illumination 
case is ruled out (see above), then the simple alternative assumption of $L_{\rm obs}\propto \mu$
implies that CF$\sim 30$ per cent requires $H/R\sim 0.65$, in contrast with $H/R=0.23$ 
derived above for plausible parameters (eq.~\ref{eq:HR1}). Are there plausible effects which can
increase the CF? 

A simple solution is that $\epsilon\simeq 0.01$, rather than 0.1, which leads to 
$H/R\simeq 0.5$ (eq.~\ref{eq:HR1}). Such a low radiative efficiency likely applies in some
objects, but is unlikely as a typical value (\citealt{Soltan82, Yu02, Marconi04, DavisLaor11}).
However, the above estimates are based on the value of $\mdot$ which reaches the centre and
produce $L_{\rm bol}$. But, the CF is set by the value of $\mdot$ at the BLR, $\mdot_{\rm BLR}$.  
Can 90 per cent of $\mdot_{\rm BLR}$ not reach the centre?
A fraction of $\mdot_{\rm BLR}$ is likely lost in a wind (see below), and may provide part of the solution.

An additional potentially significant effect is the radiation pressure of the illuminated 
dust IR emission, emitted by dust just below the photoionized surface layer.  
If the near IR optical depth of the inflated AD structure is not too large, some
of the near IR emission will diffuse through the torus structure, and escape from the other side.
The momentum flux of this radiation will provide additional vertical support, increasing $H$.
This mechanism lies at the hurt of the various AGN torus models (e.g.\ \citealt{Chan16}), 
which are supported by the IR reemission of the dust illuminated by the central AGN emission. 
Exploring this 
solution requires a detailed radiative transfer model,
and is also sensitive to the assumed optical depth of the configuration. 
A related mechanism is the momentum flux of the incident optical-UV emission, 
which is further discussed below 

Alternatively, the derived CF$\sim 0.3$ may be an overestimate. This value is based on the 
observed ionising continuum SED, which is measured only in luminous high-$z$ quasars. If the typical 
ionising SED in lower luminosity AGN is significantly harder, then the required CF will be significantly lower. 
In addition, mild reddening is likely prevalent (\citealt{SL12, Baron16}), which again implies a typically harder intrinsic ionising 
SED, and a smaller CF.

Similarly, the ionising SED we observe is at a line of sight which is typically close to face on. 
If it is significantly softer than the ionising
SED close to an edge on view, which the BLR sees, the required CF will again be lower. Such 
a situation is expected due to Doppler beaming of the radiation from the innermost disc
(e.g.\ \citealt{LaorNetzer89}), although the innermost disc may not be
geometrically thin (\citealt{LaorDavis14}).

\subsubsection{The dependence on model parameters}

The CF is inferred from the broad lines EW. A well known trend
is the Baldwin relation, where the EW of most lines (excluding the Balmer lines and \NV) decreases
with increasing luminosity (see review in \citealt{Osmer99}). The dependence of the slope of this relation
on the ionisation energy suggests it is driven, at least partly, by a softening of the ionising SED with
increasing luminosity (e.g.\ \citealt{Scott04}). This effect is not explored in the current analysis,
as we assume $L_{\rm bol}$ has a fixed SED at all luminosities. However, an additional trend which
appears to be driving some of the lines, in particular \CIV, is the value of $\dot{m}$ 
\citep*{Baskin04, Warner04, Bachev04}. 

The dynamical solution gives a relatively strong dependence of the CF on $\epsilon$
with $H/R\propto \epsilon^{-0.70}$ (eq.~\ref{eq:CF2}, Fig.~\ref{fig:prop_vs_param}), but 
unfortunately, $\epsilon$ is generally not well determined. A possible relation of the form 
$\epsilon=0.089M_8^{0.52}$ was suggested by \citet{DavisLaor11}, which together with eq.~\ref{eq:CF2}
implies
\begin{equation}
H/R\propto \dot{m}^{0.26}Z^{0.56}\ .
\label{eq:cf_mdot}  
\end{equation}
This suggests a gradual increase in the CF with $\dot{m}$, in contrast with the observed trend of a
decreasing EW of C~IV with $\dot{m}$. However, there are strong indications for a trend of increasing $Z$
with $\dot{m}$ (e.g.\ \citealt{Shemmer04}), which coupled with the expected softening of the ionising
SED with increasing $Z$ \citep{LaorDavis14}, may lead to the observed drop in the C~IV EW with increasing
$\dot{m}$.

A more robust indicator of the CF of the inflated disc torus is the ratio of the near-IR
luminosity, $L_{\rm NIR}$, to the estimated $L_{\rm bol}$. In contrast with the emission lines, which 
are set by the extreme UV SED, which is not directly observed, the dust emission is largely set by
$L_{\rm bol}$, about half of which
is directly observed (optical to near UV). In addition,
the torus is likely very optically thick, and absorbs and reradiates all the incident radiation,
independent of the dusty gas density, ionisation state, and metallicity. 

\citet{Mor11} and \citet{Mor12} studied the hot dust component in the IR SED of AGN for
$L_{\rm bol}\simeq 10^{44}-10^{47}$~erg~s$^{-1}$, and found that its strength implies a median 
CF$\sim 0.3$, consistent with our result that the BLR and the hot dust are produced by
front and back sides of the same inflated disc component. In addition, the CF of the hot dust decreases
with increasing $L_{\rm bol}$ and $\dot{m}$, as deduced for the BLR from the EW of most broad lines. 
As mentioned above, the apparent drop in the CF may also be an artefact which results from the wrong 
assumption of a fixed SED, while there is a drop in the bolometric correction factor, as the SED gets softer
with increasing luminosity.

\subsubsection{The effect of a variability}

AGN are known to be variable on timescales of years, or shorter. The disc height is driven by
the value of the local accretion rate at $R_{\rm BLR}$, $\mdot_{\rm BLR}$. As discussed above (\ref{sec:ADself}), the
accretion timescale from $R_{\rm BLR}$ is likely very long, $\sim 10^5-10^8$~yr, and may
be considered constant on years timescale. Using the numerical solution (eq.~\ref{eq:CF2}), 
with $\epsilon\propto L/\mdot$, gives
\begin{equation} 
H/R\propto L^{-0.44}M_{\rm BH}^{0.09}\mdot^{0.70}(Z/Z_{\odot})^{0.56}\ .  
\label{eq:cf_num}  
\end{equation} 
The $\propto L^{-0.44}$ drop of the CF with luminosity results from the following mechanism.
The value for $H(R)$ is independent of $L$ when $\mdot_{\rm BLR}$ is constant, 
but $R_{\rm max}\propto L^{\sim 1/2}$ due to the increasing sublimation radius.
Since CF$\propto H/R$, one expects CF$\propto L^{\sim -1/2}$. 
This mechanism is identical to the receding torus model, suggested by \citet{Lawrence91},
to explain the possible drop in the obscured fraction of AGN with increasing luminosity. 
A drop in the CF due to the receding torus effect should be evident through a drop in 
$L_{\rm NIR}/L_{\rm bol}$ with increasing $L_{\rm bol}$.

In contrast, if $\mdot_{\rm BLR}$ follows $\mdot$, and both follow $L$, then $H\propto L$,
leading to CF$\propto H/R\propto L^{\sim 1/2}$, which is not observed.

We therefore expect the emission line EW, in a given object, to go down when the luminosity increases. 
However, this should occur on timescales longer than the dynamical timescale at the 
BLR, $\tau_{\rm dyn}\sim 100R_{\rm BLR}/c=30 L_{46}^{1/2}$~yr (for 
$v\sim 3000$~km~s$^{-1}$ at the BLR). On shorter timescales, the vertical gas distribution 
remains effectively frozen, and the line EW will remain unchanged.
The well known intrinsic Baldwin effect (\citealt{Pogge92, Wilhite06}), of a drop in the lines EW
with increasing luminosity, apparently occurs on timescales faster than $\tau_{\rm dyn}$. 
It may be driven by a non linear 
dependence of the line driving UV continuum on the observed continuum.

\subsubsection{When is the BLR absent?} \label{sec:sub_CF:sub:when_BLR_absent}

Below we show that the inflated disc solution implies that below some $\dot{m}$ value, the
BLR emission will effectively disappear.
This occurs for conditions which allow the dust to survive in the illuminated face of the inflated disc.
Such objects would then become the so-called true type 2 AGN, where the BLR emission disappears,
but both the NLR emission and the continuum emission are directly observed (e.g.\ \citealt{Tran01, Bianchi12}).

The inflated disc structure inevitably terminates at $R_{\rm in}$, where the disc atmosphere is
too hot to support dust. In the examples presented above 
(Figs~\ref{fig:static_solution}--\ref{fig:BLR_illustr}), $H(R_{\rm in})$ lies well within the
dust sublimation sphere. However, what happens if $H(R_{\rm in})<z_{\rm sub}(R_{\rm in})$?
That is, the highest possible point of the inflated disc does not enter the sublimation sphere. So, although
the inflated disc is illuminated by the ionising radiation, the photoionized gas remains dusty,
and thus an inefficient line emitter. The BLR emission will be suppressed, but the NIR dust 
emission will remain. For which parameters does this happen?

We first derive $H(R_{\rm in})$, adopting the static solution. Since $\kappa_{50}=1.08Z/Z_{\odot}$ for $T=2000$~K, the
AD temperature at $R_{\rm in}$, we get (eq.~\ref{eq:H1}) 
\begin{equation}
H=0.0044\mdot_1 Z/Z_{\odot}~{\rm pc}.
\end{equation}
The dust UV opacity is carried by the smaller grains. 
Therefore, dust suppression will
start to become significant when $H<z_{\rm sub}$ for the $a=0.1$~\mic\ grains, and will
become maximal when $H<z_{\rm sub}$ for the $a=0.01$~\mic\ grains.

Using eqs.~\ref{eq:R_2000} and \ref{eq:z_sub}, we get 
\begin{equation}
z_{\rm sub}=(8.9, 2.3, 0.25)\times 10^{-5}L_{46}^{-1}M_8\mdot_1~{\rm pc} .
\end{equation}
for the $a=1$, 0.1 and $0.01$~\mic\ grains, respectively. Therefore, $H<z_{\rm sub}$ holds for 
\begin{equation}
L_{46}/M_8<(20.2, 5.2, 0.57)\times 10^{-3}(Z/Z_{\odot})^{-1}\ , 
\end{equation}
or equivalently
\begin{equation}
\dot{m}<(0.025, 0.0065, 6.5\times 10^{-4})\times (Z/Z_{\odot})^{-1}\ .
\end{equation}
Thus, at $0.025<\dot{m}\times Z/Z_{\odot}$ the illuminated gas is dustless, and line emission remains unaffected.
For $0.0065<\dot{m}\times Z/Z_{\odot}<0.025$ the illuminated BLR gas includes large, $a>0.1$~\mic, grains,
which produce little UV opacity. The line emission will not be significantly affected, 
but depletion to grains may affect the gas phase abundances,
and the gas cooling and ionisation states may also start to be affected by the grains.
At $6.5\times 10^{-4}<\dot{m}\times Z/Z_{\odot}<0.0065$, smaller grains can survive, and 
the dust effects will increase. At $\dot{m}\times Z/Z_{\odot}<6.5\times 10^{-4}$ we get the full effect
of the dust, which reduces the BLR emission by about an order of magnitude (e.g.\ fig.~5 in 
\citealt{paperII}). Since CF$\propto \dot{m}^{0.26}$ (eq.~\ref{eq:cf_mdot}) there is an additional 
decrease in the line strength due to the reduction in the CF, by an additional order of magnitude
compared to the $\dot{m}\sim 0.1-1$ objects, making the BLR emission effectively gone.

Other scenarios for the origin of the BLR gas, which invoke a general case of disc wind, 
also predict the BLR should be absent below some $\dot{m}$ value, which may depend on $M_{\rm BH}$
(e.g., \citealt*{Nicastro00, Elitzur06, Elitzur09, Elitzur14}). 
An alternative scenario for the absence of BLR emission is that the BLR gas is there, but the 
AD is too cold (at low enough $\dot{m}$), to provide the ionising photons 
to excite the lines \citep{laorDavis11}. In this case
the NLR emission will also disappear, leading to a lineless quasar, rather than a true type 2 AGN. 

On the other hand, the disappearance of the BLR may be an artefact of the inevitable broadening 
of the broad lines with decreasing $\dot{m}$, as $R_{\rm BLR}$ gets smaller, which makes the broad
lines too weak and broad to detect from the ground, due to the host dilution. High spatial resolution 
{\em HST} observations of a few nearby $\dot{m}\sim 10^{-4}-10^{-5}$ AGN indeed reveal BLR emission 
(\citealt{Bower96, Ho00, Shields00, Barth01}). The presence of a BLR, despite the low $\dot{m}$ 
in these objects, may reflect their low accretion efficiency.

A simple prediction of the dust inflated disc solution is that the required threshold $\dot{m}$ value 
is independent of $M_{\rm BH}$, but is $Z$ dependent. The BLR will be visible to lower $\dot{m}$ values 
in higher metallicity objects. In addition, in contrast with some of the above models, 
which predict that the near IR emission disappears together with the disappearance of the BLR,
the inflated disc model suggests hot dust emission is still present. 
However, given the expected weakness of the emission (since CF$\propto \dot{m}^{0.26}$)
it will be hard to detect, in particular against the host emission, which
inevitably dominates in low $\dot{m}$ objects. It may take mas-resolution near-IR interferometry to detect the hot dust in the nucleus on pc scale (e.g.\ \citealt{Kishimoto11, Kishimoto13}),
as part of the directly observed continuum in unabsorbed type 2 AGN.

\subsection{Wind}

Below we discuss the conditions where the inflated disc structure is expected to produce a wind.
We first discuss the case where the wind is driven by the local disc IR emission, assuming the dusty
gas is shielded from the central ionizing source, and when it is unshielded and 
subject to sublimation. We also briefly discuss a wind driven off the inflated disc by the incident 
optical-UV emission of the central source.

\subsubsection{How thick can the torus get?}

The static solution for $H(R)$ was found above by the requirement that $a_{\rm BH}=a_{\rm rad}$
in the z direction. Such a static solution exists for $z\ll R$, as $a_{\rm BH}\propto z$, 
and $a_{\rm rad}$ is a constant. However, at $z\gg R$, $a_{\rm BH}\propto z^{-2}$. Since
$a_{\rm BH}$ rises at small $z$ and falls at large $z$, it has a maximal value
$a_{\rm BH, max}$ at a certain $z_{\rm max}$. If $a_{\rm rad}>a_{\rm BH, max}$ at a given $R$, 
then $a_{\rm rad}$ dominates at all $z$, a static solution is not possible, and a wind is 
inevitably formed.
Using the full expression, $a_{\rm BH}=GMz/(R^2+z^2)^{3/2}$ (eq.~\ref{eq:a_gz}),
the condition $da_{\rm BH}(R,z)/dz=0$ occurs at $z_{\rm max}=R/\sqrt{2}$, which forms
the maximal possible thickness for the static disc solution (see Fig.~\ref{fig:static_solution_vs_Mdot}).
The same height limit also applies in the dynamic solution, as gas which reaches 
$z>z_{\rm max}$ at a finite positive vertical speed, will now feel
a positive vertical acceleration, and will form a wind. 
Thus, the thickest torus structure, either static or dynamic, has $H/R\le 1/\sqrt{2}$.

What is then the implied maximal CF? As noted above (following eq.~\ref{eq:HR}), for the 
$L_{\rm obs}\propto \mu$ illumination, the fraction of the absorbed continuum, i.e.\ CF, 
is $\mu_0^2$. Since $\mu_0=H/\sqrt{H^2+R^2}$, we get that CF$(z_{\rm max})=1/3$,
similar to the BLR observations which give CF$\sim 0.3$.

It therefore appears that the BLR torus typically extends to its maximal possible thickness. 
It is therefore plausible to assume that some fraction of the BLR gas will be pushed
above $z_{\rm max}$, become unbound and form a wind. 

The discussion above concerned only the shape of the region where gas can be confined. 
But, when do we expected the gas to be able to reach $z_{\rm max}$?

\subsubsection{IR driven wind with no central illumination}

The maximal vertical acceleration the BH can produce at a given $R$, 
$a_{\rm BH, max}(R)\equiv a_{\rm BH}(R, z_{\rm max})$, is
\begin{equation} 
a_{\rm BH, max}(R)=\frac{2}{3\sqrt{3}}\frac{GM}{R^2}.
\end{equation} 
If $a_{\rm rad}(R)>a_{\rm BH, max}(R)$, then the local radiation pressure 
will push the dusty gas above $z_{\rm max}$, and a wind is inevitably formed. 
Using the local disc flux (eq.~\ref{eq:ADflux}), the condition is
\begin{equation}
a_{\rm rad}(R)=\frac{3GM\mdot}{8\pi R^3}\frac{\kappa}{c}>
\frac{2}{3\sqrt{3}}\frac{GM}{R^2}\ ,
\end{equation}
and a wind is therefore expected at
\begin{equation}
R<0.31\frac{\mdot\kappa}{c}\equiv R_{\rm wind}\ , 
\end{equation}
or in convenient units
\begin{equation}
R_{\rm wind}= 0.01\mdot_1\kappa_{50}~{\rm pc}\ .
\end{equation}
However, the value of $R_{\rm wind}$  cannot be smaller than $R_{\rm in}$ (eq.~\ref{eq:Rin}), 
so that dust can be
present in the disc atmosphere. The constraint $R_{\rm wind}\ge R_{\rm in}$ (for $T_{2000}=1$ in eq.~\ref{eq:R_2000})
requires that
\begin{equation}
\mdot_1^{2/3}m_8^{-1/3}\kappa_{50}\ge 1.6\ .
\end{equation}
Note that $\kappa_{50}=1.08Z/Z_{\odot}$ for $T_{2000}=1$ (eq.~\ref{eq:kappa1}).
The above expression, together with eq.~\ref{eq:mmdot}, and the relation $L_{46}\simeq L_{\rm opt, 45}$, 
gives 
\begin{equation}
\dot{m}\ge 0.95(Z/Z_{\odot})^{-1}\ , 
\end{equation}
which is quite similar to the condition for an electron scattering driven wind ($\dot{m}>1$).
This similarity is actually just a coincidence, as the local Planck mean dust opacity is $\sim 125 Z/Z_{\odot}$
larger than the electron scattering opacity, while the driving is by the local disc IR
emission, which is only a small fraction of $L_{\rm bol}$ used in the definition
of $\dot{m}$. The two factor happen to nearly cancel out, leading to similar limits on
$\dot{m}$ for the two driving mechanisms.

The above derivation assumes dust still persists at $z_{\rm max}=R/\sqrt{2}$. This holds generally
if the dust is not illuminated by the central continuum source. This can happen if there is some obscuring structure between the centre and the outer disc, which shadows the outer disc. Alternatively, one may envision a non steady state accretion disc, where the outer accretion rate $\mdot_{\rm BLR}$, is much larger 
than $\mdot$ in the inner disc. The later case may correspond to a new accretion event, where the high
$\mdot$ reaches the BLR, but does not reach the centre yet.

In both cases, the outer disc will self-regulate the inflow rate, and 
will not allow it to exceed the Eddington rate, and the excess $\mdot$ will escape as a wind from the
outer disc. If the outer $\mdot$ corresponds to $\dot{m}\gg 1$, one expect a massive wind of dusty gas,
which may obscure the source completely.

\subsubsection{IR driven wind with central illumination}

The central source illumination will generally sublimate the dust well before it reaches $z_{\rm max}$.
For which parameters can the illuminated gas still reach $z_{\rm max}=R/\sqrt{2}$ and form a wind?
From the analytic solution for $H/R$ in the dynamic case (eq.~\ref{eq:HR11} with the correction factor of 1.67), we get 
that a wind is expected when
\begin{equation}
L_{46}^{0.18}M_8^{0.15}\epsilon^{-0.67}(Z/Z_{\odot})^{0.51}>32.6\ .
\end{equation}
For a typical $L_{46}=M_8=1$ quasar, with a plausible $Z/Z_{\odot}=5$, a wind requires $\epsilon<0.02$.
Such a low $\epsilon$ is not expected to be common, as the likely mean value is $\epsilon\sim 0.1$
(see Section~\ref{sec:CFamp}). However, if $\mdot_{\rm BLR}> 5\mdot$, a local IR driven wind 
will be formed.

\subsubsection{UV driven wind}

The optical-UV continuum, incident on the BLR, provides both energy and momentum fluxes. The energy flux 
photoionizes the gas, and heats and
potentially sublimates the grains. The momentum flux compresses and accelerates the gas. The component of the
momentum flux, which is perpendicular to the local surface element, compresses the gas (e.g., \citealt{Pier95, paperI, paperII, Namekata16}). The tangential component of the momentum flux, which is absorbed in the surface photoinized layer, can produce a force which far exceeds gravity, and leads to a wind which ablates the surface
layer of the exposed gas. 

What is the expected mass loss in this wind, $\mdot_{\rm wind}$? The photoionized surface layer is expected to
be dustless (e.g.\ Fig.~\ref{fig:BLR_illustr}). The mean induced acceleration relative to the electron scattering acceleration in that layer is $a/a_{\rm es}\sim 10-40$ (e.g.\ fig.~9, left panel, in \citealt*{paperIV}). The thickness of
the layer with $a/a_{\rm es}> 10$ is $\Sigma_{\rm ion}\simeq 1.5\times 10^{22}$~\cmmt. The acceleration is
\begin{equation} 
a(R)=\frac{GM}{R^2}f\frac{a}{a_{\rm es}}\dot{m} \ ,
\end{equation} 
where f is a geometrical factor of order unity, which depends on the incident angle of the ionising radiation,
with respect to the local surface element. The acceleration extends along the surface layer of the illuminated
face of the inflated disc, which is a fraction of $R_{\rm BLR}$. This leads to a terminal velocity of
\begin{equation} 
v_{\rm wind}\simeq v_{\rm Kepler}\sqrt{f\frac{a}{a_{\rm es}}\dot{m}} \ ,
\end{equation} 
in the radial direction, parallel to the surface layer.
Since the local escape speed is $\sqrt{2}v_{\rm Kepler}$, and the gas originates from a disc
with a tangential velocity $v_{\rm Kepler}$, the gas escapes if the radial component also
reaches $v_{\rm Kepler}$, i.e.\ if $f\frac{a}{a_{\rm es}}\dot{m}>1$. Using plausible values of 
$f=0.5$ and $a/a_{\rm es}=20$, gives 
\begin{equation} 
v_{\rm wind}\simeq 3v_{\rm Kepler}\sqrt{\dot{m}}\ .
\end{equation} 
If $\dot{m}<0.1$, most of the sheared surface layer will not escape, and will 
form a failed wind.
Since $v_{\rm Kepler}=\sqrt{GM/R}$, we get for the BLR
\begin{equation} 
v_{\rm Kepler, BLR}=2075M_8^{1/2}L_{46}^{-1/4}~~{\rm km~s}^{-1}\ .
\end{equation} 
Since $\dot{m}=0.8L_{46}M_8^{-1}$, we get the simple relation
\begin{equation} 
v_{\rm wind}\simeq 5570 L_{46}^{1/4}~~{\rm km~s}^{-1}\ . 
\end{equation} 
The wind forms an outflowing thin surface layer, moving parallel to the inflated surface
layer. The cross section area of the wind is $2\pi R_{\rm BLR}D$, where $D$ is the thickness
of the layer, and the mass flux per unit cross section area is $n m_{\rm p} v_{\rm wind}$, 
where $m_{\rm p}$ is the proton mass. Since the layer is of photoionized gas, $Dn\simeq \Sigma_{\rm ion}$.
The associated mass loss, from both sides of the AD, is therefore
\begin{equation} 
\mdot_{\rm wind}=4\pi R_{\rm BLR} \Sigma_{\rm ion} m_{\rm p} v_{\rm wind}\ .
\end{equation} 
This yields 
\begin{equation} 
\mdot_{\rm wind}\simeq 0.9 L_{46}^{3/4}~ \Msun~{\rm yr}^{-1}\ .
\end{equation} 
This value is comparable to the accretion rate $\mdot_1=0.176\epsilon^{-1}L_{46}$ (eq.~\ref{eq:epsilon}).
It indicates that the UV driven flow, off the BLR surface, can form a significant component in the system, as it involves 
a significant fraction of the accretion rate. However, most of this flow forms a failed wind, and will 
fall back to the disc, if $\dot{m}<0.1$.

This sheared surface layer wind, or ablation wind, will stream parallel to the local inflated AD 
surface. Once the stream leaves the disc surface, it will not be supported radially on its back side 
by the dense large-column disc gas, but it may still be confined by the ram pressure
of the lower density ambient medium it runs into (e.g.\ \citealt{paperIV}, sec.~2.1.1 there). However, dense gas pushed into low density gas is Rayleigh
unstable, and this stream will likely break apart within a certain distance.

This expected sheared-layer wind has properties similar to the absorbing gas seen in Broad Absorption Line Quasars (BALQs). It has
velocities typical or somewhat larger than in the BLR \citep*{Baskin15}, and lacks absorption by low ionisation lines, 
due to the sharp drop in the shearing force of the radiation acceleration in that layer \citep{paperIV}.
The observed mean CF of the BALQ outflows is estimated to be $\sim 0.15$, depending on inclination indicators and the
strength of the extreme UV emission \citep*{Baskin13}. Thus, this ablation wind can provide a significant 
additional source of line emission for the higher ionisation lines. It may provide the mechanism which fills
in the stable volume above the disc, leading to the observed typical CF$\sim 0.3$.

This ablation AD wind is also a natural source for the observed BLR wind component \citep{Richards02}, 
detected in the higher ionisation lines, as expected since the sheared layer is composed mostly of the higher 
ionisation state
gas. Also, this component is always blue shifted, as expected from a disc outflow, since the back side is 
obscured by the AD. The lower ionisation line profiles show a more symmetric profile, and are attributed
to the so called BLR disc component. This is consistent with the inflated disc model for the BLR, 
as the lower ionisation lines,
which originate mostly beyond the ionisation front, are produced on the photoionized surface layer 
of the inflated disc structure.

\subsection{The BLR emission line profiles}

There is a well known systematic change in the BLR emission profiles, which change from 
Lorentzian profiles, when the lines are narrow, to Gaussian or flat top, or even double peaked, 
when the lines are very broad (e.g.\ \citealt{Collin06, Kolla13}). A Lorentzian profile likely 
reflects a large
range of emitting radii, where the broad wings are produced from smaller radii, as also
indicated from various RM campaigns \citep{Pancoast14}. A Gaussian profile likely reflects
a smaller range of radii, and a double peaked profile is expected from a BLR with a small
range of radii, which effectively forms a ring. 

What is the expected range of BLR radii, when it is formed in a dusty inflated disc? 
The BLR extends from $R_{\rm in}$ to $R_{\rm max}$. Using 
$R_{\rm max}(a=0.1~\mic)$ as the effective outer radius for the BLR, $R_{\rm out}$, we get
using eqs.~\ref{eq:R_2000} and \ref{eq:R_max}, and the dynamic solution correction of 1.2 to $R_{\rm max}$
(paragraph following eq.~\ref{eq:z_dyn}), that 
\begin{equation} 
R_{\rm out}/R_{\rm in}=4.93(\dot{m}Z/Z_{\odot})^{0.26}\ .
\end{equation} 
The parameters $Z/Z_{\odot}\simeq 5$ and $\dot{m}\sim 1$, give $R_{\rm out}/R_{\rm in}\sim 7$,
which goes down to $R_{\rm out}/R_{\rm in}\sim 1.5$ for $Z/Z_{\odot}\sim 1$ and $\dot{m}\sim 0.01$. 

Since $M_{\rm BH}\propto v^2R_{\rm  BLR}$, where $v$ is the typical velocity at the BLR,
and $R_{\rm  BLR}\propto L^{1/2}$, we get that $\dot{m}\propto L^{1/2}v^{-2}$. Thus, objects with low $v$
generally have high $\dot{m}$, consistent with the observed trend of a Lorentzian line profiles
in narrow line AGN. If the above mechanism is valid, the main driver for a Lorentzian profile is
$\dot{m}$, rather than $v$. Given the significantly weaker dependence of $\dot{m}$ on 
$L$, one needs a large range in $L$ to find if the driving parameter is indeed $\dot{m}$,
rather than $v$, as apparently observed. 

Another prediction is a trend of the profile shape with $Z$, as
higher $Z$ should be associated with more Lorentzian profiles. However, the expected dynamic
range in $Z$ is small, and its estimated value tends to be uncertain, so it is probably not a
robust way to test the above prediction for the origin of the BLR line profiles.

\subsection{Hot Dust Poor AGN}

Some AGN clearly lack significant emission by hot dust (\citealt*{Jiang10, Hao10, Hao11, Jun13, Lyu17}). 
In the most extreme cases the observed optical-UV spectral slope extends to the near IR, with a slope
consistent with the predicted slope of $\nu^{\sim 1/3}$ from the outer accretion disc emission. This SED indicates the AD indeed extends out to 
the BLR, as assumed here, and possibly further outside.
A similar pure AD SED, from an AD which extends out to the BLR, is also seen in IR spectropolarimetry of some AGN with 
the regular near IR bump, where the polarisation excludes the hot dust emission, and allows to detect the underlying clean
AD emission  \citep{Kishimoto08}.

There are contradicting claims concerning the relation of the relative strength of the near IR emission and other AGN emission properties. For example, a positive correlation with $\dot{m}$ \citep{Jun13}, a negative correlation with $\dot{m}$ \citep{Lyu17}, or lack of a correlation 
\citep{Mor11} are reported. There are other similar apparent contradictions concerning a relation with 
$L_{\rm bol}$ and $M_{\rm BH}$, contradictions likely related to the different selection effects in the different samples used. 

The relative strength of the near IR emission is clearly set by the hot dust illumination, which depends on both the inflated disc geometrical CF, and the inclination dependence of the central source illumination. Since the broad line and near IR emission come from the same component, we expect them to be correlated. Or, equivalently, that the near IR to optical-UV
flux ratio, will be correlated with the broad lines EW, excluding extremely low $\dot{m}$ objects 
discussed above (Section~\ref{sec:sub_CF:sub:when_BLR_absent}). 

One should keep in mind that some of the lines are more sensitive to the shape of the ionising SED, 
and also to $Z$, while the dust emission is a measure of the integrated optical-UV SED only.
Also, the possible wind component, sheared off the inflated disc surface, is not expected to be associated with 
hot dust emission. In addition, although a few AGN are consistent with no detectable 
hot dust emission \citep{Jiang10}, in most Hot Dust Poor (HDP), or Hot Dust Deficient (HDD) AGN, 
the near IR emission is suppressed by less than a factor of two (e.g.\ table~3 in \citealt{Lyu17a}), so the expected suppression of the associated broad line emission will also not be dramatic.

Since CF$\propto Z^{0.56}$ (eq.~\ref{eq:cf_num}), and $Z$
appears to be related to $\dot{m}$ \citep{Shemmer04},  we expect objects with a low $\dot{m}$ to 
present weaker dust emission. 
Since $Z$ in the Universe drops with increasing $z$, 
and decreasing host galaxy mass \citep{Mannucci10}, this may explain the extreme low dust emission 
observed in a couple of $z\sim 6$ quasars, with low $M_{\rm BH}$ values, 
which may indicate a low host galaxy mass \citep{Jiang10}.

\subsection{Application to other systems}

Accretion discs are observed in other accreting systems from protoplanetary systems to X-ray binaries (XRBs).
Is the dust inflated disc solution relevant in these systems? 

In XRBs, $M_8\simeq 10^{-7}$ and $\mdot_1\sim 10^{-7}$, which implies 
$R_{\rm in}\sim 10^{12}$~cm (eq.~\ref{eq:R_2000}), or $\sim 7\times 10^5 R_{\rm g}$.
This radius is comparable, or larger, than the typical binary separation in these systems
\citep{Remillard06}. Thus, the AD in XRBs is generally too hot to harbour dust.

In other accreting stellar systems, such as symbiotic stars, young stellar objects, or protoplanetary systems,
the central object mass is $\sim M_{\odot}$, comparable to XRBs, but the accretion rate is typically a few orders of magnitude smaller, and the AD will be significantly colder. Since $R_{\rm in}\propto (M\mdot)^{1/3}$, it can be a factor 
of 10--100 smaller,
so if the central object is not a compact object, the entire disc can be dusty. 
However, as we show below based on simple considerations, dusty gas in thermal equilibrium, on the relevant scale 
of these systems, is inevitably gas pressure dominated.

Assume a gas cloud, with a size $H$, density $n$, temperature $T$, and flux weighted mean opacity $\kappa$. 
In full thermal equilibrium, the gas and the radiation temperatures are equal. The ratio of their pressures 
is therefore
\begin{equation} 
\frac{P_{\rm rad}}{P_{\rm gas}}=\frac{\sigma T^4\tau}{nk_{\rm B}T}
\simeq \frac{\sigma\kappa m_{\rm p} T^3}{c k_{\rm B}}\times H
=9.15\times 10^{-12} \kappa_{50} T_{2000}^3 H({\rm cm})\ ,
\end{equation} 
where $\tau=\kappa\Sigma m_{\rm p}$ is the optical depth of the gas cloud, and $\Sigma\simeq nH$. 
The condition $P_{\rm rad}>P_{\rm gas}$ holds when
\begin{equation} 
H> 1.09\times 10^{11} \kappa_{50}^{-1} T_{2000}^{-3}~{\rm cm}\ .
\end{equation} 
However, geometry requires $H<R_{\rm in}$, and since $R_{\rm in}\sim 10^{10}-10^{11}$~cm, 
the vertical scale in the accretion discs is not large enough to allow 
$P_{\rm rad}>P_{\rm gas}$ to build up inside the disc, even in the hottest region where
dust can survive. The AD is gas pressure dominated, and its vertical structure will not change 
dramatically once dust can survive at $R>R_{\rm in}$, . 

In contrast, the dusty gas opacity does change dramatically at $R_{\rm in}$. The disc may be optically thin
to the central source illumination at $R<R_{\rm in}$, but become optically thick at $R>R_{\rm in}$.
The central source radiation may go horizontally through an optically thin AD, until it hits a wall 
of hot dust at $R_{\rm in}$, which shields the region at $R>R_{\rm in}$. This will lead to distinct continuum
emission feature in the near IR of the hottest dust, and a spatially resolved ring at $R\sim R_{\rm in}$
of hot dust emission \citep{Dullemond10}.

In AGN, the value of $H\gg 10^{11}$~cm and allows $P_{\rm rad}/P_{\rm gas}\gg 1$, which inevitably leads
to a dramatic change in the disc vertical structure due to the opacity jump at $R_{\rm in}$. 
A similar ring of hot dust emission should also be present in AGN,
due to the inflated disc structure illuminated directly by the central source. However, even in the nearest AGN,
say at 10~Mpc, the angular size of a structure of a size of $10^{-2}$~pc ($R_{\rm BLR}$ in low luminosity AGN)
is only $10^{-9}$~rad=0.2~mas, which requires a baseline of 3~km to resolve for $\lambda=3$~\mic.
So, direct imaging of the hot dust ring in AGN, which marks the BLR, is beyond current technologies
(but see \citealt*{Stern15})

\section{Conclusions} \label{sec:conclusions}

CH11 suggested that the BLR is formed by a failed dusty AD wind. Here we explore the expected dust properties in the innermost parts of AGN, the implied accretion disc vertical structure, and 
whether it can provide the CF of the BLR in AGN. We find the following:

\begin{enumerate}

\item Large graphite grains ($a>0.3$~\mic) sublimate only at $\Tgr\simeq 2000$~K at
$n\sim 10^{11}~{\rm cm}^{-3}$. Thus, they survive the AGN illumination down to $R_{\rm BLR}$,
while at the AD surface the large graphite grains survive down to $R_{\rm in}\simeq 0.18R_{\rm BLR}$.
Their Planck mean opacity is $\sim 50$ times larger than previously assumed.

\item The AD at $R>R_{\rm in}$ is inflated by the radiation pressure of the thermal disc emission, 
which leads to an inflated disc profile of $H\propto R^{-0.87}$. 
Dust sublimation by the central optical-UV illumination, which scales as $L_{\rm obs}\propto \mu$, 
leads to a torus like structure, where the illuminated face of the inflated disc
produces the BLR emission, and the back side is a source of very hot dust emission. 

\item The peak height of the torus structure occurs at 
\[ 
R_{\rm max}\propto L^{0.59} M_{\rm BH}^{0.075}(Z/Z_{\odot})^{0.26}\epsilon^{-0.33}\ .
\] 
It gives $R_{\rm max}\simeq R_{\rm BLR}$
for $Z/Z_{\odot}=5$ and $\epsilon=0.1$, and matches the observed dependence of $R_{\rm BLR}$ on $L$.
It predicts additional weaker dependence of $R_{\rm BLR}$ on $M_{\rm BH}$, $Z$ and $\epsilon$.

\item The BLR covering factor is predicted to scale as 
\[ {\rm CF}\propto L^{0.18}M_{\rm BH}^{0.15}(Z/Z_{\odot})^{0.51}\epsilon^{-0.67}\ , \]
for the static solution. The observed typical value of CF$\sim 0.3$ requires an unrealistically 
low $\epsilon\sim 0.03$. The dynamic failed-wind solution increases the CF by $\sim 1.7$. 
Further increase is expected from ablation of the inflated surface layer by the incident 
ionising radiation, and the resulting surface layer wind.

\item The material above the disc becomes unbound when $H>R/\sqrt{2}$, which corresponds to 
CF$=1/3$. This naturally explains the observed mean CF$\sim 0.3$, if the BLR gas typically 
fills all the available volume for bound disc gas. This also suggests that some fraction of the 
BLR gas likely becomes unbound and escapes as a wind.

\item For $\dot{m}<0.025(Z/Z_{\odot})^{-1}$, large grains can survive in the illuminated BLR gas, 
which does not affect the line emission significantly. 
For $\dot{m}<0.0065(Z/Z_{\odot})^{-1}$, smaller grains survive, 
and dust absorption starts to suppress the line emission. At $\dot{m}<6.5\times 10^{-4}(Z/Z_{\odot})^{-1}$, 
the line emission is strongly suppressed by the dust. Such objects will not show BLR
emission, but will still show some hot dust emission.

\item If the disc optical-UV emission does not illuminate the outer dusty disc, then 
the local IR emission is sufficient to produce a strong dusty wind, if $\dot{m}>1$, 
which may completely obscure the source. 

\item Since the accretion time from $R_{\rm BLR}$ inwards is of order  
$\ge 10^5$~yr, the value of $\mdot_{\rm BLR}$ may differ from $\mdot$ in the inner disc. 
If $\mdot_{\rm BLR}\gg \mdot$, the SED may be dominated by the disc IR emission,
with a relatively weak BLR. If $\mdot_{\rm BLR}\ll \mdot$, then optical-UV continuum 
will be observed, but the BLR will be absent.

\item The optical-UV illumination of the inflated disc surface, inevitably leads to 
a sheared surface layer wind of ionized gas.
The implied mass flux through this surface wind is comparable to the accretion $\mdot$,
and this wind is therefore a significant component in the system. It will enhance the CF
of the higher ionisation emission lines, produce a wind component of the BLR lines,
and broad UV absorption by high ionisation lines (i.e.\ BALs). 

\item The most robust test for the dusty disc wind origin for the BLR, is the presence of
an inner boundary for the BLR at $\sim 0.18R_{\rm BLR}$, below which a dusty disc wind cannot
be formed. 

\end{enumerate}

The dusty disc wind scenario provides a useful working hypothesis for understanding the 
origin of the BLR, and potentially explaining various types of behaviours observed in the
BLR. Studies of the response of the hot dust emission to the UV continuum variability, 
in particular
studies of possible structural changes on the dynamical timescale, will provide important
constraints on this scenario (e.g.\ \citealt{Kishimoto13, Schnule15}). 

In addition, radiation hydrodynamic models are required to
solve the nature of the vertical structure of the dusty disc (e.g.\ \citealt*{Jiang13, Davis14}), the effect of the
incident central continuum on this structure (e.g.\ \citealt{Proga14}), and the structure of the 
resulting disc wind (e.g.\ \citealt{Zhang17}). 
Since radiation pressure in the BLR inevitably leads to a structure on a scale of $10^{-5}$ of the size
of the system (e.g.\ \citealt{paperII, paperIV}), the numerical schemes need to be carefully 
adapted to be able to resolve this crucial scale (e.g.\ \citealt{Namekata16}).

We note in passing that the implied column of the inflated disc in the radial direction,
$n\Delta r\sim 10^{10}\times 10^{17}$~cm$^{-2}$, is highly optically thick, and given the
observed CF$\sim 0.3$, it is likely a major source for obscured AGN.

\section*{Acknowledgements}
	
We thank B.\ Czerny, J.\ Stern and the referee for helpful comments. We thank B.\ Draine for maintaining a website that provides absorption coefficients of grains in a machine readable format. This research was supported by the Israel Science Foundation (grant no.\ 1561/13).
This research has made use of NASA's Astrophysics Data System Bibliographic Services.

\appendix
\section{Dependence of $\bmath{\kappa}$ on $\bmath{T_{\rm BB}}$} \label{sec:appendix}

We present below a further investigation of the dependence of dust opacity on $T_{\rm BB}$. In addition to the Planck mean opacity, denoted here as $\kappa_{\rm Pla}$, which is discussed in Sec.~\ref{sec:dust_op} and is suitable for an optically thin case, here we also explore the Rosseland mean opacity, $\kappa_{\rm Ros}$, 
which is relevant to an optically thick case. We adopt the Rosseland mean opacity that is generalized to account for the effect of radiation pressure (e.g., eq.~7.25 in \citealt{LamersCassinelli99})
\begin{equation}
\frac{1}{\kappa_{\rm Ros}}=\frac{\pi}{a_{\rm BB} c T_{\rm BB}^3} \frac{1}{f_{\rm d}}\int_0^{\infty} \frac{1}{\alpha_{\lambda}}\frac{\dd B_\lambda(T_{\rm BB})}{\dd T_{\rm BB}} \dd\lambda,
\end{equation}
where $a_{\rm BB}=7.56\times10^{-15}$~erg~cm$^{-3}$~K$^{-4}$ is the energy density coefficient of blackbody 
emission, and $\alpha_{\lambda}$ is given by eq.~\ref{eq:alpha_lambda}. The integration is carried out over the wavelength range $10^{-3}\leq\lambda\leq10^3$~\mic. 

\begin{figure*}
	\includegraphics[width=2\columnwidth]{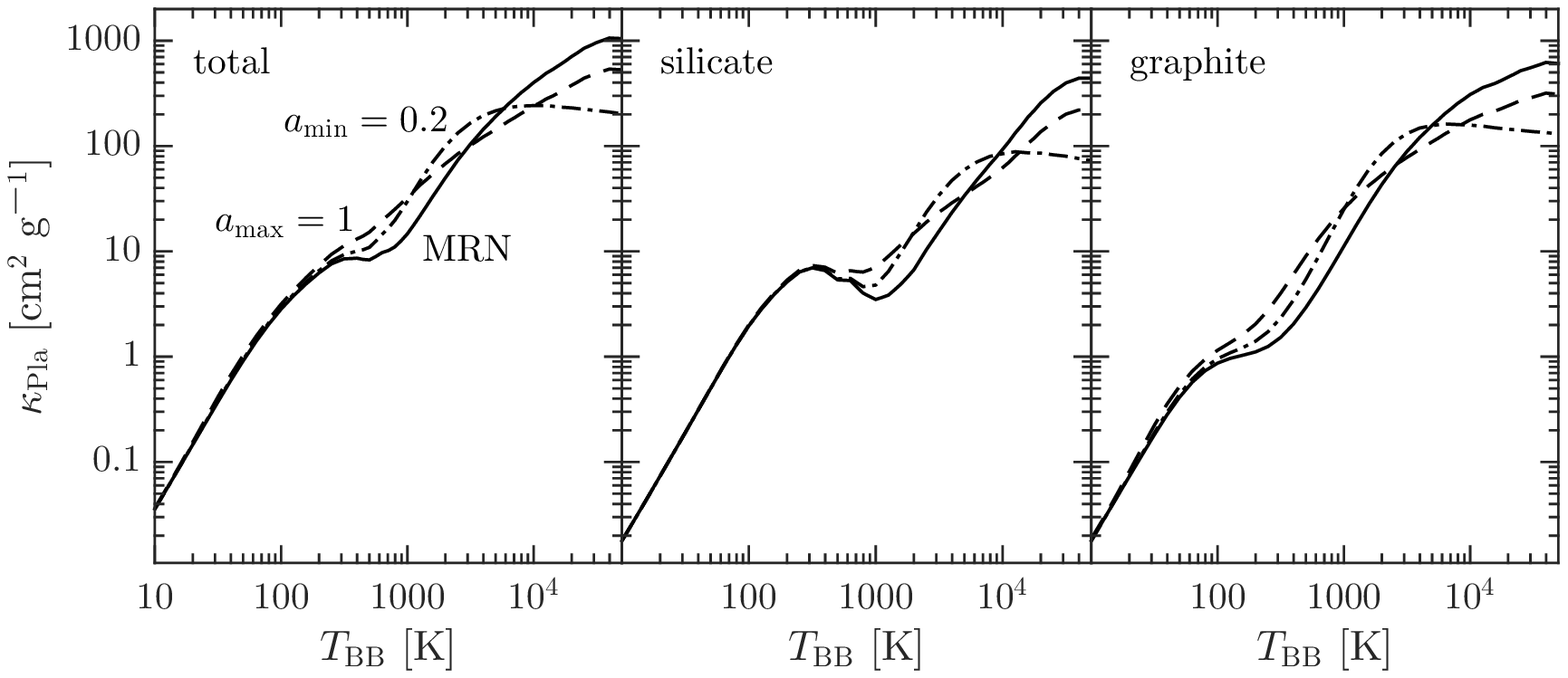}
	\caption{The Planck mean opacity for MRN dust composition (left panel), silicate grains only 
(middle panel), and graphite grains only (right panel), for different grain size distributions: 
the MRN grain-size distribution, i.e.\ $\amin=0.005$ and $\amax=0.25$~\mic\ (solid line), increasing 
\amin\ to 0.2~\mic\ (dot-dashed line) and increasing \amax\ to 1~\mic\ (dashed line), while holding other parameters at their MRN values. At low enough $T_{\rm BB}$ all grains are transparent, and the opacity becomes independent
of the grain size distribution. Increasing $\amin$ and $\amax$ reduces the high $T_{\rm BB}$ opacity, as the
fraction of small grains, which dominate the contribution to the UV opacity, is reduced.
For all three distributions, the dust opacity is dominated by graphite opacity above $T_{\rm BB}\simeq 1000$~K.}
	\label{fig:kappa_Pla_vs_T}
\end{figure*}

Figure~\ref{fig:kappa_Pla_vs_T} presents the dependence of $\kappa_{\rm Pla}$ on \Tbb, for 
$10\leq\Tbb\leq 5\times 10^4$~K. Three types of grain-size distribution are adopted: MRN ($\amin=0.005$ and $\amax=0.25$~\mic), MRN but with $\amin=0.2$~\mic, and MRN but with $\amax=1$~\mic. As noted in Sec.~\ref{sec:dust_op}, at $\Tbb\sim300$~K the dust opacity is dominated by the silicates, and for $\Tbb>1000$~K graphite dominates the opacity. This is true for all three types of grain size distribution. Increasing either \amin\ or \amax, increases the dust opacity by a few tens of per cents for $\Tbb>1000$~K. Below $\Tbb\sim50$~K, graphite and silicate opacity is similar and is roughly $\propto T_{\rm BB}^2$. For these low values of \Tbb, the blackbody emission is mostly at $\lambda>1$\mic, and $\lambda\gg a$ for most grains. Thus, the interaction between grains and radiation can be approximated by the electric dipole limit, which yields $\alpha_{\lambda}\propto \lambda^{-2}$ \citep{draine11}. Since the peak emission of black body is at $\lambda\propto T_{\rm BB}^{-1}$, the opacity should be roughly $\kappa_{\rm Pla}\propto T_{\rm BB}^2$, as indeed calculated (Fig.~\ref{fig:kappa_Pla_vs_T}).

\begin{figure*}
	\includegraphics[width=2\columnwidth, clip]{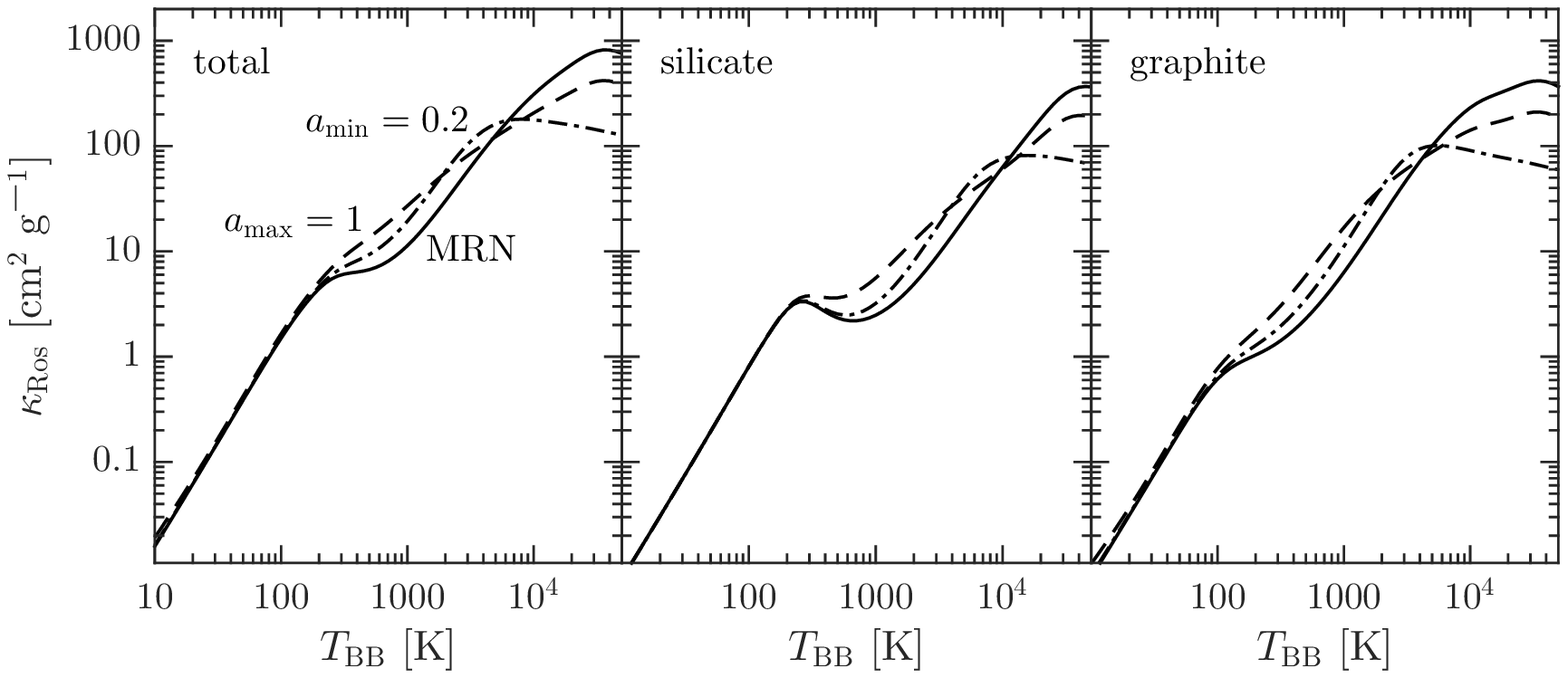}
	\caption{Same as Fig.~\ref{fig:kappa_Pla_vs_T} for the Rosseland mean opacity. The total dust opacity $\kappa_{\rm Ros}$ is dominated by the graphite opacity only from $T_{\rm BB}\simeq2000$~K, rather than from 1000~K as for $\kappa_{\rm Pla}$ (Fig.~\ref{fig:kappa_Pla_vs_T}). The only exception is the distribution with $\amin=0.2$~\mic, for which the silicate and graphite opacities become comparable again at $T_{\rm BB}\simeq 10^4$~K. Generally,
the values of $\kappa_{\rm Pla}$ and $\kappa_{\rm Ros}$ are similar, which reflects the relatively gradual change
of the absorption cross sections with $\lambda$ over the range of integration.}
	\label{fig:kappa_Ros_vs_T}
\end{figure*}

Figure~\ref{fig:kappa_Ros_vs_T} presents the dependence of $\kappa_{\rm Ros}$ on \Tbb, for the same three types of grain-size distribution as above. The dependence is similar to that of $\kappa_{\rm Pla}$ (see above). The main difference is that graphite dominates the opacity only above $\Tbb\simeq2000$~K, rather than above 1000~K. 

We fit the presented dependence of $\kappa_{\rm Pla}$ and $\kappa_{\rm Ros}$ on \Tbb\ in the range of $500<\Tbb<2500$~K which is relevant for the formation of the BLR. We adopt the following parametrization
\begin{equation}
\kappa=\xi T_{2000}^\delta \times Z/Z_{\sun}\mbox{~cm$^2$~gr$^{-1}$},  \label{eq:kappa_param}
\end{equation}
where $\xi$ and $\delta$ are the fitted coefficients. The coefficients are listed in Table~\ref{tab:kappa_coeff}.

\begin{table}
	\begin{minipage}{\columnwidth}
		\caption{The coefficients $\xi$ and $\delta$ of eq.~\ref{eq:kappa_param} for various assumptions of \amin\ and \amax\ for graphite.\fnrepeat{fn1:range}}\label{tab:kappa_coeff}
		\begin{tabular}{@{}{l}*{3}{c}{c}@{}}
			\hline
			\amin, \amax & \multicolumn{2}{c}{$\kappa_{\rm Pla}$} & \multicolumn{2}{c}{$\kappa_{\rm Ros}$}\\
			 & $\xi$ & $\delta$ & $\xi$ & $\delta$\\
			\hline
			0.005, 0.25 & 43 & 1.96\fnrepeat{fn1:slope} & 21 & 1.64 \\
			0.2, 0.25 & 91 & 1.97 & 40 & 1.80 \\
			0.005, 1 & 56 & 1.24 & 41 & 1.37 \\
			\hline
		\end{tabular}
	    \footnotetext[1]{The coefficients are fitted in the range of $500<T_{\rm BB}<2500$~K. \label{fn1:range}}
		\footnotetext[2]{The slight difference in $\delta$ compared to the value of 1.94 which is quoted in eq.~\ref{eq:kappa} is due to the different sampling in $T_{\rm BB}$ that is used for fitting the coefficients. The values in the table are derived using a sampling of 0.05~dex, while linear steps of 10~K are utilized for eq.~\ref{eq:kappa}.\label{fn1:slope}}
	\end{minipage}
\end{table}

In Table~\ref{tab:kappa_vs_TBB} we provide the values that are used to produce Figs~\ref{fig:kappa_Pla_vs_T} and \ref{fig:kappa_Ros_vs_T} for the MRN distribution (i.e.\ solid lines). Specifically, column (1) lists the value of $\log \Tbb$. Columns (2), (3) and (4) list the total, silicate and graphite $\kappa_{\rm Pla}$, respectively. Columns (5), (6) and (7) tabulate the total, silicate and graphite $\kappa_{\rm Ros}$, respectively. The values of \Tbb\ are in the range of $10\leq\Tbb\leq 5\times 10^4$~K in steps of 0.05~dex. 

\begin{table*}
	\begin{minipage}{2\columnwidth}
		\caption{The dependence of the Planck mean and the Rosseland mean dust opacity (in units of cm$^2$~gr$^{-1}$) on $T_{\rm BB}$ (in K), assuming the MRN grain-size distribution.\fnrepeat{fn2:electronic}}\label{tab:kappa_vs_TBB}
		\begin{tabular}{@{}{l}*{5}{c}{c}@{}}
			\hline
			 $\log T_{\rm BB}$ & \multicolumn{3}{c}{$\kappa_{\rm Pla}$} & \multicolumn{3}{c}{$\kappa_{\rm Ros}$}\\
			 & total & silicate & graphite &  total & silicate & graphite \\
			(1) & (2) & (3) & (4) & (5) & (6) & (7) \\
			\hline
			1.00 & 3.583E$-$02 & 1.788E$-$02 & 1.795E$-$02 & 1.584E$-$02 & 7.789E$-$03 & 8.049E$-$03 \\
			1.05 & 4.522E$-$02 & 2.255E$-$02 & 2.268E$-$02 & 1.982E$-$02 & 9.762E$-$03 & 1.006E$-$02 \\
			1.10 & 5.707E$-$02 & 2.843E$-$02 & 2.865E$-$02 & 2.483E$-$02 & 1.224E$-$02 & 1.258E$-$02 \\
			1.15 & 7.212E$-$02 & 3.595E$-$02 & 3.619E$-$02 & 3.112E$-$02 & 1.537E$-$02  & 1.576E$-$02 \\
			1.20 & 9.113E$-$02 & 4.546E$-$02 & 4.570E$-$02 & 3.904E$-$02 & 1.930E$-$02 & 1.975E$-$02 \\
			\hline
		\end{tabular}
		\footnotetext[1]{The full table is available online. \label{fn2:electronic}}
	\end{minipage}
\end{table*}

\bsp	% typesetting comment
\label{lastpage}
\end{document}